\documentclass{article}
\usepackage[top=1in, bottom=1in, left=1.25in, right=1.25in]{geometry}

\usepackage{lineno}

\newcommand*\patchAmsMathEnvironmentForLineno[1]{%
  \expandafter\let\csname old#1\expandafter\endcsname\csname #1\endcsname
  \expandafter\let\csname oldend#1\expandafter\endcsname\csname end#1\endcsname
  \renewenvironment{#1}%
     {\linenomath\csname old#1\endcsname}%
     {\csname oldend#1\endcsname\endlinenomath}}
\newcommand*\patchBothAmsMathEnvironmentsForLineno[1]{%
  \patchAmsMathEnvironmentForLineno{#1}%
  \patchAmsMathEnvironmentForLineno{#1*}}
\AtBeginDocument{%
  \patchBothAmsMathEnvironmentsForLineno{equation}%
  \patchBothAmsMathEnvironmentsForLineno{align}%
}

\usepackage[utf8]{inputenc} % allow utf-8 input
\usepackage[T1]{fontenc}    % use 8-bit T1 fonts
\usepackage{lmodern}
\usepackage{hyperref}       % hyperlinks
\usepackage{url}            % simple URL typesetting
\usepackage{booktabs}       % professional-quality tables
\usepackage{amsfonts}       % blackboard math symbols
\usepackage{microtype}      % microtypography
\usepackage[table,x11names]{xcolor}         % colors
\usepackage{tcolorbox}
\usepackage{graphicx} 
\usepackage{enumitem}
\usepackage{amsmath}
\usepackage{tabularx}
\usepackage{float}
\usepackage{multirow}
\usepackage{booktabs}
\usepackage{nameref,hyperref}
\usepackage[font=small,labelfont=bf,skip=4pt]{caption}

\title{On the Promises and Limits of Multi-omics Integration for Deconvolution: The HADACA3 Community Benchmark}
\author{%
\begin{minipage}{\textwidth}
 \centering
 \normalfont
 Hugo Barbot$^{*,1}$,
  Elise Amblard$^{*,2}$
  Nicolas Homberg$^{*,2}$,
  Lucie Lamothe$^{2}$,
  Morgane Térézol$^{3}$,
  HADACA3 Consortium$^{\dagger}$,
  Mira Ayadi $^{4}$,
  Aurélia Baurès $^{4}$,
  Lucile Armenoult $^{4}$,
  Yasmina Kermezli $^{2}$,
  Audrey Cassard $^{5}$,
  Jérôme Cros $^{5}$,
  Carl Herrmann $^{6}$,
  Sebastien Dejean$^{7}$,
  Lionel Spinelli$^{8}$,
  David Causeur$^{1}$,
    Florent Chuffart$^{9}$
  Anaïs Baudot$^{\S,3}$,
  Yuna Blum$^{\S,10}$,
  Magali Richard$^{\S,2}$
  \\[0.5em]
  \normalfont\small
  $^{1}$IRMAR CNRS, Institut Agro Rennes Angers\;
  $^{2}$Univ. Grenoble-Alpes, CNRS, LIG\;
  $^{3}$Aix-Marseille Univ, INSERM, MMG\;
  $^{4}$CIT, Ligue Nationale Contre le Cancer, Paris\;
  $^{5}$AP-HP, Beaujon Hospital, Univ. Paris Cité\;
  $^{6}$BioQuant, IPMB, Heidelberg Univ.\;
  $^{7}$IMT CNRS, Univ. Toulouse\;
  $^{8}$Aix-Marseille Univ, CNRS, CIML\;
  $^{9}$Univ. Grenoble-Alpes, INSERM, IAB\;
  $^{10}$IGDR CNRS, Univ. Rennes\;
  \\[0.3em]
  $^{*}$Co-first authors \quad $^{\S}$Co-senior authors \quad
  $^{\dagger}$Full consortium list available at the end of the article
   \end{minipage}
}
\usepackage{amssymb}
\usepackage{array}

\begin{document}

\maketitle
%\modulolinenumbers[5]
%\linenumbers  % désactivé pour arXiv

\begin{abstract}
  %The abstract paragraph should be indented \nicefrac{1}{2}~inch (3~picas) on both the left- and right-hand margins. Use 10~point type, with a vertical  spacing (leading) of 11~points.  The word \textbf{Abstract} must be centered, bold, and in point size 12. Two line spaces precede the abstract. The abstract must be limited to one paragraph.

Understanding the cellular composition of complex tissues, such as tumours, is a key challenge in biology and medicine. Deconvolution methods estimate cell-type proportions from bulk molecular measurements, and the increasing availability of multiple molecular modalities has raised the expectation that their integration should improve estimation accuracy. However, whether and under which conditions multi-omics integration actually improves deconvolution remains unclear.
Here we present HADACA3, a community-driven competition followed by a modular benchmark of more than 18,000 pipelines combining pre-processing, feature selection, deconvolution and integration of RNA and DNA methylation (DNAm), across nine pancreatic cancer datasets. 
With the available references, DNAm alone achieves the highest median performance, outperforming RNA alone and most integration strategies.
Dataset-specific optimisation of generic integration pipelines rarely improves upon DNAm alone: the best configurations mostly reduce the weight of RNA, and their gains do not generalise across datasets. 
The only consistent improvement, confirmed on unseen data, comes from an integration rule that estimates cancer cell types from DNAm alone and averages both modalities for non-cancer cell types. 
These findings challenge the assumption that integrating more molecular data necessarily improves deconvolution. 
Instead, effective multi-omics integration relies on prior knowledge of the cell-type-specific strengths of each modality, rather than on generic statistical integration strategies.
\end{abstract}

\section*{Introduction} %9pages max expect title and figures

Complex tissues are composed of heterogeneous cell populations, and their cellular composition shapes tissue function and disease. In cancer, tumours combine cancer cells with stromal and immune cells, whose relative proportions influence tumour progression, prognosis and response to therapy~\cite{hanahan_hallmarks_2026}. Bulk molecular measurements, however, capture an average signal over all these populations, which masks their individual contributions. Cellular deconvolution aims to estimate the proportions of the underlying cell populations from such bulk data, including transcriptomic (RNA) and DNA methylation (DNAm) profiles.

Supervised deconvolution methods estimate cell-type proportions by fitting bulk molecular profiles against reference profiles obtained from purified cell populations. Early reference-based methods for RNA, such as CIBERSORT~\cite{newman_robust_2015}, and for DNAm, such as the approach introduced by Houseman et al.~\cite{houseman_dna_2012}, laid the foundations of the field. 
Since then, extensive methodological developments have produced a broad range of approaches, spanning constrained linear regression, Bayesian models and machine learning methods~\cite{de_ridder_benchmarking_2024, avila_cobos_benchmarking_2020, ferro_dos_santos_computational_2024}; for DNAm alone, approximately 25 methods are now available, spanning supervised and unsupervised strategies~\cite{ferro_dos_santos_computational_2024}. 
For RNA, references are increasingly derived from single-cell RNA-seq data~\cite{dietrichOmnideconvUnifyingFramework2026a}, whereas comparable single-cell references remain scarce for DNAm.  
Despite this methodological diversity, deconvolution performance remains strongly dependent on several interrelated factors, including reference data quality, pre-processing choices, feature selection and collinearity between reference profiles~\cite{vathrakokoili_pournara_catd_2024, de_ridder_benchmarking_2024}. Comparing deconvolution methods in isolation is therefore insufficient: the choices made at each step of the analysis pipeline must be evaluated jointly. Benchmarks must in addition use performance metrics that reflect biologically meaningful estimation accuracy, and datasets that capture realistic biological variability, so that methods are evaluated on data whose noise structure, cellular composition and experimental setting differ from those on which they were developed~\cite{wolfram-schauerte_approaching_2025, amblard_robust_2024}.

Integrating multiple omics modalities has been proposed as a way to improve performance across a range of computational biology tasks. 
The underlying premise is that partially independent molecular layers carry complementary information, which an integration strategy can combine into a more accurate estimate than either layer alone. Integration strategies are broadly categorized into three paradigms: early integration, where both modalities are combined at the feature level prior to analysis; intermediate integration, where the integration is tightly coupled to the task-specific model; and late integration, where modality-specific results are combined at the output level~\cite{cai_machine_2022}. Algorithmic approaches for multi-omics integration range from covariance and matrix factorization methods to probabilistic models, kernel-based methods, and deep learning~\cite{baiao_technical_2025}. 
Whatever their algorithmic form, most of them combine modalities according to generic statistical criteria, derived from the data alone rather than from biological knowledge, such as the variance shared between modalities, a common scaling, or the overall reconstruction error of each modality. In cellular deconvolution, however, the informativeness of each modality may differ across cell types, and whether such generic criteria can exploit this complementarity remains an open question. At the time of our study, multi-omics integration for bulk tissue deconvolution had not been systematically benchmarked: previous benchmarks evaluated RNA~\cite{avila_cobos_benchmarking_2020} or DNAm~\cite{de_ridder_benchmarking_2024} deconvolution methods alone, or both in parallel without integrating them~\cite{amblard_robust_2024}.

The HADACA initiative has progressively addressed the evaluation of deconvolution methods through community-driven scientific competitions. The first edition focused on unsupervised approaches for DNAm data, providing initial guidelines and identifying key factors influencing performance~\cite{hadaca_consortium_guidelines_2020}. 
The second edition extended the challenge to unsupervised methods for both RNA and DNAm data, and led to the DECONbench platform~\cite{decamps_deconbench_2021}. DECONbench was designed to support continuous benchmarking: its datasets and scoring remained available after the challenge, so that new methods could be compared with previous ones under identical conditions. 
However, neither edition systematically addressed RNA–DNAm integration, and the scope of the benchmarks remained limited in terms of dataset diversity and exploration of pre-processing strategies. 
HADACA3 extends the previous editions by introducing i) a multi-omics integration module, ii) a four-module pipeline enabling systematic evaluation of option combinations, and iii) a more diverse collection of datasets. HADACA3 focuses on early and late integration strategies, two widely used paradigms that had not yet been systematically compared in the context of bulk tissue deconvolution. 
Since the HADACA3 competition, several methods dedicated to multi-omics deconvolution have emerged, including EMixed~\cite{caiEMixedProbabilisticMultiOmics2025}, MOADE~\cite{sunMOADEMultimodalAutoencoder2025}, and DECODE~\cite{zhaoDECODEDeepLearningbased2026}. Of these, EMixed and MOADE are directly applicable to RNA and DNAm data while DECODE does not yet support DNAm. These recent developments further highlight the growing interest in multi-omics deconvolution and the need for systematic evaluation of integration strategies. 

Our contributions can be summarized as follows. We introduce HADACA3 (Figure~\ref{fig:competition}), a community-driven competition followed by a modular benchmark of bulk tissue deconvolution, which compares RNA--DNAm integration strategies with uni-omic strategies and identifies which modules drive performance (Box~1). 
To this end, we assemble nine pancreatic ductal adenocarcinoma (PDAC) datasets~\cite{halbrook_pancreatic_2023} with different levels of ground truth: \textit{in silico} simulations with known cell-type proportions, \textit{in vitro} mixtures of purified cell populations, and \textit{in vivo} tumour samples whose composition is estimated from histology. PDAC is a demanding test case for deconvolution, as its abundant and diverse stroma and its closely related cancer cell types make cellular composition particularly difficult to resolve from bulk profiles.
With the RNA and DNAm reference profiles available for PDAC, deconvolution based on DNAm data alone achieves the highest median performance. Integration strategies based on generic statistical criteria outperform it only when the pipeline is re-optimised for each dataset, and the only consistent gain, confirmed on unseen data, comes from an integration strategy that incorporates prior knowledge of cell-type-specific modality strengths.
To support future work, we provide a reproducible Nextflow workflow implementing the four modules, and publicly release all datasets, pipelines and benchmarking infrastructure through Codabench. The benchmark remains open after the competition, providing an extensible resource for evaluating new methods under the same conditions and for continuous, community-driven method development.

%Since our competition, several methods have been proposed specifically for multi-omics deconvolution, including EMixed~\cite{caiEMixedProbabilisticMultiOmics2025}, MOADE~\cite{sunMOADEMultimodalAutoencoder2025}, and DECODE~\cite{zhaoDECODEDeepLearningbased2026}. Of these, EMixed and MOADE are directly applicable to RNA and DNA methylation data while DECODE does not yet support DNA methylation. These recent developments further highlight the growing interest in multi-omics deconvolution and the need for systematic evaluation of integration strategies. 

\section*{Results}

\subsection*{HADACA3: a community-driven competition to assess multi-omics integration for cell-type deconvolution.}
\label{sec:competition}

%\paragraph{Task and data overview}

\begin{figure}[!ht]
%\centering
\includegraphics[width=\linewidth]{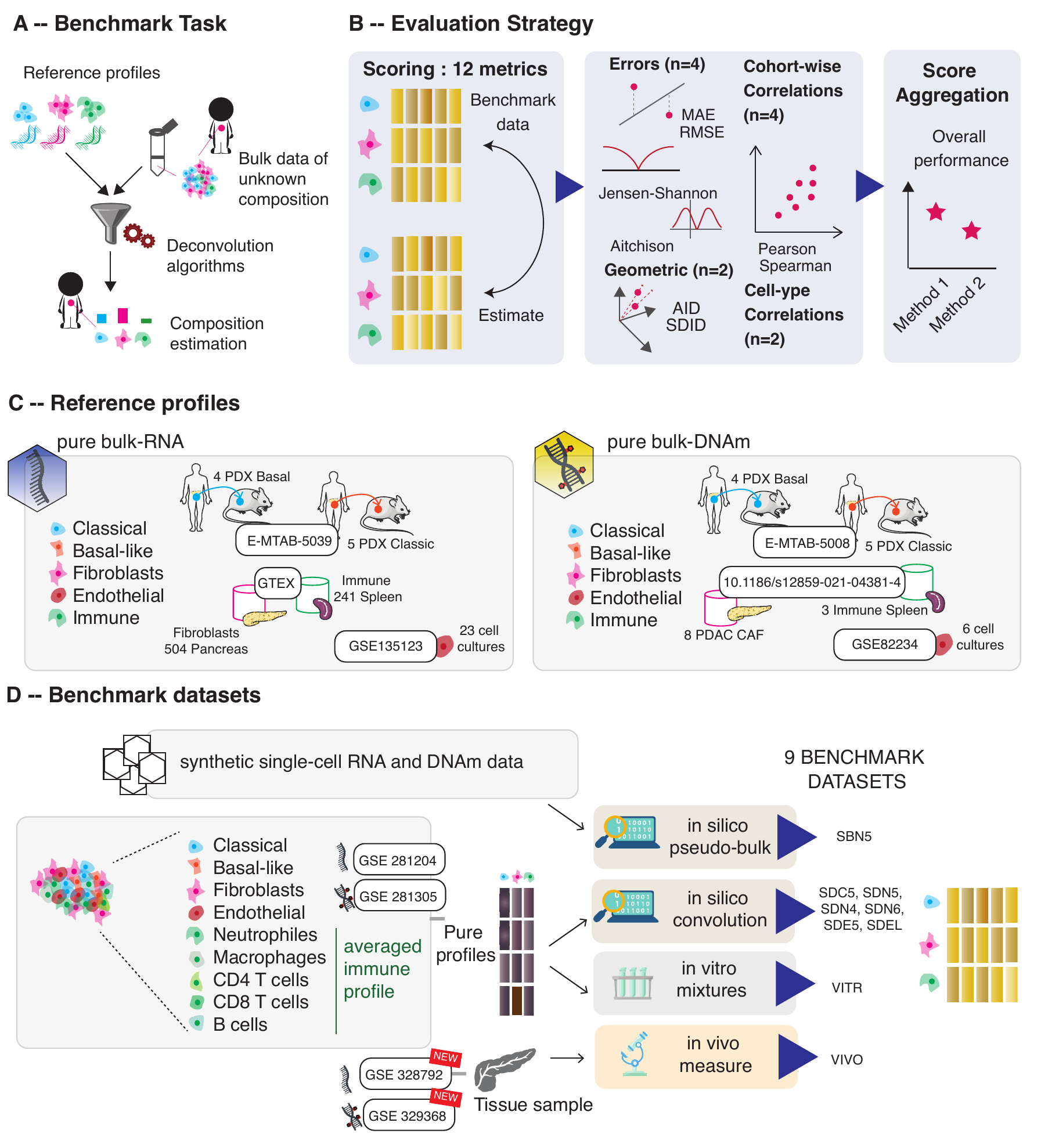}
\caption{\textbf{Schematic representation of the competition setup.} \textbf{A.} Deconvolution problem: a bulk molecular profile (RNA or DNAm) of a tissue sample is modelled as a mixture of cell-type-specific reference profiles, and deconvolution estimates the proportion of each cell type in the sample. \textbf{B.}  Estimated proportions are compared with the ground-truth proportions using twelve metrics grouped into four families. Each metric is normalized to $[0,1]$, metrics are combined by geometric mean within each family, and the four families are combined with equal weights into the aggregate score. \textbf{C.} Bulk RNA and DNAm profiles of the five cell types (classical and basal-like tumour cells, fibroblasts, endothelial cells and immune cells) were derived from publicly available datasets of purified or sorted cell populations  and are distinct from the profiles used to simulate the benchmark datasets. \textbf{D.} Benchmark datasets and their ground truth. \emph{In silico} datasets are simulated by mixing cell-type profiles with known proportions, under different noise models and compositions; \emph{in vitro} datasets are mixtures of purified cell populations with controlled proportions; \emph{in vivo} tumour samples have ground-truth proportions of the two cancer cell types estimated from histology. See ~\nameref{methods} for more details on metrics and datasets.}
\label{fig:competition}
\end{figure}

HADACA3 was a four-day on-site competition hosted on the Codabench platform (December 2--6 2024), which brought together ten teams of four participants \cite{hadaca3_codabench}. 
The objective was to perform supervised deconvolution of bulk multi-omics data (Figure~\ref{fig:competition}A and Equation~\eqref{eq:deconv}): for each sample $i$, given its bulk profiles $Y_i^{(m)} \in \mathbb{R}^{F^{(m)}}$ and the reference profiles $X^{(m)} \in \mathbb{R}^{F^{(m)} \times k}$, where $m \in \{\mathrm{RNA}, \mathrm{DNAm}\}$ denotes the modality, $F^{(m)}$ its number of features and $k$ the number of reference cell types, teams were asked to estimate a proportion vector $p_i \in \mathbb{R}^k$ shared by the two modalities, whose entries $p_{ic}$ give the proportion of cell type $c$ and satisfy:
\begin{equation}
Y_i^{(m)} \approx X^{(m)} p_i, \quad \forall m, 
\quad p_i \geq 0, \quad \sum_{c=1}^{k} p_{ic} = 1.
\label{eq:deconv}
\end{equation}

Deconvolution performance was assessed using twelve complementary metrics grouped into four families (Figure~\ref{fig:competition}B). The different families of evaluation metrics capture distinct aspects of the performance, reflecting  (i) the relative intra-sample variation (global and sample-wise correlations, both Pearson and Spearman); (ii) the inter-sample variation (cell-type Pearson and Spearman correlation); (iii) the ability to accurately estimate the magnitude of true proportions (error metrics: RMSE, MAE, Aitchison distance, Jensen-Shannon divergence); and (iv) the alignment between predicted and true compositional vectors independently of their overall magnitude (geometric metrics: AID, SDID). Each metric was normalized to $[0,1]$ and combined into a single aggregate score via a weighted geometric mean, with each family contributing equally one quarter of the total score (Supplementary Material ~\ref{suppmat:metric} and Supplementary Table~\ref{supptab:metrics}).

Participants were provided with pure bulk-RNA, pure bulk-DNAm of pure cell types representing five major cell-types found in pancreatic tumours: three non-cancer cell types (immune cells, fibroblasts, endothelial 
cells) and two cancer cell types (classical and basal-like) 
(Figure~\ref{fig:competition}C and Supplementary Material~\ref{suppmat:ref}). 

We assembled nine benchmark datasets spanning three experimental settings 
(Figure~\ref{fig:competition}D, Table~\ref{tab:bench_datasets}), 
designed to evaluate both accuracy and robustness of deconvolution methods 
under realistic biological and technical conditions. The feature space was restricted to the intersection of RNA and DNAm features available across all benchmark datasets and reference profiles, yielding a common set of 15,908 RNA and 23,724 DNAm features (Supplementary Material~\ref{suppmat:feature}).

\begin{table*}[!h]
\caption{Summary of the benchmark datasets. $\alpha$: proportion sampling 
scheme ($\alpha_{\text{real}}$: Dirichlet with realistic parameters; 
$\alpha_{\text{rare}}$: rare cell type scenario). 
$\varepsilon$: noise model applied independently to each modality 
(—: not applicable for real datasets).
\label{tab:bench_datasets}}
\tabcolsep=0pt
\begin{tabular*}{\textwidth}{@{\extracolsep{\fill}}ccccccc@{\extracolsep{\fill}}}
\toprule
Name & Source & Samples & Cell types & $\alpha$ & 
$\varepsilon^{\text{RNA}}$ & $\varepsilon^{\text{DNAm}}$ \\
\midrule
VITR & \textit{in vitro}  & 15 & 5 & — & — & — \\
VIVO & \textit{in vivo}   & 24 & 2 & — & — & — \\
SBN5 & \textit{in silico} & 30 & 5 & $\alpha_{\text{real}}$ & pseudo-bulk & pseudo-bulk \\
SDN5 & \textit{in silico} & 30 & 5 & $\alpha_{\text{real}}$ & $\chi^2$ & Gaussian \\
SDN4 & \textit{in silico} & 30 & 4 & $\alpha_{\text{real}}$ & $\chi^2$ & Gaussian \\
SDN6 & \textit{in silico} & 30 & 6 & $\alpha_{\text{real}}$ & $\chi^2$ & Gaussian \\
SDE5 & \textit{in silico} & 30 & 5 & $\alpha_{\text{real}}$ & EM & EM \\
SDEL & \textit{in silico} & 30 & 5 & $\alpha_{\text{rare}}$ & EM & EM \\
SDC5 & \textit{in silico} & 30 & 5 & $\alpha_{\text{real}}$ & Copula+NB & Copula+Beta \\
\bottomrule
\end{tabular*}
\end{table*}

Among the 627 submissions received during the four days of the competition, the JOKER submission achieved the best overall performance (results available on Codabench~\cite{hadaca3_codabench}). JOKER deconvolved each modality separately and then combined the two sets of estimates with a simple rule: it used the DNAm estimates for the two cancel cell types (classical and basal-like) and averaged the RNA and DNAm estimates for the other cell types (Supplementary Material~\ref{suppmat:JOKER}). The team designed this rule after observing, on the training data available during the competition, that DNAm gave more stable estimates for the cancer cell types. JOKER performed consistently across benchmark datasets, with similar distributions of absolute bias (Supplementary Figure~\ref{suppfig:bestsubmission}A), but its errors varied across cell types, being higher for the basal-like subtype and fibroblasts (Supplementary Figure~\ref{suppfig:bestsubmission}B). As a single competition submission, JOKER couples its pre-processing, feature selection, deconvolution algorithm and integration rule, whose respective contributions cannot be assessed separately. This naturally raises two questions: which of these choices actually drives performance, and does the benefit of its integration rule generalise to other settings?

\subsection*{Modular benchmarking of multi-omics integration for bulk tissue deconvolution}
\label{sec:benchmark}

\begin{figure}[!ht]%
\centering
\includegraphics[width= \linewidth]{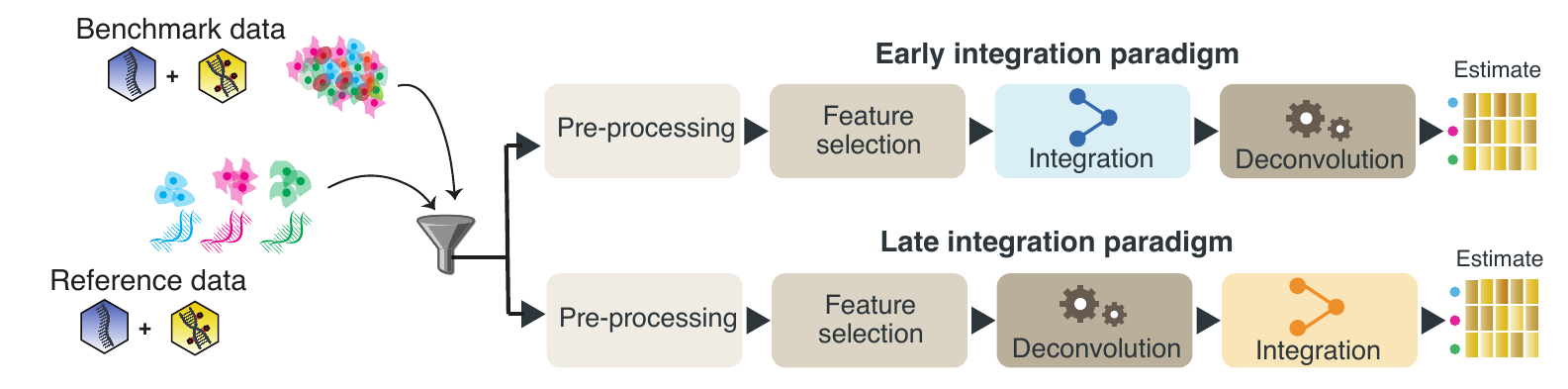}
\caption{Overview of the modular benchmark framework. The competition submissions were decomposed into four main modules: pre-processing, feature selection, deconvolution, and integration (early or late). The two integration paradigms differ in the order of the last two modules. In \emph{early integration} (top), the RNA and DNAm data are first combined by the integration strategy into a joint representation, which is then deconvolved once. In \emph{late integration} (bottom), each modality is deconvolved separately, and the two sets of estimated proportions are then combined by the integration strategy. All compatible combinations of module options were systematically evaluated.}
\label{benchmark}
\end{figure}

%\paragraph{Objective and design}

To fairly evaluate multi-omics integration strategies and identify which modules drive overall performance, we decomposed JOKER and other top submissions into four key modules (Figure~\ref{benchmark} and Box~1): (i) pre-processing, (ii) feature selection, (iii) deconvolution and (iv) multi-omics integration. Each pipeline combines one integration strategy with one option per modality for pre-processing, feature selection and, in late integration, deconvolution, or a single deconvolution option in early integration (Table~\ref{tab:modules}); within each modality, the mixtures and the reference profiles undergo the same pre-processing and feature selection. We evaluated all compatible combinations on the nine benchmark datasets (Table~\ref{tab:bench_datasets}) with the aggregate score defined above, totalling 4,050 early integration and 14,580 late integration pipelines, using an open-source Nextflow workflow that we developed and made available on GitHub~\cite{hadaca3_framework}. This design enables us to quantify the contribution of each module to performance, to identify robust strategies across settings, and to assess whether multi-omics integration consistently improves over uni-omic strategies.

\begin{table}[ht]
\centering
\caption{Options available in each module of the benchmark. Pre-processing and feature selection are applied separately to each modality. A pipeline combines one integration strategy with one option per module for each modality. $^{\dagger}$Deconvolution is applied once to the combined data in early integration and to each modality in late integration. All options are described in Supplementary Material~\ref{supp:benchmark}.}
\label{tab:modules}
\footnotesize
\begin{tabularx}{\linewidth}{p{2.5cm} p{1.2cm} p{2.5cm} p{7.3cm}}
\toprule
Module & Modality & Option & Brief description \\
\midrule
\multirow{3}{=}{Pre-processing} & \multirow{3}{=}{RNA, DNAm} & \texttt{ppID} & None (raw values) \\
 & & \texttt{SumNorm} & Column-sum normalization \\
 & & \texttt{LogNorm} & Log-normalization (Seurat) \\
\midrule
\multirow{8}{=}{Feature selection} & \multirow{3}{=}{RNA} & \texttt{fsID} & None (all features kept) \\
 & & \texttt{Toastbulknbfs} & Top 1,000 markers (via TOAST on reference) \\
 & & \texttt{Toastvst} & Top 1,000 markers (via TOAST on VST-transformed ref.) \\
\cmidrule(l){2-4}
 & \multirow{5}{=}{DNAm} & \texttt{fsID} & None (all features kept) \\
 & & \texttt{Toastpercent} & Top 80\% probes (TOAST) \\
 & & \texttt{mostmethylated} & Probes above 75th percentile per cell type \\
 & & \texttt{maxdiscriminant} & Most discriminant non-overlapping probes \\
 & & \texttt{splsda} & Sparse PLS-DA on logit reference \\
\midrule
\multirow{6}{=}{Deconvolution} & \multirow{6}{=}{RNA, DNAm$^{\dagger}$} & \texttt{lm} & Ordinary least squares \\
 & & \texttt{nnls} & Non-negative least squares \\
 & & \texttt{epic} & Constrained least squares (Non-neg. and sum-to-one)\\
 & & \texttt{RLR} & Robust linear regression (IRLS) \\
 & & \texttt{wRLR} & RLR with Poisson-inspired weights \\
 & & \texttt{RLRnnls} & RLR or NNLS, chosen by reconstruction RMSE \\
\midrule
\multirow{13}{=}{Integration} & \multirow{13}{=}{---} & \multicolumn{2}{l}{\emph{Early integration strategies}} \\
 & & \texttt{Concat} & Raw feature concatenation \\
 & & \texttt{ConcatStd} & Normalized concatenation \\
 & & \texttt{MCIA} & Latent linear embedding (MCIA) \\
 & & \texttt{MKL} & Non-linear kernel embedding (kernel PCA) \\
 & & \texttt{OT} & Optimal transport-based representation (uniPort) \\
 & & \multicolumn{2}{l}{\emph{Late integration strategies}} \\
 & & \texttt{Mean} & Uniform averaging \\
 & & \texttt{wMean} & Error-weighted averaging \\
 & & \texttt{PriorMean} & Rule-based selective averaging \\
 & & \multicolumn{2}{l}{\emph{Uni-omic strategies}} \\
 & & \texttt{onlyRNA} & RNA only \\
 & & \texttt{onlyDNAm} & DNAm only \\
\bottomrule
\end{tabularx}
\end{table}

% préambule : \usepackage[most]{tcolorbox} et \usepackage{tabularx}
\begin{tcolorbox}[title={\textbf{Box 1. Vocabulary}}, colback=gray!5, colframe=gray!60,
  fonttitle=\footnotesize, fontupper=\footnotesize, boxrule=0.4pt, arc=1pt,
  left=3pt, right=3pt, top=2pt, bottom=2pt]
\begin{tabularx}{\linewidth}{@{}>{\bfseries}l X@{}}
Module & Pre-processing, feature selection, deconvolution or integration. \\
%Sub-module & A module applied to one modality (e.g.\ RNA pre-processing). \\
Option & A choice within a module (e.g.\ \texttt{SumNorm}, \texttt{RLR}; Table~\ref{tab:modules}). \\
Paradigm & Early (one joint deconvolution) or late (one per modality) integration. \\
Strategy & An integration option (\texttt{Concat} \ldots\ \texttt{PriorMean}) or a uni-omic one (\texttt{onlyDNAm}, \texttt{onlyRNA}). \\
Pipeline & A complete combination of one option per module and one strategy \\
RNA / DNAm side & The RNA (resp.\ DNAm) options of a pipeline. \\
Counterpart & The \texttt{onlyDNAm} pipeline sharing the DNAm side of an integrated pipeline. \\
Best pipeline & The top-scoring pipeline of a strategy in a given dataset. \\
Generalist pipeline & The top-scoring pipeline of a strategy on average across datasets. \\
Unseen data & Datasets marked $'$, never released to participants nor used for selection. \\
\label{box}
\end{tabularx}
\end{tcolorbox}

%\paragraph{Early integration methods}

Early integration strategies  consist of combining RNA and DNAm data at the feature level prior to deconvolution (Eq.~\eqref{eq:early}). In this setting, a joint representation $\tilde{X}$ and $\tilde{Y}$ is constructed:
\begin{equation}
\tilde{Y}, \tilde{X} = \mathcal{F_\text{early}}\left(Y^{\text{RNA}}, Y^{\text{DNAm}}, X^{\text{RNA}}, X^{\text{DNAm}}\right),
\label{eq:early}
\end{equation}
where $\mathcal{F_\text{early}}$ denotes a transformation specific to each strategy. Deconvolution is then performed on this joint representation, as in Eq.~\eqref{eq:deconv} but with $\tilde{X}$ and $\tilde{Y}_i$ in place of $X^{(m)}$ and $Y_i^{(m)}$. Early integration strategies fall into feature-level, latent and transport-based families, which all aim to improve the estimation of the shared proportion vector $p_i$ by reducing cross-modal discrepancies between RNA and DNAm (Supplementary Table \ref{supptab:early_integration_summary} and Supplementary Material~\ref{suppmat:integrationEI}).

%\paragraph{Late integration methods}

Late integration strategies (Eq.~\eqref{eq:late}) instead solve Eq.~\eqref{eq:deconv} separately for each modality and then combine the two estimates.
\begin{equation}
Y_i^{\text{RNA}} \approx X^{\text{RNA}} \hat{p}_i^{\text{RNA}}, 
\quad
Y_i^{\text{DNAm}} \approx X^{\text{DNAm}} \hat{p}_i^{\text{DNAm}},
\quad
\hat{p}_i = \mathcal{F_\text{late}}\big(\hat{p}_i^{\text{RNA}}, \hat{p}_i^{\text{DNAm}}\big),
\label{eq:late}
\end{equation}

where $\mathcal{F_\text{late}}$ is an aggregation function that varies across methods (Supplementary Table \ref{supptab:late_integration_summary} and Supplementary Material~\ref{suppmat:integrationLI}). The three strategies differ only in how the two estimates are combined. \texttt{Mean} averages them with equal weights; \texttt{wMean} weights each modality by the inverse of its reconstruction error, a generic statistical criterion; and \texttt{PriorMean}, derived from the winning competition submission (JOKER), applies a rule based on biological prior knowledge: it takes the proportions of the two cancer cell types (classical and basal-like) from DNAm alone, and averages both modalities for the other cell types.

\subsection*{DNAm only deconvolution achieves the highest median performance across datasets.}

Because VIVO is scored on a partial ground truth (Supplementary Material~\ref{suppmat:metric}), we first describe results on the 8 fully-annotated datasets and discuss VIVO separately.

Figure~\ref{fig:robust}A shows that \texttt{onlyDNAm} and \texttt{wMean} achieve the highest median aggregate scores across pipelines (0.70), whereas the other integration strategies and \texttt{onlyRNA} reach lower medians. The proximity of \texttt{wMean} to \texttt{onlyDNAm} is expected from its design: weighting each modality by its reconstruction error leaves little weight to the RNA. This advantage is stable across datasets: \texttt{onlyDNAm} achieves the highest median performance on 7 of the 8 fully-annotated datasets (Figure~\ref{fig:robust}B). 

It is also statistically supported: on the fully-annotated datasets, \texttt{onlyDNAm} is significantly better than the integration strategies in 54 of the 64 comparisons (paired Wilcoxon test, Figure~\ref{fig:robust}C), and 9 others are not significant. Only one integration strategy significantly outperforms it, \texttt{wMean} in SDE5, and by a small margin (+0.011 in median aggregate score).
Strategy rankings based on median performance are similarly stable across the fully-annotated datasets, despite their different noise models, cell-type compositions and abundances, and simulation procedures (Figure~\ref{fig:robust}D).
\texttt{onlyDNAm} ranks first or second in each of the 8 fully-annotated datasets, and the rank of every strategy varies by one position or less across them. This suggests that strategy rankings are robust to substantial changes in simulation design and noise structure. 

Rankings do, however, depend on the metric family considered (Supplementary Figure~\ref{suppfig:rank4}): \texttt{onlyDNAm} remains competitive on most of them (cohort-wise correlation, error, and geometric metrics) which explains its strong overall performance. Cell-type correlation behaves differently: the best strategies lie within a few hundredths of each other, so that small differences reorder them. This family is also the hardest to optimise, with lower inter-sample correlations in datasets with more complex noise structures, reflecting the difficulty of recovering the variation of each cell type across samples. Together with the Aitchison distance, it is the most discriminative component of the aggregate score (Supplementary Figure~\ref{suppfig:radar}).

This metric dependence is key to interpreting VIVO, where the ranking is reversed: seven of the eight integration strategies significantly outperform \texttt{onlyDNAm}, with \texttt{onlyRNA} ranking first and \texttt{onlyDNAm} last. VIVO's partial ground truth restricts scoring to inter-sample cell-type correlation for the two cancer cell types, that is, to how well each strategy orders the samples according to their proportion of each subtype. Its ranking therefore reflects a single, narrow aspect of performance, and precisely the one where RNA helps: in VIVO, RNA alone orders the samples better than DNAm alone, even though DNAm alone estimates their absolute proportions more accurately in the other datasets. Consistently, scoring the other datasets with cell-type correlation only also reshuffles their rankings, which then differ depending on whether this correlation is computed on the two cancer cell types or on all five cell types (Supplementary Figure~\ref{suppfig:rankcor}). The VIVO ranking should therefore be interpreted as complementary to, rather than representative of, overall performance.

\begin{figure}[!ht]%
\centering
\includegraphics[width=\linewidth]{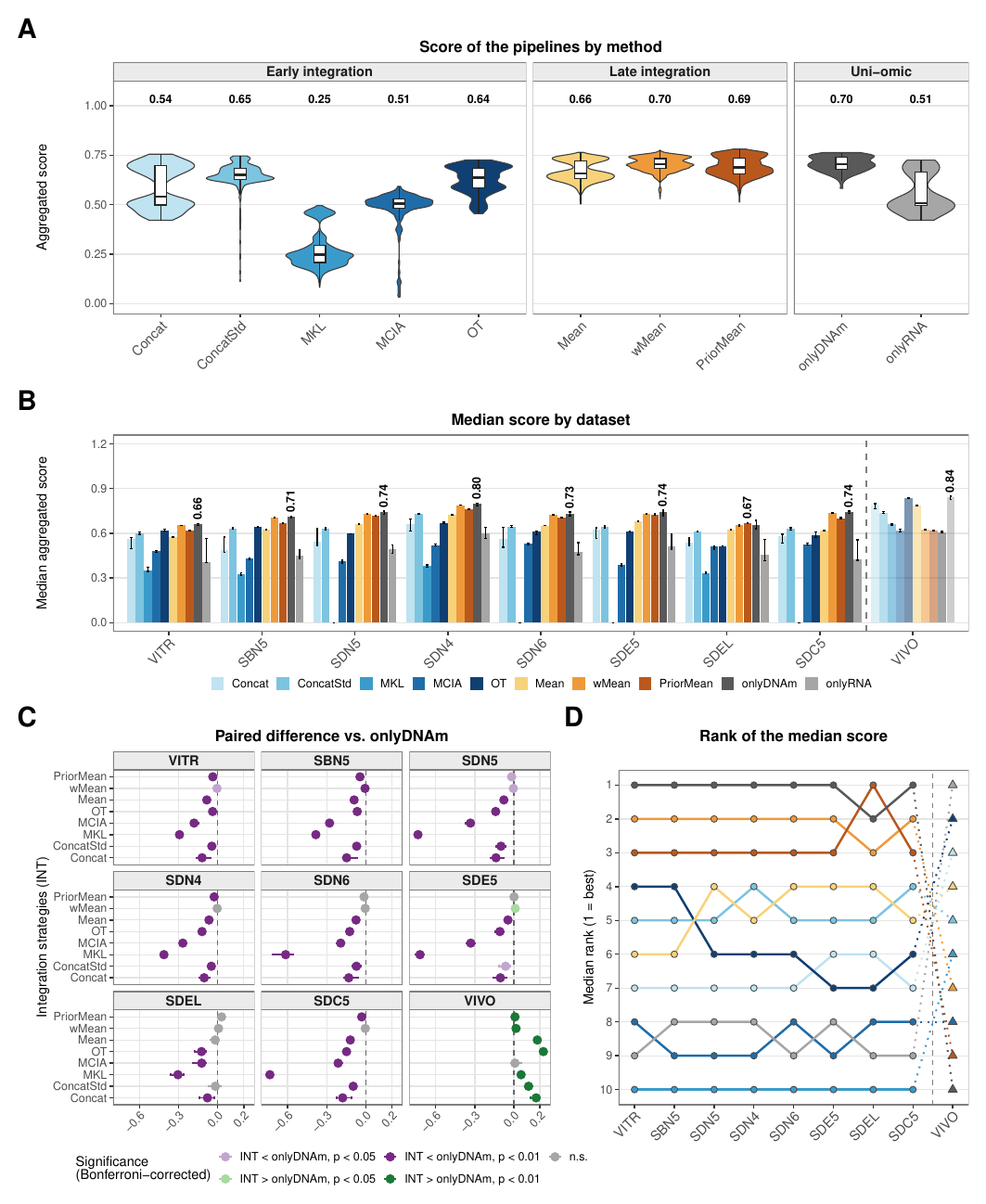}
\end{figure}

\begin{figure}[!ht]%
\caption{\textbf{Median performance of integration strategies across datasets.} 
\textbf{A.} Median aggregated performance across integration strategies. Early integration strategies include $N = 810$ pipelines per strategy, late integration strategies include $N = 4,860$ pipelines per strategy, and uni-omic performs deconvolution using a single omic independently,  with $N =90$ combinations for \texttt{onlyDNAm} and $N =54$ combinations for \texttt{onlyRNA}. Median aggregate scores are indicated. 
\textbf{B.} Median aggregate benchmark scores for each integration method across datasets. Error bars show 95\% bootstrap on the median across configuration (see Supplementary Material~\ref{suppmat:bootstrap}). 
For each dataset, the median score of the best-performing strategy is reported above the corresponding bar. 
\textbf{C.} Paired comparison of \texttt{onlyDNAm} versus each RNA-DNAm integration strategy, per dataset (see Supplementary Material~\ref{suppmat:wilcoxon}). 
Coloured points indicate significant differences after Bonferroni correction. Bars show the 95\% confidence interval from the paired Wilcoxon signed-rank test.
\textbf{D.} Rank of each strategy among the ten strategies, based on its median aggregate score, in each dataset. In \textbf{B} and \textbf{D}, VIVO is separated from the other 8 datasets by a dashed vertical line, since its aggregate score is computed with a different formula than the other datasets (see Supplementary Material \ref{suppmat:partialgroundtruth}). }
\label{fig:robust}
\end{figure}

\subsection*{Early and late integration differ in their main performance determinants.}

\begin{figure}[!ht]%
\centering
\includegraphics[width=\linewidth]{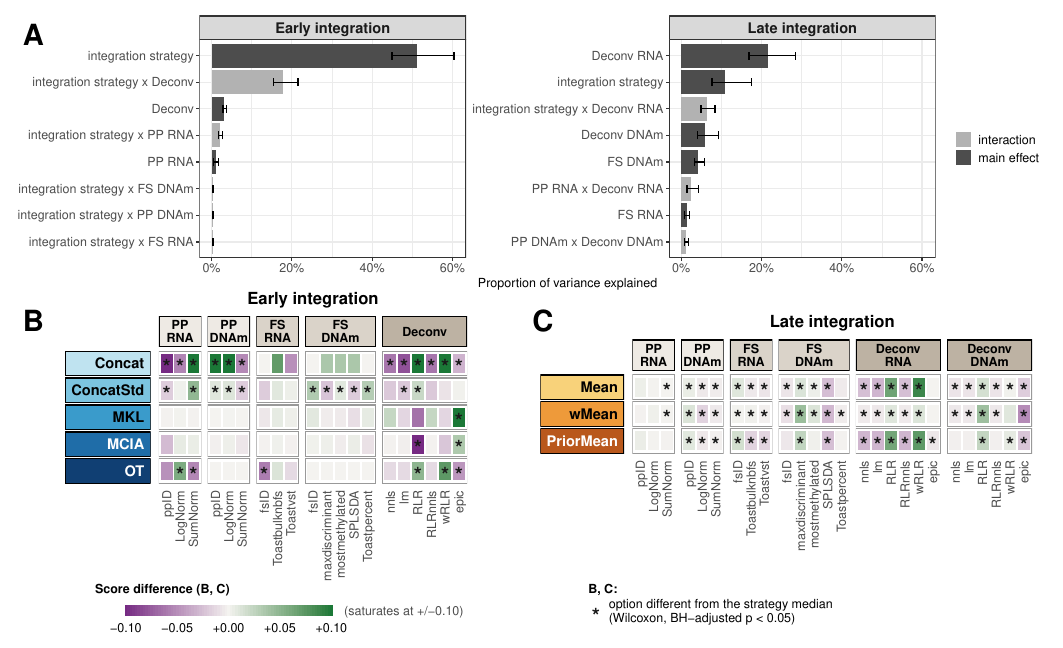}
  \caption{\textbf{Effect of the module options on the performance of early and late integration.}
  \textbf{A.} Proportion of variance of the aggregate score explained by the integration strategy, the module options and their two-way interactions ($\eta^2$, eight largest terms; dark grey: main effects, light grey: interactions; error bars: 95\% bootstrap intervals over the datasets).
\textbf{B--C.} For each early (B) and late (C) integration strategy (rows), median aggregate score of the pipelines using each option of each module (columns), relative to the median of all pipelines of the same strategy (green: improvement; purple: degradation). Stars: options whose deviation from the strategy median is consistent across datasets (Wilcoxon signed-rank test on the eight per-dataset deviations, BH-adjusted $P < 0.05$).
All scores are computed on the eight datasets scored with the full set of metrics (VIVO excluded)}
  \label{fig:combination}
\end{figure}

We next examined how the options chosen in each module shape performance (Figure~\ref{fig:combination}). To quantify the contribution of each module, we decomposed the variance of the aggregate score into the contributions of the modules and of their pairwise interactions (Figure~\ref{fig:combination}A). This decomposition shows that the two paradigms differ in which modules drive performance. In early integration, the integration strategy explains most of the score variance (51\%), followed by its interaction with the deconvolution algorithm (18\%). In late integration, the RNA deconvolution dominates (22\%), whereas integration strategies explain only 11\%.

Consistent with the strong interaction between the integration strategy and the deconvolution algorithm, early integration strategies differ in their preferred options (Figure~\ref{fig:combination}B). For instance, \texttt{RLR} is among the best deconvolution algorithms for \texttt{Concat} and \texttt{OT} but the worst for \texttt{MCIA}, and \texttt{SumNorm} RNA pre-processing benefits \texttt{Concat}, whereas \texttt{OT} performs best with \texttt{LogNorm}. In contrast, the three late integration strategies share the same preferred options (Figure~\ref{fig:combination}C): unprocessed DNAm, no RNA feature selection \texttt{maxdiscriminant} DNAm feature selection, and RLR-based deconvolution for both modalities, consistent with previous benchmarks~\cite{amblard_robust_2024}. 

Additional interaction analyses show that cross-modal pre-processing and feature-selection interactions occur only for the two concatenation-based early integration strategies (Supplementary Figure~\ref{suppfig:interaction}A). For \texttt{Concat}, raw RNA counts are several orders of magnitude larger than DNAm beta values, so that the estimated proportions are determined almost entirely by the RNA features. Only \texttt{SumNorm} RNA, which shrinks RNA values below DNAm beta values, performs well (+0.18 to +0.20 above the median). With \texttt{ConcatStd}, the Gaussian transformation bounds both modalities, but DNAm-dominated combinations still perform best, either through pre-processing or by pairing many CpGs with a 1,000-gene RNA selection. \texttt{MKL}, \texttt{MCIA} and \texttt{OT}, which normalise each modality separately, show no significant interaction.
In late integration, interactions between RNA and DNAm pre-processing and feature-selection choices are small (absolute contrasts below 0.002), as are interactions between their deconvolution algorithms (at most 0.02; Supplementary Figure~\ref{suppfig:interaction}B).  Together, these results indicate that early integration requires strategy-specific optimization, with concatenation-based strategies particularly sensitive to the coordination of RNA and DNAm pre-processing. Late integration strategies, by contrast, show more consistent methodological preferences and weaker dependencies between the two modalities.

%Four of the DNAm deconvolutions (\texttt{nnls}, \texttt{lm}, \texttt{nnlslargeref}, \texttt{RLRnnls}) yield identical predictions, because with five cell types they all reduce to the non-negative least-squares solution.

\subsection*{Integration can outperform DNAm alone when re-optimised for each dataset.}

\begin{figure}[!ht]%
\centering
\includegraphics[width=\linewidth]{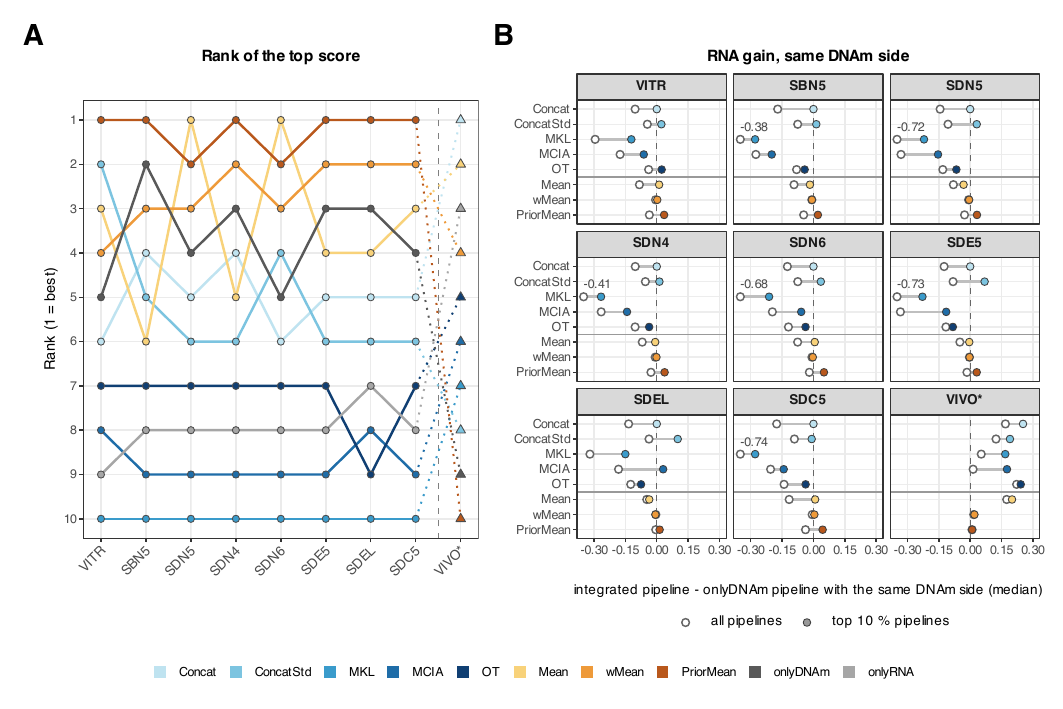}
\caption{\textbf{Best pipelines and contribution of RNA.}  
VIVO* is shown apart as its aggregate score uses a different formula (cell-type correlations only).
\textbf{A.} Rank (1 = best) of the best pipeline of each strategy, uni-omic strategies included, in each dataset. 
\textbf{B.} RNA gain in each dataset (panels): for each integrated pipeline, its score minus that of its DNAm-only counterpart, the \texttt{onlyDNAm} pipeline with the same DNAm pre-processing, feature selection and deconvolution (Box~1). Open circles: median gain over all pipelines of each integration strategy; filled circles: median gain over its top 10\% pipelines, ranked by aggregate score. Values below $-0.35$ are drawn at the left edge, with their value.}
\label{fig:top}
\end{figure}

We next turned from median to top performance, by comparing the best pipeline of each strategy in each dataset. Although \texttt{onlyDNAm} has the highest median performance, its best pipeline is never the best overall: it ranks between second and fifth, behind the best pipeline of \texttt{PriorMean} in six datasets and of \texttt{Mean} in two (Figure~\ref{fig:top}A). The best pipeline of \texttt{PriorMean} exceeded the best \texttt{onlyDNAm} pipeline in all eight fully-annotated datasets, and that of \texttt{wMean} in seven, compared with four for \texttt{Mean} and two for \texttt{ConcatStd}, whereas \texttt{Concat}, \texttt{MKL}, \texttt{MCIA} and \texttt{OT} never did (Supplementary Figure~\ref{suppfig:peakscore}).

This comparison is, however, biased in favour of integration strategies, which have far more pipelines than \texttt{onlyDNAm} (up to 4,860 versus 90 per dataset): the more pipelines are compared, the more likely one of them is to reach a high score by chance, even if none is truly better. To isolate the contribution of RNA independently of the size of the search space, we compared each integrated pipeline with its DNAm-only counterpart, that is, the \texttt{onlyDNAm} pipeline with exactly the same DNAm pre-processing, feature selection and deconvolution (Figure~\ref{fig:top}B; Box~1). Because both pipelines share the same DNAm side, their score difference reflects only the effect of adding RNA, through the RNA-side options and the integration strategy.
Across all pipelines, this difference is generally negative, strongly so for latent-embedding and optimal-transport strategies: with an arbitrary RNA side, adding RNA degrades the DNAm pipeline. Among the top 10\% pipelines of each strategy, however, the difference becomes positive for \texttt{PriorMean} and, in most datasets, \texttt{ConcatStd}, whereas it remains close to zero for \texttt{wMean}, \texttt{Concat} and, in most datasets, \texttt{Mean}. RNA can therefore improve a DNAm pipeline, but only with specific RNA sides.

Why do some integration strategies exceed the best \texttt{onlyDNAm} pipeline and others not? The score of an integrated pipeline can be split into two parts: the score of its DNAm side alone (its DNAm-only counterpart) and the gain brought by adding RNA. To exceed the best \texttt{onlyDNAm} pipeline, an integrated pipeline therefore needs both a DNAm side close to the best one and a positive RNA gain.
The integration strategies with a high peak performance in each paradigm meet these conditions differently (Supplementary Figure~\ref{suppfig:topuniomic}).
\texttt{ConcatStd}, the best early integration strategy on median performance, has a positive RNA gain but not a near-optimal DNAm side: its best pipelines keep many CpGs (\texttt{Toastpercent}), probably to balance RNA in the concatenated matrix, and remain below the best \texttt{onlyDNAm} pipeline in most of the datasets. 
\texttt{Mean} has a near-optimal DNAm side, but its RNA gain is inconsistent: it produces the best pipeline overall in SDN5 and SDN6, but falls below the best \texttt{onlyDNAm} pipeline in other datasets.
\texttt{PriorMean}, however, obtains a positive RNA gain in every dataset, large enough to exceed the best \texttt{onlyDNAm} pipeline even where its DNAm side is not optimal. \texttt{PriorMean} uses the DNAm estimates for the two cancer cell types. As this rule is the only difference between \texttt{Mean} and \texttt{PriorMean}, this suggests that the benefit of integration comes from restricting RNA to the cell types where it does not degrade the estimates.

Finally, we asked whether the best pipeline of each strategy is a robust configuration or an isolated high score obtained by chance. To this end, we examined how the score decreases from the best pipeline to the next-ranked ones (Supplementary Figure~\ref{suppfig:leaderboard_decaytop100}). The best pipelines of most strategies lie on a plateau of pipelines with almost the same score, which indicates a robust optimum, the flattest being that of \texttt{PriorMean}. In contrast, the best pipelines of \texttt{MKL} and \texttt{MCIA}, and of \texttt{Mean} in SDN5 and SDN6, where it produces the best pipeline overall, are isolated peaks that overstate typical performance.

Even when robust, however, these best pipelines are selected separately in each dataset, and their composition changes from one dataset to another: for most strategies, the selected pre-processing, feature selection and deconvolution options differ across datasets (Supplementary Figure~\ref{suppfig:bestpip}). A strategy may therefore appear to outperform \texttt{onlyDNAm} only because it is re-tuned for each dataset, as the isolated peaks of \texttt{Mean} suggest. This raises two questions: whether a single pipeline, applied unchanged to all datasets, can still outperform DNAm alone, and whether this advantage holds on new data not used for its selection.

\subsection*{Only prior-knowledge integration consistently outperforms DNAm alone, including on unseen data.}

To answer these questions, we defined for each strategy a generalist pipeline, i.e.\ one pipeline applied unchanged to all datasets, selected as the pipeline with the highest mean relative gain over the best uni-omic pipeline across the eight fully-annotated datasets (Methods). We applied this test to three integration strategies, chosen to contrast generic statistical integration with integration based on biological prior knowledge. For early integration, we selected \texttt{ConcatStd}, the best strategy on median performance and the only early integration strategy to outperform \texttt{onlyDNAm} in some datasets (VITR and SDN6, Figure~\ref{fig:top}A). For late integration, we selected \texttt{Mean}, which produced the best pipeline overall in two datasets (SDN5 and SDN6, Figure~\ref{fig:top}A), and \texttt{PriorMean}, which differs from \texttt{Mean} only by its prior-knowledge rule. The \texttt{onlyDNAm} and \texttt{onlyRNA} generalist pipelines serve as references. We evaluated these generalist pipelines on the benchmark datasets and on unseen datasets that were not used for their selection (Figure~\ref{fig:validation}).

The composition of the selected generalist pipelines is shown in Figure~\ref{fig:validation}A. Notably, the \texttt{PriorMean} generalist pipeline uses exactly the same DNAm side as the \texttt{onlyDNAm} generalist pipeline, combined with an RNA side that keeps all genes (\texttt{LogNorm}, \texttt{fsID}, \texttt{wRLR}). Because \texttt{PriorMean} takes the two cancer cell types from DNAm only, any difference between these two generalist pipelines is due to RNA estimates of the non-tumour cell types. 
In contrast, the \texttt{Mean} and \texttt{ConcatStd} generalist pipelines use a different DNAm feature selection from the \texttt{onlyDNAm} generalist pipeline, and keep many more CpGs (8,780 and 18,979, respectively, against 320).

On the benchmark datasets, the \texttt{PriorMean} generalist pipeline outperforms the \texttt{onlyDNAm} generalist pipeline in each of the eight fully-annotated datasets  (Figure~\ref{fig:validation}B). The \texttt{Mean} and \texttt{ConcatStd} generalist pipelines fall below it on average: \texttt{Mean} generalist pipeline gains in the datasets where it produced the best pipeline (SDN5, SDN6) but loses heavily in others, and \texttt{ConcatStd} generalist pipeline gains only in SDN6. The advantage of \texttt{PriorMean} generalist pipeline is significant for the aggregate score and for three of the four metric families (cohort correlation, error and geometric metrics), but not for cell-type correlation (Figure~\ref{fig:validation}C). %This is consistent with its design. For the two cancer cell types, \texttt{PriorMean} uses the same DNAm estimates as the \texttt{onlyDNAm} generalist pipeline, so that the two pipelines differ only for the three non-cancer cell types. Metrics computed separately for each cell type therefore show no difference for two of the five cell types, whereas metrics evaluating the composition of each sample as a whole capture the improvement brought by RNA for the other three. 

Because the generalist pipelines were selected on the benchmark datasets themselves, their performance is optimistically biased and must be confirmed on data not used for selection. We therefore evaluated them on unseen data: new simulated datasets generated with the same designs as the benchmark ones (SBN5$'$ to SDC5$'$), and 38 unseen samples of VITR and VIVO (15 VITR$'$ and 23 VIVO$'$; Methods). The advantage of \texttt{PriorMean} generalist pipeline was confirmed, again in every dataset (Figure~\ref{fig:validation}D), and remained significant for the same metric families (Figure~\ref{fig:validation}E). In contrast, \texttt{Mean} generalist pipeline still fell below \texttt{onlyDNAm}, and \texttt{ConcatStd} generalist pipeline remained below it in every dataset, with no significant gain in any metric family. 

Taken together, these results support a cautious use of RNA-DNAm integration for deconvolution. Without prior knowledge, \texttt{onlyDNAm} remains the best default choice: it achieves the highest median performance and requires no modality-specific tuning. \texttt{PriorMean} is the only integration strategy whose advantage holds in every dataset, with a single generalist pipeline, and on unseen data.

\begin{figure}[!ht]%
\centering
\includegraphics[width=\linewidth]{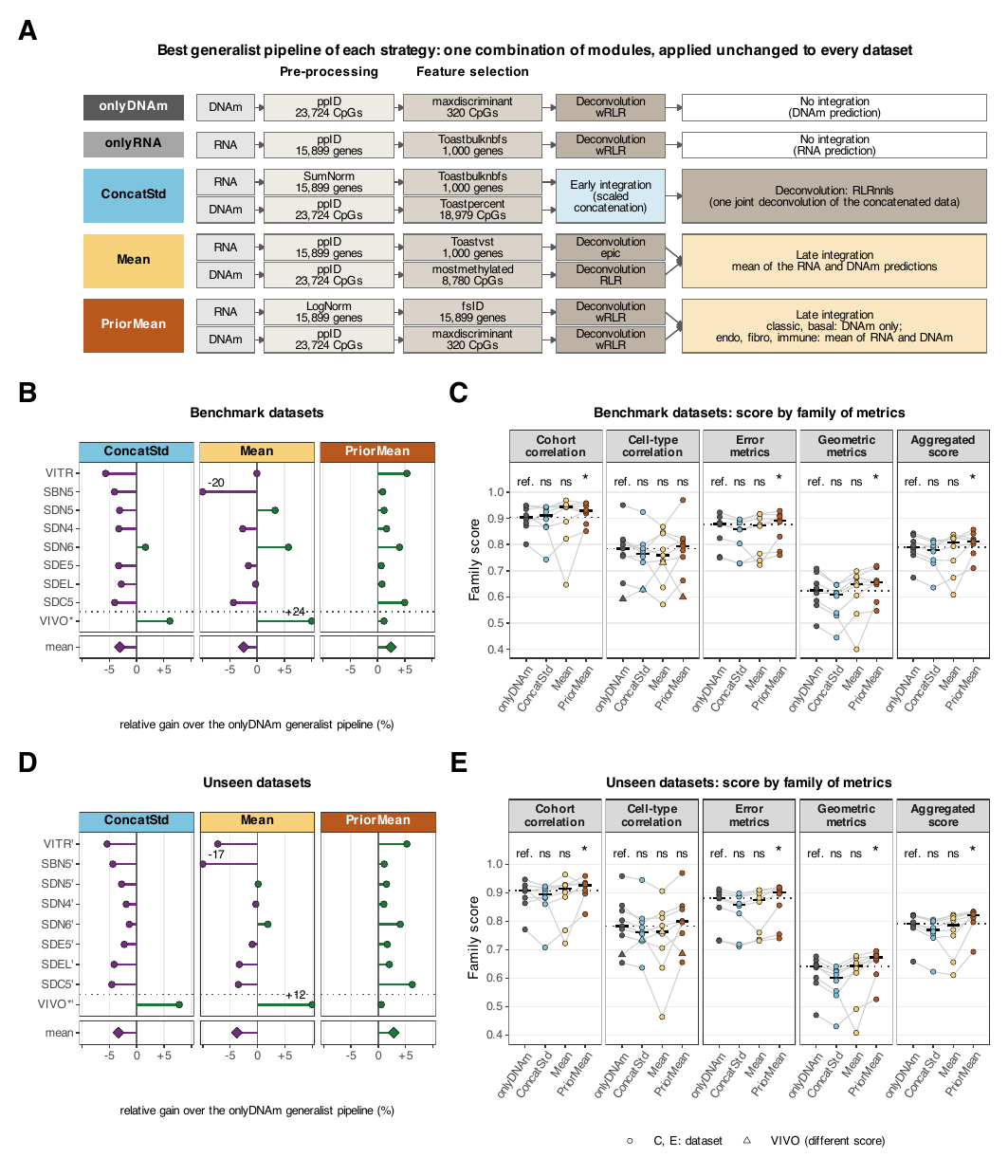}
\end{figure}

\begin{figure}[t]
\caption{\textbf{Validation of the top-performing pipelines on unseen datasets.}
\textbf{A} Generalist pipeline of each retained strategy. A generalist pipeline is a combination of functional modules (pre-processing, feature selection, deconvolution, integration) yielding to the best mean relative gain over the best uni-omic pipeline of each dataset. Numbers: features of the reference after pre-processing and after feature selection. 
\textbf{B, D} Relative gain (\%) of the aggregate score of each generalist pipeline over the \texttt{onlyDNAm} generalist pipeline, in each dataset of the benchmark (B) and on unseen data (D), one column per integration strategy. Each stem goes from 0 (the \texttt{onlyDNAm} generalist pipeline) to the gain of the generalist pipeline; green: better than onlyDNAm; purple: worse. The axis is cut at $\pm$10\,\%; larger gains are drawn at the edge, with their value written next to the stem. Unseen data (datasets marked with a prime, e.g. VITR', SBN5'). Diamond (bottom row): mean gain over the 8 other datasets.
\textbf{C, E} Score of each generalist pipeline by family of metrics (cohort correlation, cell-type correlation, error metrics, geometric metrics) and aggregate score, on the benchmark (C) and on unseen data (E). One point per dataset (triangle: VIVO, cell-type correlation only); grey lines link the points of a dataset across strategies; black bar: median; dotted line: median of \texttt{onlyDNAm}. Tests: one-sided paired Wilcoxon signed-rank test, strategy $>$ \texttt{onlyDNAm}, on the 8 datasets without VIVO, Benjamini–Hochberg adjusted over the 3 strategies of each family; * adjusted p $<$ 0.05; ns: not significant.}
\label{fig:validation}
\end{figure}

%\begin{figure}[p]%
%\centering
%\includegraphics[width=\linewidth]{Figures/main_natcom/figure6.pdf}
%\end{figure}

%\begin{figure}[t]
%\caption{\textbf{Composition of the top-performing pipelines.}
%\textbf{A, B.} Best pipeline of each strategy in each dataset, for early integration (\textbf{A}) and for late integration and uni-omic pipelines (\textbf{B}). Each row is one strategy--dataset pair; the left colour strip indicates the strategy and the next column the aggregate score of the best pipeline. Remaining columns indicate the option selected in each module (black), grouped by module: pre-processing (PP), feature selection (FS) and deconvolution (Deconv) for RNA and DNAm. In \textbf{A}, deconvolution is applied once to the integrated data; in \textbf{B}, it is applied to each modality separately. Hatched cells mark the best pipeline of a given dataset.
%\textbf{C, D.} Five best fixed recipes of \texttt{ConcatStd} (\textbf{C}), \texttt{Mean} and \texttt{PriorMean} (\textbf{D}). A recipe is a single pipeline applied unchanged to the eight fully-annotated datasets; recipes are ranked by their mean percentile rank across datasets (mean top \%; lower is better). Coloured cells give, for each dataset, the relative gain of the recipe over the best uni-omic pipeline of that dataset (red: gain; blue: loss); the last column gives the mean gain across datasets.
%}
%\label{fig:topdetailed}
%\end{figure}

\section*{Discussion}
\label{sec:limitations}

HADACA3 provides the first systematic, community-driven evaluation of RNA-DNAm integration strategies for bulk tissue deconvolution, spanning over 167,000 evaluations, across 18,774 pipelines and nine datasets. Our results support two conclusions of different scope. The first is specific to our setting: with the RNA and DNAm reference profiles available for PDAC, deconvolution based on DNAm data alone achieves the highest median performance across datasets and pipelines. The second is more general: integration strategies based on generic statistical criteria can outperform DNAm alone when the pipeline is re-optimised for each dataset, but this gain disappears when a single pipeline is applied unchanged to all datasets, and the only consistent gain comes from an integration strategy that incorporates prior knowledge of cell-type-specific modality strengths.

%\paragraph{Statistical integration does not denoise one modality with the other.}
The premise of multi-omics integration is that two partially independent signals can be combined to recover a more accurate estimate than either alone. Our modular benchmark shows that, for deconvolution, generic pre-processing, feature selection and aggregation criteria do not achieve this. Instead, the best-tuned pipelines of every statistical integration strategy converge on reducing the contribution of the weaker modality. In early integration, concatenation performs well only when pre-processing and feature selection prevent RNA from dominating the fitted matrix, at the cost of a sub-optimal DNAm side, and latent-embedding and optimal-transport strategies, which do not preserve the linear mixing structure, consistently degrade performance. In late integration, error-weighted averaging (\texttt{wMean}) is robust precisely because it assigns little weight to RNA, so that its best pipelines essentially reproduce the best \texttt{onlyDNAm} pipeline. In other words, without dataset-specific tuning, optimising a statistical integration pipeline comes down to selecting the better modality rather than combining both.

This does not mean that RNA carries no complementary information. When pipelines are optimised separately for each dataset, statistical integration strategies do find configurations in which adding RNA improves performance. These configurations, however, change from one dataset to another, and their advantage disappears when a single generalist pipeline is applied to all datasets or to unseen data. 
The complementary signal is therefore present in the data, but integration strategies based on generic statistical criteria fail to exploit it reliably. 
We propose that this failure is inherent to how generic statistical criteria operate, rather than a matter of tuning.
Pre-processing, feature selection and error-based weighting operate at the level of features or modalities, and are agnostic to cell types. The complementarity between RNA and DNAm, however, depends on the cell type. In our data, DNAm alone predicts the absolute proportions of basal-like and classical cancer cell types more accurately, whereas averaging the two modalities improves estimates for some non-cancer populations. A global criterion such as reconstruction error or feature variance cannot detect this cell-type-specific trade-off, and therefore cannot decide, for each cell type, whether combining the two modalities helps or hurts.

\texttt{PriorMean}, derived from the winning competition submission, resolves this trade-off: it uses the DNAm estimates for the proportions of the two cancer cell types and averages both modalities for the other cell types.
It is the only strategy that outperformed DNAm alone in every dataset, both with its best pipeline and with its generalist pipeline, including on unseen data. These gains remain modest, around 3\% of the aggregate score, but they are consistent.
This validation shows that the gain of \texttt{PriorMean} is not an artefact of selecting its generalist pipeline on the benchmark datasets. It does not show, however, that the gain holds on different data: the unseen data share the tissue, simulation designs and experimental batches of the benchmark, and confirmation on independent cohorts remains necessary.

The two conclusions of this study should be generalised differently. The superiority of DNAm alone is tied to our setting, for three reasons. First, all datasets focus on pancreatic cancer, whose cancer cell subtypes are strongly distinguished epigenetically~\cite{lomberk2018distinct}. Second, supervised deconvolution depends on the reference profiles, which remain limited for PDAC in both modalities. As each modality was evaluated with a single set of references, our benchmark cannot fully separate the effect of the modality from that of the quality of its reference. Third, the lower performance of RNA alone may partly reflect properties of RNA data, such as sparsity and zero inflation, which penalise some metrics, including the Aitchison distance. This conclusion may thus not hold in other tissues or with improved RNA references, although DNAm alone remains a robust default whenever a high-quality DNAm reference is available. By contrast, the difficulty of purely statistical integration and the benefit of cell-type-specific prior knowledge follow from a structural property of deconvolution, namely that modality reliability varies across cell types, and we expect them to extend to other tissues. What does not transfer directly is the rule itself: which modality to trust for which cell type must be established for each system, from biological knowledge or from calibration data such as in vitro mixtures or realistic simulations. Two caveats bound this generalisation: our simulations do not model cross-modal correlations arising from shared regulation between RNA and DNAm, and our conclusions concern RNA-DNAm integration and may not extend directly to other modalities such as proteomics or metabolomics.

These results open two directions for future research. The first is to identify selection criteria that can recognise, without ground truth, the pipelines in which RNA complements DNAm, for instance from the consistency between modality-specific estimates or from properties of the references. The second is to learn cell-type-specific modality weights from calibration data rather than fixing them by hand, which would generalise the principle behind \texttt{PriorMean}. Intermediate integration, in which the two modalities are combined within the deconvolution model itself, is a natural candidate for learning such weights. In this paradigm, however, the integration mechanism is intrinsically linked to the architecture of the deconvolution model, so that it could not be represented as an option of the integration module of our workflow; its evaluation will then require a dedicated study. The HADACA3 continuous benchmark~\cite{hadaca3_codabench} is designed to support these developments: all datasets, pipelines and evaluation procedures are openly available through Codabench, so that new pipelines can be evaluated in the same framework, as we illustrate with the recently published EMixed~\cite{caiEMixedProbabilisticMultiOmics2025}, which performed comparably to \texttt{wMean}. More broadly, our results suggest a general lesson for multi-omics integration: combining modalities improves inference not when they are merged on generic statistical grounds, but when the integration reflects what each modality is known to measure well.

\section*{Methods}
\label{methods}

\subsection*{HADACA3 datasets}
\label{sec:dataset}

\emph{Reference profiles used for supervised deconvolution.} Two types of reference profiles were provided, both derived from publicly available datasets: bulk RNA and bulk DNAm profiles of pure cell types (Supplementary Table~\ref{supptab:bulk_reference}). Single-cell RNA-seq reference profiles (Supplementary Table~\ref{supptab:scrna_reference}) were also available during the competition, but were not included in the subsequent modular benchmark because they did not improve performance in our setting.

\emph{\textit{In vivo} and \textit{in vitro} benchmark datasets.} The gold standard (VIVO) consists of 47 PDAC surgical samples with matched RNA and DNAm data (24 in the benchmark, 23 held out for validation), for which classical and basal-like tumour components were estimated from 
histology slides using PACpAInt~\cite{saillard_pacpaint_2023} (not a proportion; only its correlation with predicted proportions is evaluated). This dataset is original and is released with the benchmark. The silver standard (VITR) is a previously published \textit{in vitro} mixture dataset of nine purified cell types mixed in known proportions~\cite{amblard_robust_2024}, grouped into five cell types (the immune cell types being pooled). Further details are provided in Supplementary Material~\ref{suppmat:benchmarkdt}.

\textit{In silico} benchmark datasets. Six simulated datasets were generated using a linear mixing model, by sampling cell-type proportions from a Dirichlet distribution and combining them with pure cell-type profiles~\cite{amblard_robust_2024}:
\begin{equation}
Y_i^{(m)} = X_{\mathrm{sim}}^{(m)} p_i + \varepsilon_i^{(m)}, \quad m \in \{\mathrm{RNA}, \mathrm{DNAm}\},
\label{eq:simulation}
\end{equation}
where $Y_i^{(m)} \in \mathbb{R}^{F^{(m)}}$ is the bulk profile of sample $i$ in modality $m$, $p_i \in \mathbb{R}^k$ its vector of cell-type proportions, shared by the two modalities, $X_{\mathrm{sim}}^{(m)} \in \mathbb{R}^{F^{(m)} \times k}$ the pure cell-type profiles used for simulation, and $\varepsilon_i^{(m)}$ a modality-specific noise term. Importantly, $X_{\mathrm{sim}}^{(m)}$ is entirely distinct from the reference profiles $X^{(m)}$ provided to participants, which avoids circularity. 
Simulation scenarios cover heteroscedastic noise (SDN5), rare cell types (SDEL), mismatched compositions (SDN4, SDN6), and noise with structured feature correlations modeled via EM (SDE5) or copula-based approaches (SDC5).
An additional pseudo-bulk dataset (SBN5) was generated by aggregating single-cell profiles, providing a more realistic noise structure. Full simulation details are provided in Supplementary Material~\ref{supp:simu}.

\emph{Feature space.} The feature space was restricted to the intersection of RNA and DNAm features available across all benchmark datasets and reference profiles, yielding a common set of 15,908 RNA and 23,724 DNAm features (Supplementary Material~\ref{suppmat:feature}).

\subsection*{Competition Starting kit and baselines}

The competition ran in three phases: an introductory phase on a simple simulated dataset, a robustness phase on multiple datasets with varying noise and composition, and a generalization phase on private unseen data, with no feedback (Supplementary Table~\ref{supptab:competphase}). Participants received a starting kit with five baseline implementations (Supplementary Table~\ref{supptab:baseline}), covering Non-Negative Least Squares (NNLS)-based deconvolution on RNA alone and on both modalities with late averaging, and utility scripts for package installation and external file loading. Baselines were implemented in both R and Python. Submissions were made through Codabench, which provided real-time feedback via a public leaderboard. Datasets were split into public sets available during development and private sets used for final evaluation to prevent overfitting. 

\subsection*{The JOKER method}
For RNA, data were normalized, variance-stabilized, and the 5,000 most variable genes were selected, followed by gene-wise scaling. For DNAm, low-variance CpG sites were removed. Deconvolution was performed independently on each modality using NNLS. 
Final estimates were obtained via rule-based aggregation: RNA and DNAm predictions were averaged for most cell types, while onlyDNAm estimates were retained for the two cancer cell types (basal-like and classical). This choice was motivated by empirical observations that DNAm provided more stable and consistent estimates for these subtypes across datasets.

\subsection*{Modular benchmark}
Each pipeline combines one strategy with one option per module for each modality it uses (Table~\ref{tab:modules}). In early integration, the RNA and DNAm data are pre-processed and feature-selected separately, combined by the integration strategy, and deconvolved in a single step. In late integration, each modality is pre-processed, feature-selected and deconvolved separately, with its own deconvolution option, and the two estimates are then aggregated by the integration strategy. Uni-omic strategies use a single modality, pre-processed, feature-selected and deconvolved without integration: \texttt{onlyDNAm} uses DNAm only and \texttt{onlyRNA} RNA only. Within each modality, the same pre-processing and feature selection options were applied to the mixtures and to the reference profiles, so that both are expressed on the same scale and restricted to the same features before deconvolution.
All compatible combinations were evaluated on the nine benchmark datasets. With 3 pre-processing options per modality, 3 RNA and 5 DNAm feature selection options, and 6 deconvolution options, this yields, per dataset, 4,050 early integration pipelines ($3 \times 3 \times 3 \times 5 \times 6$ options for 5 strategies), 14,580 late integration pipelines ($3 \times 3 \times 3 \times 5 \times 6 \times 6$ options for 3 strategies), and 144 uni-omic pipelines (90 for \texttt{onlyDNAm} and 54 for \texttt{onlyRNA}), i.e.\ 167,322 scored evaluations (168,966 planned; a few pipelines are missing in VITR, SDN4, SDN5 and SDN6). Each pipeline was scored with the aggregate score defined in Supplementary Material~\ref{suppmat:metric}.

\subsection*{Matched comparison with DNAm-only counterparts}
To isolate the contribution of RNA, each integrated pipeline was compared with its \texttt{onlyDNAm} counterpart, defined as the \texttt{onlyDNAm} pipeline using the same DNAm pre-processing, feature selection and deconvolution (for early integration, the same deconvolution algorithm as the joint deconvolution). For each integrated pipeline $p$ and dataset $d$, the RNA gain was computed as $g_{p,d} = s_{p,d} - s_{c(p),d}$, where $s$ is the aggregate score and $c(p)$ the counterpart of $p$. Gains were summarised, for each strategy and dataset, by their median over all pipelines of the strategy and over its top 10\% pipelines ranked by aggregate score (Figure~\ref{fig:top}B). Peak scores were defined as the score of the best pipeline of a strategy in a dataset minus that of the best \texttt{onlyDNAm} pipeline of the same dataset (Supplementary Figure~\ref{suppfig:peakscore} ).

\subsection*{Selection of the generalist pipelines} For each strategy (\texttt{onlyDNAm}, \texttt{onlyRNA}, \texttt{ConcatStd}, \texttt{Mean} and \texttt{PriorMean}), the generalist pipeline was selected on the benchmark datasets, among the pipelines evaluated on all eight fully-annotated datasets (VIVO excluded). For each pipeline and dataset, the relative gain over the best uni-omic pipeline of the dataset was computed as $(s - s^{*}_{d}) / s^{*}_{d}$, where $s$ is the aggregate score of the pipeline and $s^{*}_{d}$ the higher of the best \texttt{onlyDNAm} and best \texttt{onlyRNA} scores in dataset $d$. The generalist pipeline of each strategy is the one with the highest mean relative gain over the eight datasets (Supplementary Material~\ref{suppmat:fixed_recipes}).

\subsection*{Unseen validation data}
To assess whether the generalist pipelines generalise beyond the data used to select them, we evaluated them on validation data that were not used at any stage of pipeline selection (datasets marked with a prime). 
For the simulated settings, we used the test halves of the competition datasets (SBN5$'$ to SDC5$'$, 30 samples each), which were generated together with the benchmark datasets but never released to participants (Supplementary Material~\ref{suppmat:benchmarkdt}).  Each simulated validation dataset contains 30 samples, and was generated from the same pure cell-type profiles as the benchmark datasets. For the experimental settings, 15 in vitro mixtures (VITR$'$) and 23 tumour samples (VIVO$'$) were set aside from the same experiments as VITR and VIVO and were not used in the benchmark. These samples were never released to participants at any phase of the competition, and were used neither in the benchmark nor for generalist pipeline selection (Supplementary Material~\ref{suppmat:unseen} and Supplementary Table~\ref{supptab:unseen}). Generalist pipelines were applied to these data without any re-tuning, using the same reference profiles and the same aggregate score as for the benchmark datasets (VIVO$'$ being scored on cell-type correlations of the two cancer cell types, as VIVO). The validation datasets are available on Zenodo~\cite{hadaca3}.

\subsection*{Statistics and reproducibility}
Unless stated otherwise, the statistical unit was the dataset, and tests were performed on the eight datasets scored with the full set of metrics ($n = 8$; VIVO, whose score uses a different formula, was analysed separately). For each effect, a per-dataset statistic was computed and the eight values were tested with an exact Wilcoxon signed-rank test. Multiple-testing correction used Bonferroni for the comparisons of each strategy with \texttt{onlyDNAm} within a dataset, which support the main claim, and Benjamini–Hochberg for the larger, exploratory families (pairs of options, interaction contrasts, cell types, generalist pipelines).

\emph{Median performance.} For each dataset and strategy, the median aggregate score was computed over all evaluated pipelines, with 95\% intervals from 2,000 bootstrap resamplings of the pipelines (Figure~\ref{fig:robust}B). As pipelines share modules by construction, these intervals describe the stability of the median over the evaluated configuration space rather than sampling uncertainty. Differences between each integration strategy and \texttt{onlyDNAm} were tested within each dataset with a two-sided paired Wilcoxon signed-rank test over the 15 shared DNAm pre-processing $\times$ feature selection configurations, each summarised by the median score of the strategy over its other modules, with Bonferroni correction over the eight comparisons of each dataset; effect sizes are Hodges--Lehmann estimates with 95\% confidence intervals (Figure~\ref{fig:robust}C; Supplementary Material~\ref{suppmat:median}).

\emph{Module effects.} The proportion of variance explained by each module and two-way interaction was estimated with a linear model including a dataset effect, all main effects and all two-way interactions, as the type I sum of squares of each term divided by the total sum of squares ($\eta^2$); 95\% intervals were obtained from 100 bootstrap resamplings of the datasets, and no P-values were computed for this model (Figure~\ref{fig:combination}A). The effect of each option was the median score of the pipelines using it minus the median of all pipelines of the same strategy, tested with a two-sided Wilcoxon test on the eight per-dataset deviations, with BH correction over all cells of each panel (Figure~\ref{fig:combination}B-C). RNA $\times$ DNAm interactions were measured, within each integration strategy and dataset $d$, by the interaction contrast $I^d_{ik} = \mu^d_{ik} - \mu^d_{i\cdot} - \mu^d_{\cdot k} + \mu^d_{\cdot\cdot}$, where $\mu^d_{ik}$ is the mean score of the pipelines combining RNA option $i$ with DNAm option $k$, and a dot denotes averaging over the corresponding index. Contrasts were tested in the same way, with BH correction over the cells of each heatmap (Supplementary Material~\ref{suppmat:module}).

\emph{Generalist pipelines.} The score of each generalist pipeline was compared with that of the \texttt{onlyDNAm} generalist pipeline with a one-sided paired Wilcoxon signed-rank test across datasets (alternative: strategy better than \texttt{onlyDNAm}), separately for the aggregate score and for each metric family, on the benchmark datasets and on the unseen simulated and in vitro datasets, with BH correction over the three strategies for the aggregate score and over the three integration strategies for each metric family (Figure~\ref{fig:validation}B-E).

\section*{Code availability} 
All analyses were performed in R version 4.3.1, using dplyr (1.1.4) and tidyr (1.3.1) for data handling, the stats package for statistical tests (Wilcoxon signed-rank tests, Benjamini–Hochberg and Bonferroni corrections), multcompView (0.1.9) for compact letter displays, rhdf5 (2.46.1) to read the HDF5 files of the workflow, and ggplot2 (3.5.2), patchwork (1.3.1), ggh4x (0.3.1), ggforce (0.5.0), scales (1.4.0) and RColorBrewer (1.1.3) for the figures.  The code used to simulate datasets and analyse data is publicly available on GitHub \cite{hadaca3_framework}. 
 
\section*{Data availability} 
The \textit{in vivo} datasets are an original contribution of this work, released publicly on NCBI GEO portal \cite{GSE328792, GSE329368}. The \textit{in vitro} dataset is derived from a publicly available study on NCBI GEO \cite{GSE281204, GSE281305}. Simulated (\textit{in silico}) datasets were generated using a custom pipeline and are available (with croissant format documentation) on Zenodo \cite{hadaca3}. 
%Anonymized code is provided \cite{hadaca3_framework}, full anonymity cannot be guaranteed due to the use of public datasets \cite{hadaca3} and online platforms \cite{hadaca3_codabench}, but a best-effort attempt was made. Complete resources will be released upon acceptance.

\section*{Competing interests}
The authors declare no competing interests.

\section*{Author contributions}
H.B., E.A.\ and N.H.\ contributed equally. Conceptualization: Y.B., M.R. Methodology: H.B., E.A., N.H., L.L., F.C., Y.B., M.R. Software: N.H., F.C. Data generation (in vivo and in vitro datasets): M.A., A.Ba., L.A., Y.K., A.C., J.C. Formal analysis: H.B., M.R. Competition design (scientific committee): M.T., C.H., S.D., L.S., D.C., A.B, Y.B, M.R. Writing -- original draft: H.B., A.B., Y.B., M.R. Writing -- review and editing: all authors. Supervision:  Y.B., M.R. Funding acquisition: M.R., A.B. Creative thinking: HADACA3 Consortium.

\section*{Acknowledgments}
%ANONYM:
We thank all supporting organizations: GDR BIMMM, Région Auvergne-Rhône-Alpes, M4DI PEPR Santé Numérique, ITMO Cancer Aviesan, La Ligue Française contre le cancer, LabEx PERSYVAL-2, RT Math Bio Santé (CNRS), CLARA, RIS, EFELIA-MIAI, and the GRICAD mesocenter at UGA. We thank the CAES Paul Langevin facility in Aussois for hosting, and Isabelle Guyon, Franck Picard and Charles Lecellier for fruitful discussions. This work was supported by ANR (CauseHet, ANR-22-CE45-0030), France 2030 (ANR-22-PESN-0013, ANR-23-IACL-0006), and ITMO Cancer of Aviesan / Inserm (ACACIA, AAP-MIC-2021).

%\section*{References}
%\newpage

\section*{HADACA3 Consortium Members}
We thank all members of the HADACA3  consortium for helpful discussion and contributions during the data challenge (December 2024, Aussois, France). 

\textbf{HADACA3 collaborating authors:}

%The full list of collaborating 
%authors will be provided upon acceptance, as it has been omitted here 
%for anonymization purposes.

\noindent
Al-Shahrour Fatima (Bioinformatics Unit, CNIO, Madrid, SP);
Amblard Elise (Univ. Grenoble Alpes, CNRS, LIG, Grenoble, FR);
Appé Guillaume (Epigene Labs, Paris, FR);
Agonkoui Christelle (Université de Lille / Inria MODAL, FR);
Barbot Hugo (IRMAR CNRS, Institut Agro Rennes Angers, FR);
Baudot Anaïs (Aix Marseille Univ, INSERM, MMG, Marseille, FR);
Beaufils Audrey (Université Paris Cité, CRI, INSERM U1149, FR);
Becht Etienne (Université Paris Cité, CRI, INSERM U1149, FR);
Benoit Clément (Institut de Biologie et Pathologie, CHUGA, FR);
Blum Yuna (IGDR CNRS, Université de Rennes, FR);
Burban Ewen (Univ Rennes, Inserm, Irset UMR\_S 1085, FR);
Carpentier Océane (Univ Rennes, Inria, CNRS, IRISA, FR);
Causeur David (IRMAR CNRS, Institut Agro Rennes Angers, FR);
Chassagnol Bastien (Aix Marseille Univ, INSERM, MMG, Marseille, FR);
Chepeleva Maryna (Luxembourg Institute of Health, LU);
Chuffart Florent (Univ. Grenoble Alpes, INSERM, CNRS, IAB, FR);
Dejean Sebastien (IMT CNRS, Université de Toulouse, FR);
De Wit Renske (Netherlands Cancer Institute / Utrecht University, NL);
Deligne Margaux (CentraleSupélec / MICS, FR);
Derouin Margot (INSERM UMR1141 NeuroDiderot, Paris, FR);
Dhar Gaurav (Toulouse School of Economics, FR);
Essabbar Abdelmounim (CRCT, Université de Toulouse, INSERM, CNRS, FR);
Garreau Jules (IGDR CNRS, Université de Rennes, FR);
Ghestem Florence (Université Paris-Saclay, UVSQ, Inserm, CESP, FR);
Gorse Marine (Univ Rennes, Inserm, Irset UMR\_S 1085, FR);
Herrmann Carl (BioQuant, IPMB, Heidelberg University, DE);
Homberg Nicolas (Univ. Grenoble Alpes, CNRS, LIG, FR);
Hughes Arthur (Bordeaux Population Health, INSERM U1219, FR);
Hurtado Marcelo (CRCT, Université de Toulouse, INSERM, CNRS, FR);
Kubica Jędrzej (Univ. Grenoble Alpes, CNRS, TIMC, FR);
Lamothe Lucie (Univ. Grenoble Alpes, CNRS, LIG, FR);
Laval Quentin (INSERM, Bordeaux Population Health, FR);
Le Clech Renée (HeKa UMR1346, Inserm, Inria, Université Paris Cité, FR);
%Lecellier Charles (CNRS, Univ. Montpellier, IGMM and LIRMM, FR);
Loire Benjamin (Aix Marseille Univ, INSERM, MMG, FR);
Louistisserand Juliette (Univ. Grenoble Alpes, CNRS, LIG, FR);
Lukic Vesna (CentraleSupélec / MICS, FR);
Malleval Inès (CentraleSupélec / MICS, FR);
Marcou Quentin (Inria–Inserm COMPO, CRCM, Marseille, FR);
Menard Thomas (CentraleSupélec / MICS, FR);
Najm Matthieu (Institut Pasteur, Université Paris Cité, FR);
Nosirov Bakhtiyor (Luxembourg Institute of Health, LU);
Pal Annesh (Bordeaux Population Health, INSERM U1219, FR);
Pignolet Camille (Université Paris Cité, CRI, INSERM U1149, FR);
Privé Florian (Aarhus University, DK);
Rau Andrea (Université Paris-Saclay, INRAE, AgroParisTech, GABI, FR);
Richard Magali (Univ. Grenoble Alpes, CNRS, LIG, FR);
Spinelli Lionel (Aix-Marseille Univ, CNRS, INSERM, CIML, FR);
Tabbakh Danial (Luxembourg Institute of Health, LU);
Térézol Morgane (Aix Marseille Univ, INSERM, MMG, Marseille, FR);
Toffano Antoine (Université de Montpellier, LIRMM UMR5506, FR);
Valentin Lorena (Aix Marseille Univ, INSERM, MMG, Marseille, FR);
Ventre Elias (Inria–Inserm COMPO, CRCM, Marseille, FR);
Vignes Matthieu (School of Mathematical and Computational Sciences, Massey University, NZ);
Von Grafenstein Klaus (Univ Rennes, Inserm, Irset UMR\_S 1085, FR);
Weill Solene (Epigene Labs, Paris, FR).

\bibliographystyle{unsrt}
\bibliography{reference}

%%%%%%%%%%%%%%%%%%%%%%%%%%%%%%%%%%%%%%%%%%%%%%%%%%%%%%%%%%%%

%\Supplementary Material

%%%%%%%%%%%%%%%%%%%%%%%%%%%%%%%%%%%%%%%%%%%%%%%%%%%%%%%%%%%%

%\newpage

%\section{Supplementary Material : Supplementary Material, Tables and Figures}
%\label{supp}
%Technical appendices with additional results, figures, graphs, and proofs may be submitted with the paper submission before the full submission deadline (see above). You can upload a ZIP file for videos or code, but do not upload a separate PDF file for the Supplementary Material. There is no page limit for the technical appendices. 

%Note: Think of the Supplementary Material as ``optional reading'' for reviewers. The paper must be able to stand alone without the Supplementary Material; for example, adding critical experiments that support the main claims to an Supplementary Material is inappropriate. 

%\newpage
%\input{checklist.tex}

%%%%%%%%%%%%%%%%%%%%%%%%%%%%%%%%%%%%%%%%%%%%%%%%%%%%%%%%%%%%
%%%%  SUPPLEMENTARY MATERIAL
%%%%%%%%%%%%%%%%%%%%%%%%%%%%%%%%%%%%%%%%%%%%%%%%%%%%%%%%%%%%
\clearpage
\newgeometry{margin=1in}
\nolinenumbers
\setcounter{section}{0}\setcounter{figure}{0}\setcounter{table}{0}\setcounter{equation}{0}
\captionsetup[figure]{name=Supplementary Figure,labelsep=colon}
\captionsetup[table]{name=Supplementary Table,labelsep=colon}
\renewcommand{\theHfigure}{S\arabic{figure}}
\renewcommand{\theHtable}{S\arabic{table}}
\renewcommand{\theHequation}{S\arabic{equation}}
\renewcommand{\theHsection}{S\arabic{section}}

\begin{center}
{\LARGE\bfseries Supplementary Material}
\end{center}
\vspace{1em}

\tableofcontents
\clearpage

\section{HADACA3 datasets}
\label{subsec:dataset}

\subsection{Reference data used for supervised deconvolution}
\label{suppmat:ref}

\subsubsection{Bulk DNAm and RNA} Participants were provided with reference profiles of five major cell types found in pancreatic tumours: immune cells, fibroblasts, endothelial cells, and two tumour subtypes: classical and basal-like. All reference data were obtained from all publicly available datasets. 
For pure-bulk RNA and pure-bulk DNAm, replicates corresponding to a given cell type were aggregated into a single meta-reference profile by averaging all available cell-type-specific profiles (Supplementary Table~\ref{supptab:bulk_reference}). 
\begin{itemize}
    \item Transcriptomic profiles of immune cells and fibroblasts were obtained from the GTEx Analysis \cite{coorens_human_2025}, endothelial cell profiles from GEO \cite{bastounis_subendothelial_2019}, and tumour subtype profiles from patient-derived xenografts (PDX) \cite{nicolle_pancreatic_2017}. Only genes shared across all datasets were retained, resulting in a final set of 27,786 genes.
    \item  For DNAm profiles, immune and fibroblast profiles were obtained from Decamps et al. \cite{decamps_deconbench_2021}, endothelial cell profiles from GEO \cite{franzen_senescence-associated_2017}, and tumour subtype profiles from the same PDX study as the bulk RNA-seq reference \cite{nicolle_pancreatic_2017}. All collected reference datasets were generated using the Illumina Infinium HumanMethylation450 BeadChip platform. CpG sites were restricted to the intersection across datasets (n = 416,830). Features were then restricted to a subset of 23,724 probes during feature-space harmonization with the benchmark dataset.
\end{itemize}

\begin{table}[!ht]
\centering
\caption{Bulk reference datasets used to construct cell-type signatures for PDAC deconvolution.}
\begin{tabularx}{\linewidth}{p{1cm} p{2cm} p{1cm} p{2.2cm} p{8cm}}
\toprule
Omic & Cell type & Samples (n) & Source & Description \\
\midrule

RNA-seq & Endothelial & 23 & \cite{bastounis_subendothelial_2019} (GSE135123) &
Cultured endothelial cells (HMEC-1 or HUVEC) grown on soft or stiff matrices. One empty sample removed from the original 24 samples. \\

RNA-seq & Basal tumour & 4 & \cite{nicolle_pancreatic_2017} &
Patient-derived xenograft (PDX) samples (PDAC001T, PDAC004T, PDAC028T, PDAC024T). Extreme basal samples selected according to the PAMG gradient \cite{nicolle_establishment_2020}. \\

RNA-seq & Classical tumour & 5 & \cite{nicolle_pancreatic_2017} &
PDX samples (PDAC018T, PDAC011T, PDAC026T, PDAC006T, PDAC030T) selected according to the PAMG gradient. \\

RNA-seq & Immune & 260 & GTEx (v8) &
“Spleen” samples from the GTEx portal (RNASeQCv1.1.9). \\

RNA-seq & Fibroblasts & 527 & GTEx (v8) &
“Cells - Cultured fibroblasts” samples from the GTEx portal. \\

\midrule

DNAm & Endothelial & 6 & \cite{franzen_senescence-associated_2017} (GSE82234) &
Cultured HUVEC donor samples. \\

DNAm & Basal tumour & 4 & \cite{nicolle_pancreatic_2017} &
PDX samples corresponding to the RNA-seq reference dataset. \\

DNAm & Classical tumour & 5 & \cite{nicolle_pancreatic_2017} &
PDX samples corresponding to the RNA-seq reference dataset. \\

DNAm & Fibroblasts/ Immune & 8 / 3 & \cite{decamps_deconbench_2021} &
Aggregated beta values obtained without batch correction. \\

\bottomrule
\end{tabularx}
\label{supptab:bulk_reference}
\end{table}

\subsubsection{scRNA-seq.} 
We provided scRNA-seq reference profiles during the HADACA3 competition, that could be used as reference for deconvolution of bulk RNA-seq data. 
The scRNA-seq reference integrated data from three studies: Peng et al., which includes immune, fibroblast, endothelial, and cancer cells \cite{peng_single-cell_2019}; Baron et al., which includes immune, fibroblast, and endothelial cells \cite{baron_single-cell_2016}; and Raghavan et al., which includes immune, endothelial, and cancer cells \cite{raghavan_microenvironment_2021}. In the Peng and Raghavan datasets, cancer cells were further stratified into basal-like and classical subtypes using the \texttt{PurIST} classifier \cite{rashid_purity_2020}. Briefly, tumour cells were normalized using \texttt{SCTransform}, subtype predictions were obtained using a pre-trained PurIST model, and cells with ambiguous (``lean'') subtype assignments were excluded. For each datasets, $~$ 5,000 single-cell profiles were provided (Supplementary Table~\ref{supptab:scrna_reference}).
Although scRNA-seq reference profiles were provided during the competition, no submission achieved competitive performance by using them as deconvolution references. We therefore did not retain scRNA-seq-based deconvolution in the modular benchmark, which also substantially reduced the size and complexity of the combinatorial search space.

\begin{table}[!ht]
\centering
\caption{Single-cell RNA-seq reference datasets used to derive cell-type profiles.}
\begin{tabularx}{\linewidth}{
>{\centering\arraybackslash}p{1.2cm}
>{\centering\arraybackslash}p{1.2cm}
p{5.5cm}
X}
\toprule
Study & Samples (n) & Data source & Cell-type mapping used in this study \\
\midrule

\cite{peng_single-cell_2019} &
35 &
PRJCA001063 (CRA001160) &
\textit{Mapping:} \newline
Endothelial $\rightarrow$ endothelial; \newline
B cell, T cell, Macrophage $\rightarrow$ immune; \newline
Stellate, Fibroblast $\rightarrow$ fibroblasts; \newline
Ductal cell type 2 $\rightarrow$ cancer cells. \newline
\textit{tumour subtypes:} basal/classical inferred using \texttt{PurIST} \cite{rashid_purity_2020}. \\

\addlinespace

\cite{raghavan_microenvironment_2021} &
7 &
Broad Institute Single Cell Portal (SCP1644) &
\textit{Mapping:} \newline
Endothelial $\rightarrow$ endothelial; \newline
B\_Cells, Macrophage, T\_NK, T\_Regs, DC, pDC\_cell $\rightarrow$ immune; \newline
tumour $\rightarrow$ cancer cells. \newline
\textit{Note:} no fibroblasts available. \\

\addlinespace

\cite{baron_single-cell_2016} &
4 &
GSE84133 &
\textit{Mapping:} \newline
endothelial $\rightarrow$ endothelial; \newline
macrophage, mast, t\_cell $\rightarrow$ immune; \newline
activated\_stellate, quiescent\_stellate $\rightarrow$ fibroblasts. \newline
\textit{Note:} no cancer cells available. \\

\bottomrule
\end{tabularx}
\label{supptab:scrna_reference}
\end{table}

\subsection{Benchmark datasets}
\label{suppmat:benchmarkdt}

\subsubsection{Generation of the original \textit{in vivo} dataset}
\label{suppmat:invivo}

The \textit{in vivo} multi-omics dataset comprises transcriptomic and DNA methylation (DNAm) profiles from 47 tumour samples, that were generated exclusively for this study. Of these 47 samples, 24 form the VIVO benchmark dataset and 23 were held out as unseen validation data (VIVO$'$; Supplementary Material~\ref{suppmat:unseen}). RNA and DNA were simultaneously extracted from formalin-fixed, paraffin-embedded (FFPE) samples using the Manual-Quick DNA/RNA FFPE Miniprep Kit (Zymo Research).
Transcriptomic profiling was performed using a 3$'$ RNA-seq protocol, and DNA methylation was assessed using the Infinium MethylationEPIC 850K array (Illumina). For this dataset, ground truth information is only partially available. Specifically, we used estimates of the relative proportions of classical and basal-like tumour cells as inferred by PACpAInt~\cite{saillard_pacpaint_2023}, a deep learning-based tool for automated histological subtype classification from H\&E-stained slides (hematoxylin and eosin staining).

\textit{3$'$ RNA-seq protocol.} The protocol was applied as follows. Total RNA was fragmented, and the 3$'$ end of mRNA molecules was captured using a poly(T) reverse transcription primer containing a Unique Molecular Identifier (UMI). During reverse transcription, Illumina adapters were incorporated via a template-switching mechanism. The resulting fragments were amplified through two rounds of PCR to complete the adapter sequences and introduce sample-specific indices. The read structure consisted of paired-end 100~bp reads sequenced on an Illumina NovaSeq platform. Read~1 included a 26~bp UMI followed by the start of the poly(T) region. Read~2 began with a GGG motif (to be trimmed), followed by a variable-length 3$'$ tag insert and a poly(A) tail. Image analysis and base calling were performed using Illumina Real-Time Analysis (RTA) software version 3.4.4 with default settings. The data were preprocessed as follows:  Fastq file were aligned using STAR (2.7.1a) on UCSC hg38 genome, STAR  --genomeDir {STAR.index}  --readFilesCommand gunzip -c --readFilesIn {input.rev} --runThreadN {threads}  --sjdbGTFfile {params.gtf} --outFilterMismatchNoverLmax 0.08 --outSAMtype BAM SortedByCoordinate  --genomeLoad NoSharedMemory. BAM file where counted using featureCounts (v2.0.0) with options   -T 15  -Q  -t exon -g gene\_name. Gene counts were normalized using standard DESeq2 procedure

\textit{DNAm protocol.} 
The library and data were generated using standard Illumina Protocol(Infinium Methylation EPIC BeadChip). The raw DNA methylation intensity data files (IDAT) were processed with the lumi and methylumi R packages. We performed pre-normalization filtering (removing probes containing SNP, high intensity probes, not detected probes). We performed normalization using color balance adjustment and between sample normalisation by the "quantile" method.

\textit{Data accessibility.} The gene expression data and the DNA methylation data have been deposited on GEO under accession codes GSE328792 and GSE329368.

\subsubsection{In vitro dataset}
\label{suppmat:invitro}

We used a previously published \textit{in vitro} multi-omic dataset (GSE281305 DNA methylation MethEPIC 850K, GSE281204 RNA-seq) \cite{amblard_robust_2024}.  This dataset contains 30 controlled mixtures of 9 pure cell types commonly found in pancreatic ductal adenocarcinoma (PDAC): classical-like tumour cells, basal-like tumour cells, cancer-associated fibroblasts, endothelial cells and immune cells (B cells, CD4$^+$ cells, CD8$^+$ cells, neutrophils and M2-macrophages). Of these 30 mixtures, 15 form the VITR benchmark dataset and 15 were held out as unseen validation data (VITR$'$; Supplementary Material~\ref{suppmat:unseen}). In order to reduce the number of less abundant cell types, all immune cell types were grouped in a single population termed Immune. Cells were mixed in known proportions that are coherent with proportions in human PDAC samples.

\subsubsection{Simulated datasets}
\label{supp:simu}

\paragraph{Cell-type proportions.} For all simulated datasets, the vector of cell-type proportions $p_i$ of each sample $i$ was sampled in two steps, to control the balance between the two cancer cell types. First, the proportions of endothelial cells, fibroblasts, immune cells and cancer cells were drawn from a Dirichlet distribution, with parameter $\alpha_{\mathrm{real}} = (1.5, 4.5, 1, 3)$  or, for the rare-immune design, drawn from $\alpha_{\mathrm{rare}}$ is such that the immune type is rare. Second, the cancer fraction was split between the classical and basal-like subtypes according to a Dirichlet distribution with parameter $(0.5, 0.5)$, $(0.75, 0.25)$ or $(0.9, 0.1)$, each applied to one third of the samples, so that the datasets cover balanced as well as classical-dominated tumours. Each simulated dataset comprises $n = 60$ samples.

\paragraph{Linear mixing model.} Except for SBN5 (see below), simulated bulk profiles were generated by mixing pure cell-type profiles $X_{\mathrm{sim}}^{(m)} \in \mathbb{R}^{F^{(m)} \times k}$, with $F^{(m)}$ the number of features of modality $m \in \{\mathrm{RNA}, \mathrm{DNAm}\}$, and adding noise:
\begin{equation*}
Y_i^{(m)} = X_{\mathrm{sim}}^{(m)} p_i + \varepsilon_i^{(m)},
\end{equation*}
where $X_{\mathrm{sim}}^{(m)} \in \mathbb{R}^{F^{(m)} \times k}$ contains the pure profiles of the $k$ cell types, $F^{(m)}$ is the number of features of modality $m$, and $\varepsilon_i^{(m)}$ is a modality-specific noise term. The simulation profiles $X_{\mathrm{sim}}^{(m)}$ are distinct from the reference profiles $X^{(m)}$ provided to participants (Supplementary Material~\ref{suppmat:ref}). The datasets differ by their noise model and by the cell types of the simulation profiles, as described below.

\paragraph{Training and test halves.} For each simulated dataset, 30 of the 60 samples were drawn at random to form the training half, released to participants during the competition; the remaining 30 samples formed the test half, never released to participants and on which they were evaluated. In the modular benchmark, the training half forms the benchmark dataset, and the test half the unseen validation dataset (Supplementary Material~\ref{suppmat:unseen}).

\paragraph{Simulations SDN4, SDN5 and SDN6.} The first noise model adds a heteroscedastic $\chi^2$ noise to RNA data and a Gaussian noise to DNAm data. It was used in three simulations, which differ in the cell types of the simulation profiles compared with the reference profiles: the same five cell types (SDN5), one fewer (SDN4) or one more (SDN6). In SDN4, the basal-like subtype was removed from the simulation profiles, and its proportion was transferred to the classical subtype to keep the same total proportion of tumour cells. In SDN6, the profile of the additional cell type was generated by taking, for each feature, the value of one of the five existing cell types chosen uniformly at random; its proportion was set to 10\% on average, by subtracting 2\% from each of the five original cell types.

\paragraph{Simulations SDE5 and SDEL.} The second noise model accounts for the dependence between features. This dependence was inferred, conditionally on the cell types, from the residuals of the \textit{in vitro} dataset:
\begin{equation*}
\varepsilon_{\mathrm{vitro}, i}^{(m)} = Y_{\mathrm{vitro}, i}^{(m)} - X_{\mathrm{sim}}^{(m)} p_{\mathrm{vitro}, i},
\end{equation*}
where $p_{\mathrm{vitro}, i}$ is the vector of true proportions of sample $i$ of the \textit{in vitro} dataset. For DNAm, the number of features was first reduced by averaging the CpG probes of each gene promoter into a gene-level value, yielding 18,735 gene-level features. The correlation matrix of the residuals of all \textit{in vitro} samples was approximated by a low-rank factor decomposition estimated with an Expectation--Maximisation procedure~\cite{friguet_factor_2009}, from which a noise was generated, centred and scaled. Each feature of this noise was then multiplied by the standard deviation of the corresponding residuals, to reproduce its heteroscedasticity. For DNAm, probe-level noise was obtained by assigning to each probe the noise of its gene, plus a small centred uniform noise; intergenic probes received the noise of a gene-level feature sampled uniformly. This noise model was used for SDE5, with proportions drawn with $\alpha_{\mathrm{real}}$, and for SDEL, with proportions drawn with $\alpha_{\mathrm{rare}}$.

\paragraph{Simulation SDC5.} The third noise model infers the dependence between features with an empirical copula fitted on the centred and scaled residuals of the \textit{in vitro} dataset, using the R package \texttt{copula}. From this copula, a dependent noise was generated, with negative binomial margins for RNA and Beta margins for DNAm.

\paragraph{Simulation SBN5: pseudo-bulk samples from simulated single cells.} To obtain a dataset independent of the linear mixing model, we generated pseudo-bulk samples by aggregating synthetic single-cell RNA and DNAm profiles, simulated for 100 cells of each of the five cell types.

\begin{enumerate}
\item \emph{Single-cell RNA data.} For each gene $g$ shared by the bulk profiles and the single-cell data, and each cell type $c$, the count of gene $g$ in cell $j$ of type $c$ was sampled from a negative binomial distribution:
\begin{equation*}
z_{gj}^{(c)} \sim \mathrm{NB}(\mu_{gc}, \theta_{gc}), \qquad \mu_{gc} = E_{gc}^{\mathrm{bulk}} / N_{\mathrm{cells}},
\end{equation*}
where $E_{gc}^{\mathrm{bulk}}$ is the bulk expression of gene $g$ in the profile of cell type $c$, and $N_{\mathrm{cells}} = 5{,}000$ approximates the number of cells in a bulk sample, based on the ratio between the average number of transcripts per bulk sample and per single cell. The dispersion parameters $\theta_{gc}$ were estimated from the single-cell RNA-seq data of Peng et al.~\cite{peng_single-cell_2019}, by fitting, for each gene, an intercept-only negative binomial generalized linear model on the cells of type $c$. For genes with only zero counts, or for which the fit failed, $\theta_{gc}$ was sampled at random among the values estimated for the other genes of the same cell type.

\item \emph{Single-cell DNAm data.} For each probe $q$ and cell type $c$, the methylation value of probe $q$ in cell $j$ of type $c$ was sampled among three states (unmethylated, partially methylated and fully methylated):
\begin{equation*}
z_{qj}^{(c)} \in \{0, 0.5, 1\}, \qquad z_{qj}^{(c)} \sim \mathrm{Categorical}(\pi_0, \pi_{0.5}, \pi_1),
\end{equation*}
with probabilities matching the mean methylation $\mu_{qc}$ of probe $q$ in the profile of cell type $c$: $\pi_0 + \pi_{0.5} + \pi_1 = 1$ and $0 \cdot \pi_0 + 0.5 \cdot \pi_{0.5} + 1 \cdot \pi_1 = \mu_{qc}$ . As these two constraints do not determine the three probabilities, $\pi_1$ was drawn, for each probe and cell type, uniformly in $[\max(0, 2\mu_{qc} - 1), \mu_{qc}]$, the range in which all probabilities are non-negative, and $\pi_{0.5} = 2\mu_{qc} - 2\pi_1$ and $\pi_0 = 1 - 2\mu_{qc} + \pi_1$ were derived from it.

\item \emph{Aggregation.} Proportions $p_i$ were sampled with the two-step procedure described above. Pseudo-bulk samples were then generated with an adaptation of the generator of Avila Cobos et al.~\cite{avila_cobos_benchmarking_2020}, based on a pool of $N = 100$ cells. For each sample $i$ and cell type $c$, $\lfloor p_{ic} N \rfloor$ cells were sampled with replacement among the 100 simulated cells of type $c$, without correction of the total. Each distinct selected cell was then included once in the pseudo-bulk profile: the RNA profile of sample $i$ is the sum of the counts of these cells, and the DNAm profile the mean of their methylation values:
\begin{equation*}
Y_{gi}^{\mathrm{RNA}} = \sum_{j \in S_i} z_{gj}, \qquad
Y_{qi}^{\mathrm{DNAm}} = \frac{1}{|S_i|} \sum_{j \in S_i} z_{qj},
\end{equation*}
where $S_i$ is the set of distinct cells selected for sample $i$, all cell types combined.
\end{enumerate}

\subsection{Unseen validation datasets}
\label{suppmat:unseen}

To assess whether the generalist pipelines selected on the benchmark datasets generalise to new data, we used validation datasets that were never released to participants, on which no feedback was provided, and which played no role in the modular benchmark or in generalist pipeline selection.
Supplementary Table~\ref{supptab:unseen} summarises how the samples of each setting are distributed across the competition, the benchmark and the validation.

\paragraph{Simulated validation datasets.} Each simulated dataset was generated with 60 samples and split into two halves (Supplementary Material~\ref{supp:simu}). The training half, released to participants during the competition, forms the benchmark dataset; the test half, never released to participants, forms the corresponding validation dataset (SBN5$'$ to SDC5$'$). The two halves therefore share the same simulation design, noise model and pure cell-type profiles, and differ only by their samples.

\paragraph{Experimental validation datasets.} The in vitro and in vivo experiments yielded more samples than those included in the benchmark. VITR$'$ consists of the 15 in vitro mixtures of the GSE281305/GSE281204 experiment that are not part of VITR, with cell types mixed in the same range of proportions. VIVO$'$ consists of the 23 tumour samples of the original in vivo cohort that are not part of VIVO; as for VIVO, only the proportions of the classical and basal-like tumour subtypes inferred by PACpAInt are available, and VIVO$'$ is scored on cell-type correlations of these two subtypes only.

\paragraph{Evaluation.} The generalist pipelines of each strategy were applied to the validation datasets without any re-tuning, with the same reference profiles, the same feature space (Supplementary Material~\ref{suppmat:feature}) and the same aggregate score as for the benchmark datasets. As for the benchmark, statistical tests were performed on the eight fully-annotated validation datasets, VIVO$'$ being reported separately.

%\paragraph{Scope of the validation.} Because the validation datasets were never seen by participants and were not used for recipe selection, they control for the selection bias arising from the large number of evaluated pipelines and from the design of competition strategies. They share, however, the tissue, the simulation designs, the pure cell-type profiles and the experimental batches of the benchmark datasets, and therefore do not assess transfer to independent cohorts, other tissues or other reference profiles.

All validation datasets are available on Zenodo~\cite{hadaca3}, in the same format as the benchmark datasets.

\begin{table}[ht]
\centering
\caption{Correspondence between the competition phases and the data used in the modular benchmark. Each setting was split into two disjoint sets of samples. The training set, released to participants during phases 1 and 2, forms the benchmark datasets. The test set, used only to score the single final submission of phase 3 and never released to participants, forms the unseen validation datasets, so that neither competition strategies nor generalist pipelines could be learned or tuned on it.}
\label{supptab:unseen}
\small
\begin{tabular}{p{4.2cm}p{5cm}p{5cm}}
\toprule
 & \textbf{Training set} & \textbf{Test set} \\
\midrule
\multicolumn{3}{l}{\emph{Competition}} \\
Phases & Phases 1 (introduction) and 2 (robustness) & Phase 3 (generalization) \\
Access for participants & Released to participants & Never released to participants \\
Feedback & Leaderboard scores & None; single final submission \\
\midrule
\multicolumn{3}{l}{\emph{Modular benchmark}} \\
Role & Benchmark datasets (pipeline evaluation and generalist pipeline selection) & Unseen validation datasets (evaluation of generalist pipelines, without re-tuning) \\
\midrule
\multicolumn{3}{l}{\emph{Number of samples}} \\
Simulated (each of 7 designs) & 30 (SBN5, \ldots, SDC5) & 30 (SBN5$'$, \ldots, SDC5$'$) \\
In vitro & 15 (VITR) & 15 (VITR$'$) \\
In vivo & 24 (VIVO) & 23 (VIVO$'$) \\
\bottomrule
\end{tabular}
\end{table}

\subsection{Feature space harmonization}
\label{suppmat:feature}
To ensure comparability across benchmark datasets and reference profiles, 
we restricted the feature space to the intersection of features within each omic, across the nine benchmark datasets and the reference profiles for that omic (e.g., the set of RNA genes common to all RNA datasets and the RNA reference; separately, the set of DNAm CpG probes common to all DNAm datasets and the DNAm reference). The within-modality intersection only ensures a given omic's feature space is consistent across the nine datasets and the reference, which is necessary for any single pipeline to be applied uniformly across our benchmark; it does not discard cross-modal complementarity.

To limit the size of the distributed datasets and facilitate the 
execution of computationally intensive methods, an additional 
restricted probe set was defined as the intersection of the authorized 
probes with the Illumina HumanMethylation27k annotation, 
yielding $F_{\text{DNAm}} = 23,724$ probes. This subset retains probes 
consistently covered across dataset while preserving 
biologically informative methylation variation. All deconvolution 
methods were evaluated on this harmonized feature space to ensure fair comparison.

\section{Competition setup}
\label{suppmat:competition}

%\subsubsection{Task and data overview}

\begin{table}[!ht]
\centering
\small
\caption{Overview of the competition phases.}
\label{supptab:competphase}
\begin{tabular}{@{}ll@{}}
\toprule
\textbf{Phase} & \textbf{Key characteristics} \\
\midrule
1 -- Introduction   & Single dataset: 5 cell types; matched RNA-seq and DNAm + baseline scripts \\
2 -- Robustness     & Multiple datasets: Noise, missing/extra cell types, pseudo-bulk, \textit{in vitro/vivo} mixtures \\
3 -- Generalization & Unseen data; no feedback; single final submission \\
\bottomrule
\end{tabular}
\end{table}

\subsection{Starting kit and baselines}
\label{suppmat:starting}

\begin{table}[ht]
\centering
\small
\caption{Baseline scripts provided in the starter kit.}
\label{supptab:baseline}
\begin{tabular}{@{}p{0.32\linewidth}p{0.63\linewidth}@{}}
\toprule
\textbf{Script} & \textbf{Description} \\
\midrule
\texttt{submission\_script.R} &
NNLS deconvolution on RNA-seq data. \\
\texttt{\_nnlsmultimodal.R} &
NNLS on each modality independently, with late averaging. \\
\texttt{\_nnlsmultimodalSource.R} &
Demonstrates loading external R scripts or \texttt{.rds} files. \\
\texttt{\_installpkgcran.R} &
Demonstrates CRAN package installation within a submission. \\
\texttt{Submission\_script.py} &
Python equivalent of the NNLS baseline. \\
\bottomrule
\end{tabular}
\end{table}

\subsection{Evaluation strategy}
\label{suppmat:metric}

\subsubsection{Individual metrics}

Let $A \in \mathbb{R}^{k \times n}$ denote the ground-truth proportion matrix 
and $\hat{A} \in \mathbb{R}^{k \times n}$ the predicted proportion matrix, 
where $k$ is the number of cell types and $n$ the number of samples.
Twelve metrics are computed, grouped into four families (Supplementary Table~\ref{supptab:metrics}). 

We measured the quality of the deconvolution based on various metrics. We used error metrics such as the Root Mean Square Error (RMSE), the Mean Absolute Error (MAE), the Aitchison distance and the Jensen-Shannon divergence (JSD) between the predicted and the real proportions matrix. We looked also at the angle between predicted and real samples and derived two metrics from this angle: the Angular Inequality/Disproportionality (AID) and the Sine-Diagonal ID (SDID). Finally, we included Pearson's and Spearman's correlations on the whole vectorized matrix (global), on the rows (cell types) or on the columns (samples).  All these metrics are not fully independent, as some of these properties are mathematically related. For the columns' and the rows' correlations, we did the arithmetic mean along the respective axis. In the case of an extra cell type in the reference matrix used for the deconvolution compared to the ground truth, we added a row of zeroes in the ground truth matrix to match the number of cell types in the predicted matrix. In the case of a missing cell type in the reference, we kept in the ground truth matrix only the cell types that have been estimated.

\begin{table}[H]
\centering
\small
\caption{Summary of the twelve evaluation metrics. Dir. indicated directionality of best score.}
\label{supptab:metrics}
\begin{tabularx}{\linewidth}{p{2.5cm} p{2.5cm} p{1cm} p{1cm} p{4.5cm}}
\toprule
\textbf{Family} & \textbf{Metric} & \textbf{Dir.} & \textbf{Weight} & \textbf{Formula} \\
\midrule

\multirow{4}{*}{Cohort Correlation}
& Pearson (global)      & $\uparrow$ & $1/16$ & $\rho_P(A, \hat{A})$ \\
& Spearman (global)     & $\uparrow$ & $1/16$ & $\rho_S(A, \hat{A})$ \\
& Pearson (sample)      & $\uparrow$ & $1/16$ & $\frac{1}{n}\sum_i \rho_P(A_{\cdot i}, \hat{A}_{\cdot i})$ \\
& Spearman (sample)     & $\uparrow$ & $1/16$ & $\frac{1}{n}\sum_i \rho_S(A_{\cdot i}, \hat{A}_{\cdot i})$ \\

\midrule
\multirow{2}{*}{Cell-type correlation}
& Pearson (cell type)   & $\uparrow$ & $1/8$  & $\frac{1}{k}\sum_c \rho_P(A_{c\cdot}, \hat{A}_{c\cdot})$ \\
& Spearman (cell type)  & $\uparrow$ & $1/8$  & $\frac{1}{k}\sum_c \rho_S(A_{c\cdot}, \hat{A}_{c\cdot})$ \\

\midrule

\multirow{4}{*}{Error}
& RMSE     & $\downarrow$ & $1/16$ & $\sqrt{\frac{1}{kn}\sum_{c,i}(A_{ci} - \hat{A}_{ci})^2}$ \\
& MAE      & $\downarrow$ & $1/16$ & $\frac{1}{kn}\sum_{c,i}|A_{ci} - \hat{A}_{ci}|$ \\
& Aitchison & $\downarrow$ & $1/16$ & $\frac{1}{n}\sum_i d_A(A_{\cdot i}, \hat{A}_{\cdot i})$, values clipped at $10^{-9}$ \\
& JSD      & $\downarrow$ & $1/16$ & $\frac{1}{n}\sum_i \mathrm{JSD}(A_{\cdot i} \| \hat{A}_{\cdot i})$ \\

\midrule

\multirow{2}{*}{Geometric}
& AID  & $\downarrow$ & $1/8$ & $\frac{1}{n}\sum_i \frac{90}{\pi/2} \arccos\!\left(\frac{A_{\cdot i} \cdot \hat{A}_{\cdot i}}{\|A_{\cdot i}\|_2 \|\hat{A}_{\cdot i}\|_2}\right)$ \\[6pt]
& SDID & $\downarrow$ & $1/8$ & $\frac{1}{n}\sum_i \sqrt{\sin\!\left(\arccos\!\left(\frac{A_{\cdot i} \cdot \hat{A}_{\cdot i}}{\|A_{\cdot i}\|_2 \|\hat{A}_{\cdot i}\|_2}\right)\right)}$ \\

\bottomrule
\end{tabularx}
\end{table}

Correlation between metrics are presented in Supplementary Figure \ref{suppfig:cormetrics}

\begin{figure}[!ht]
\centering
\includegraphics[width=0.6\linewidth]{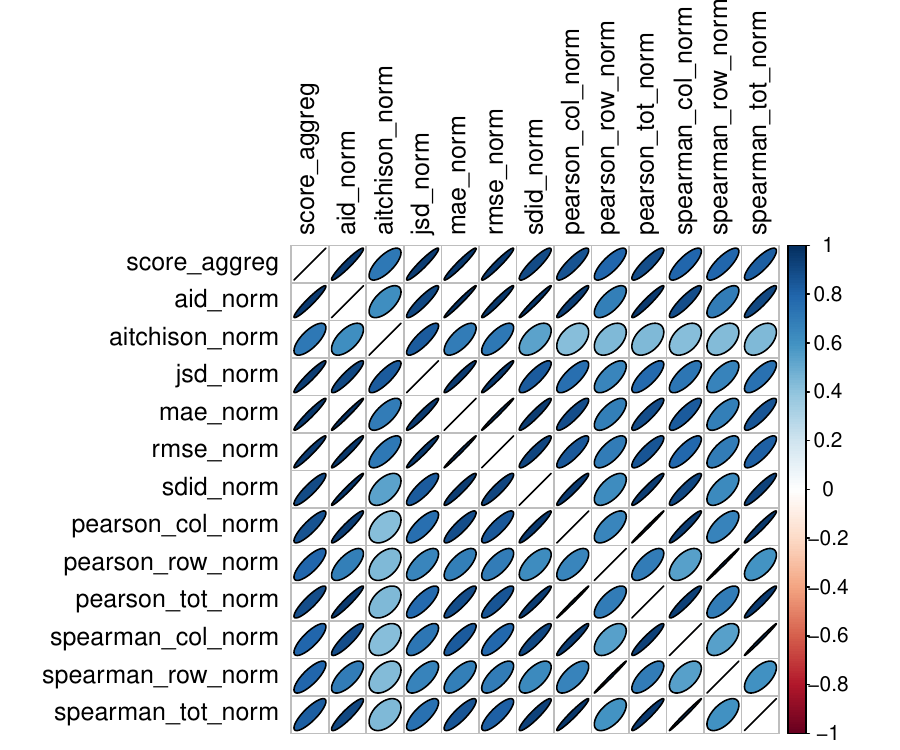}
\caption{Correlations observed across the 12 metrics included in the benchmark after center-scaling transformation for all deconvolution methods evaluated, as well as the aggregate score.}
\label{suppfig:cormetrics}
\end{figure}

\subsubsection{Normalization}
Each metric $m$ is normalized to $[0,1]$ using a linear shift between its 
worst-case and best-case values:
\[
\tilde{m} = \frac{m - m_{\text{worst}}}{m_{\text{best}} - m_{\text{worst}}},
\]
where $m_{\text{best}}$ and $m_{\text{worst}}$ are computed from an ideal 
prediction ($\hat{A} = A$) and a deliberately bad prediction that assigns 
all probability mass to the least abundant cell type, respectively. All 
normalized metrics are oriented so that $\tilde{m} = 1$ is the best possible 
score. For the correlations metrics, the best score is 1 and the worst is -1. For the angle-based metrics, the best and worst scores are defined by the best and worst angles: the best angle is 0 and the worst is $\frac{\pi}{2}$ since we have compositional data. For the error metrics, the best score is 0 and the worst has been computed based on the worst possible prediction. The worst prediction is a prediction with a proportion of 1 for the lowest abundant cell type and zeroes everywhere else. After the center-scaling procedure, we transformed the error and angle-based metrics by subtracting them from 1, such that the best score is 1 and the worst is 0 for all metrics. Metrics worse than the worst-case value are set to 0, so that the aggregate score of such a pipeline is 0. 

\subsubsection{Aggregate score}
The twelve normalized metrics are combined into a single aggregate score via 
a weighted geometric mean:
\[
s = \prod_{m} \tilde{m}^{w_m}, \quad \sum_m w_m = 1,
\]
with the following weight scheme: cell-type correlation metrics (2), other correlation metrics (4), error metrics (4) and geometric metrics (2) each account for one fourth of the score, with equal weights within each group.

\subsubsection{Special case: partial ground truth for VIVO}
\label{suppmat:partialgroundtruth}
For the \textit{in vivo} dataset (VIVO), ground truth is only available for two cell types (basal-like and classical). Basal-like and classical tumour cell proportions can be quantified from histology via the PACpAInt classifier, but there is currently no comparably validated, quantitative estimate for fibroblast, endothelial, and immune cell proportions in these tumour sections (see \ref{suppmat:invivo} for details on the VIVO dataset's construction). For VIVO, only cell-type-wise Pearson and Spearman correlations are computed, each weighted equally at $w = 1/2$, while the other 10 metrics and the 3 unavailable cell types are not scored.

\subsection{Description of the winning method JOKER}
\label{suppmat:JOKER}

The proposed method follows a omic-specific pre-processing strategy, combined with independent deconvolution and late integration of estimates derived from RNA-seq and DNAm data.

\paragraph{RNA-seq pre-processing.}
Raw RNA-seq counts were first normalized for sequencing depth using counts per million (CPM), obtained by dividing each column by its total counts and scaling by $10^6$. To reduce noise and focus on informative features, the 5,000 most variable genes were selected based on their variance after applying a variance-stabilizing transformation defined as $\log_2(x + c)$, where $c$ is the median of non-zero CPM values. To ensure comparability between mixture and reference profiles, gene expression values were further scaled by the mean expression across both datasets. Specifically, for each gene $g$, values were divided by $\mu_g = \text{mean}(X_g^{\text{mix}}) + \text{mean}(X_g^{\text{ref}}) + \epsilon$, where $\epsilon$ is a small constant to avoid division by zero.

\paragraph{DNA methylation pre-processing.}
For DNAm data, CpG sites were filtered based on their variability across the reference profiles. Only sites with variance greater than 0.1 were retained, resulting in a subset of informative features used for deconvolution.

\paragraph{Deconvolution model.}
For each modality, cell-type proportions were estimated independently using a non-negative least squares (NNLS) model. For a given sample $i$, the bulk profile $Y_i$ is modeled as a linear combination of reference profiles $X$ weighted by a vector of proportions $p_i$:
\[
\hat{p}_i = \arg\min_{p_i \geq 0} \| Y_i - X p_i \|_2^2,
\]
followed by normalization to enforce the sum-to-one constraint:
\[
\tilde{p}_i = \frac{\hat{p}_i}{\sum_{c=1}^{k} \hat{p}_{ic}}.
\]

\paragraph{Late integration of modalities.}
Final cell-type proportions were obtained using a late integration strategy. For most cell types, proportions estimated from RNA-seq and DNAm were averaged:
\[
p_i^{\text{final}} = \frac{1}{2} \left( p_i^{\text{RNA}} + p_i^{\text{DNAm}} \right).
\]
However, for basal-like and classical tumour cell types, RNA-seq-based estimates were deemed less reliable and were therefore replaced by DNAm-based estimates. The resulting proportions were finally renormalized to ensure that they sum to one for each sample.

\begin{figure}[!ht]%
\centering
\includegraphics[width= \linewidth]{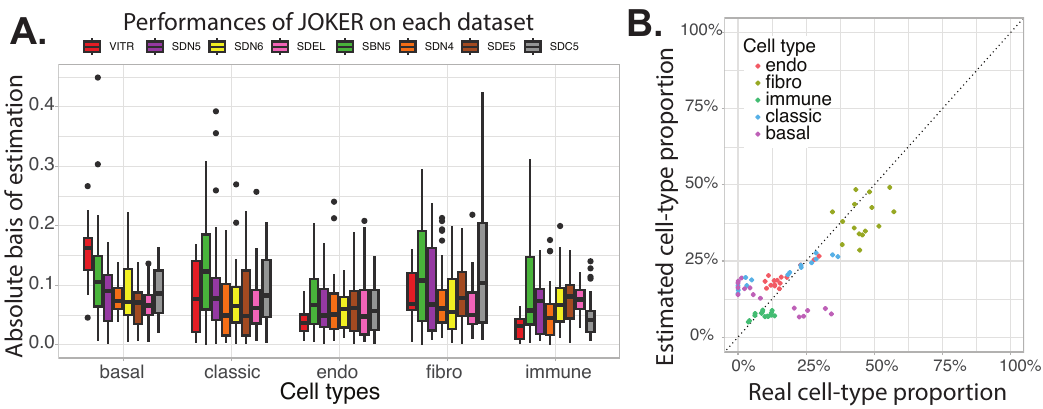}
\caption{Performances of the JOKER method on cell type proportion estimation. \textbf{A.} Boxplots of absolute estimation bias across cell types and benchmark datasets. 
The absolute bias corresponds to the absolute difference between the estimated and ground-truth proportions for each cell type in each sample. 
\textbf{B.} Example scatterplot comparing estimated and ground-truth proportions on the VITR dataset, where each point corresponds to one sample and colors indicate cell types. The JOKER method was developed by R. De Wit, A. Toffano, R. Le Clech, and F. Privé.}
\label{suppfig:bestsubmission}
\end{figure}

%\newpage

\section{Benchmark}
\label{supp:benchmark}

\subsection{Algorithms description}

\subsubsection{Pre-processing}
\label{supp:benchmark_pp}

Three types of pre-processing were tested (Supplementary Table~\ref{supptab:pp_methods}. 

\begin{table}[!h]
\centering
\small
\caption{Comparison of pre-processing options, applied to both RNA-seq and 
DNA methylation data. All options transform the mixture and reference profiles 
prior to feature selection and deconvolution.}
\label{supptab:pp_methods}
\begin{tabularx}{\linewidth}{p{1.2cm} p{4cm} p{10cm}}
\toprule
\textbf{Method} & \textbf{Principle} & \textbf{Key idea} \\
\midrule

\texttt{ppID} &
Identity (no pre-processing) &
Returns the data unchanged. Serves as a baseline to assess the benefit 
of pre-processing. Applied identically to RNA-seq and DNAm data. \\

\texttt{SumNorm} &
Column-sum normalization &
Normalizes each sample by its total count, so that all samples sum to 1:
\[
\tilde{X}_{gi} = \frac{X_{gi}}{\sum_{g'} X_{g'i}}.
\]
Makes samples comparable regardless of sequencing depth or global methylation 
level differences. Applied identically to RNA-seq and DNAm data. \\

\texttt{LogNorm} &
Log-normalization &
Applies Seurat's \texttt{LogNormalize} function, which scales each sample 
by its total count, multiplies by a scale factor of $10^4$, and applies 
a log transformation:
\[
\tilde{X}_{gi} = \log\!\left(1 + \frac{X_{gi}}{\sum_{g'} X_{g'i}} \times 10^4\right).
\]
For bulk RNA-seq and DNAm data, an exponential transformation is applied 
post hoc to recover a linear scale: $\hat{X}_{gi} = \exp(\tilde{X}_{gi})$, 
which is equivalent to a $10^4$-scaled column-sum normalization plus 1, without 
log transformation. For scRNA-seq data, the log-transformed values are 
retained directly. \\

\bottomrule
\end{tabularx}
\end{table}

%\newpage

\subsubsection{Feature selection}
\label{supp:benchmark_fs}

Feature selection consists in identifying a subset of genes or CpG probes that are most informative for distinguishing cell types, prior to deconvolution. 
We evaluated diverse strategies for both RNA-seq and DNAm data, described in Supplementary Tables~\ref{suptab:fs_rna_methods} and~\ref{suptab:fs_met_methods}.

\begin{table}[H]
\centering
\small
\caption{Comparison of RNA-seq feature selection options. All options select a 
subset of genes from the mixture and reference profiles prior to deconvolution.}
\label{suptab:fs_rna_methods}
\begin{tabularx}{\linewidth}{p{2.2cm} p{4cm} p{9cm}}
\toprule
\textbf{Method} & \textbf{Principle} & \textbf{Key idea} \\
\midrule

\texttt{fsID} &
Identity (no selection) &
Returns the data unchanged. All available genes are passed to the deconvolution 
module. Serves as a baseline to assess the benefit of feature selection. \\

\texttt{Toastbulknbfs} &
Marker gene selection from bulk reference &
Applies \texttt{TOAST::findRefinx}~\cite{li_toast_2019} to the bulk RNA-seq 
reference matrix to select the top $n = 1{,}000$ marker genes. Marker genes are 
defined as genes with high cell-type specificity in the reference profiles. 
Both mixture and reference are then restricted to this gene set. \\

\texttt{Toastvst} &
Marker gene selection from VST-transformed bulk reference &
Extends \texttt{Toastbulknbfs} by applying a variance-stabilizing transformation 
(VST) to the bulk reference prior to marker selection.
\texttt{TOAST::findRefinx} is then applied to select the top 
$n = 1{,}000$ marker genes. The VST reduces the influence of highly expressed 
genes on marker selection. \\

%\texttt{SCcluster} &
%Differential expression markers from clustered scRNA-seq &
%Uses Seurat's \texttt{FindAllMarkers} (Wilcoxon rank-sum test) on clustered scRNA-seq data to identify cell-type-specific marker genes. Genes are retained if they satisfy: adjusted $p$-value $< 0.05$, fraction expressed in the target cluster $> 0.6$, and fraction expressed in other clusters $< 0.3$. This combines statistical significance with specificity criteria. \\

%\texttt{scpseudobulk} &
%pseudo-bulk references using scRNA-seq with a cell-type specific gene set&
%For each pair of cell types $(c, c')$ and each scRNA-seq dataset $d$, computes a $t$-statistic  for each gene $g$. The top $n_{\text{top}} = 20$ genes with highest and lowest $t$-statistics are retained per pair, across all datasets. The final gene set is the union of genes consistently selected across all pairwise comparisons. A pseudo-bulk reference is then constructed by averaging single-cell profiles per cell type over the selected genes. \\

\bottomrule
\end{tabularx}
\end{table}

\begin{table}[!h]
\centering
\small
\caption{Comparison of DNA methylation feature selection options. All options 
select a subset of CpG probes from the mixture and reference profiles prior 
to deconvolution.}
\label{suptab:fs_met_methods}
\begin{tabularx}{\linewidth}{p{2.2cm} p{4cm} p{9cm}}
\toprule
\textbf{Method} & \textbf{Principle} & \textbf{Key idea} \\
\midrule

\texttt{fsID} &
Identity (no selection) &
Returns the data unchanged. All available CpG probes are passed to the 
deconvolution module. Serves as a baseline to assess the benefit of 
feature selection. \\

\texttt{Toastnbfs} &
Full probe ranking via TOAST &
Applies \texttt{TOAST::findRefinx}~\cite{li_toast_2019} to the DNAm 
reference matrix with $n_{\text{marker}} = p$ (all probes), returning all 
probes ranked by cell-type specificity. Equivalent to \texttt{fsID} in terms 
of probe set, but reorders probes by discriminative power. \\

\texttt{Toastpercent} &
Partial probe selection via TOAST &
Applies \texttt{TOAST::findRefinx} to the DNAm reference matrix retaining 
the top $\lfloor 0.8 \times p \rfloor$ probes, where $p$ is the total number 
of probes. Removes the least cell-type-specific 20\% of probes. \\

\texttt{mostmethylated} &
Biologically informed probe selection &
Retains probes with high methylation levels based on biological prior knowledge 
that cancer cells exhibit high methylation. For each cell type $c$, probes 
above the 75th percentile of methylation are selected. \\

\texttt{maxdiscriminant} &
Maximally discriminant non-overlapping probes &
For each cell type $c$, probes are ranked by their absolute deviation from 
the global mean. The largest $n$ such that the top-$n$ probe sets across cell types are 
mutually non-overlapping is determined iteratively (up to $n_{\max} = 100$ 
per cell type). This ensures maximal cell-type specificity with no 
redundancy across types. \\

\texttt{splsda} &
Sparse PLS-DA on logit-transformed reference &
First applies \texttt{TOAST::findRefinx} to retain the top $10{,}000$ most 
variable probes. A logit transformation is then applied to stabilize 
beta-valued methylation data.
Sparse partial least squares discriminant analysis 
(sPLS-DA)~\cite{le_cao_sparse_2011} is then fitted on reference data with 
cell type as the response, retaining $n = 1{,}000$ probes per component over 
2 components. The final probe set is the union of selected variables from 
both components. \\

\bottomrule
\end{tabularx}
\end{table}

\subsubsection{Deconvolution}
\label{supp:benchmark_deconv}

\begin{table}[p]
\centering
\small
\caption{Comparison of deconvolution options. All options estimate cell-type 
proportions $p_i$ from a mixture profile $Y_i$ and a reference matrix $X$.}
\label{suptab:deconv_methods}
\begin{tabularx}{\linewidth}{p{2.2cm} p{4cm} p{9cm}}
\toprule
\textbf{Method} & \textbf{Principle} & \textbf{Key idea}  \\
\midrule

\texttt{lm} &
Ordinary least squares &
Minimizes $\|Y_i - X p_i\|_2^2$ without intercept; negative coefficients are 
set to zero and proportions are renormalized  \\

\texttt{nnls} &
Non-negative least squares &
Minimizes $\|Y_i - X p_i\|_2^2$ subject to $p_i \geq 0$, enforcing 
non-negativity by construction (Lawson-Hanson algorithm)~\cite{mullen_nnls_2007}, non-negativity by construction; proportions are renormalized post hoc 
to satisfy $\mathbf{1}^\top p_i = 1$.  \\

%\texttt{nnlslargeref} &
%NNLS with reference truncation &
%Extends NNLS by iteratively removing one reference cell type at a time and 
%retaining the truncated model if it yields lower reconstruction RMSE; handles potential over-specification of the reference. \\

\texttt{epic} &
Constrained least squares with internal normalization &
Solves a simplex-constrained least squares problem~\cite{racle_simultaneous_2017}:
\[
\hat{p}_i = \arg\min_{\substack{p \geq 0 \\ \mathbf{1}^\top p = 1}}
\left\| \bar{Y}_i - \bar{X} p \right\|_2^2,
\]
where $\bar{Y}_i$ and $\bar{X}$ denote internally TPM-normalized versions of the 
mixture and reference profiles. The simplex constraint $\mathbf{1}^\top p = 1$ is 
enforced during optimization, unlike NNLS where it is applied post hoc. No 
uncharacterized cell population is included (\texttt{withOtherCells=F}). In the original EPIC formulation, gene-specific weights $w_g \propto 1/V_{gc}$ 
down-weight genes with high variability across reference replicates. However, as 
our reference consists of a single meta-profile per cell type (no replicates 
available), all weights are equal ($w_g = 1$), reducing the weighted objective 
to a standard least squares criterion. \\

\texttt{epic} &
Constrained least squares (EPIC) &
Uses the \texttt{EPIC} function of the EPIC package~\cite{racle_simultaneous_2017}, which solves, for each sample, a least squares problem under non-negativity and sum constraints:
\[
\hat{p}_i = \arg\min_{\substack{p \geq 0 \\ 0.99 \,\leq\, \mathbf{1}^\top p \,\leq\, 1}}
\left\| Y_i - X p \right\|_2^2 ,
\]
solved by constrained optimisation (\texttt{constrOptim}); the estimate is then rescaled to sum to one. Unlike NNLS, where the sum-to-one constraint is applied post hoc, the sum is constrained during the optimisation. All the features of the mixture are used as signature genes, without internal rescaling of the data (\texttt{scaleExprs = FALSE}), and no uncharacterised cell population is included (\texttt{withOtherCells = FALSE}). EPIC can weight each gene by the inverse of its variability across reference samples; as no reference variability is provided (a single meta-profile per cell type), all genes have the same weight and the criterion is an unweighted least squares. The returned estimates are the mRNA proportions, without correction for the mRNA content of each cell type. \\

\texttt{RLR} &
Robust linear regression &
Applies robust regression via the RPC algorithm 
implemented in EpiDISH~\cite{zheng_epidish_2020}; down-weights outlier 
features via iteratively reweighted least squares (IRLS). \\

\texttt{wRLR} &
Robust linear regression with Poisson weights &
Extends RLR by weighting features inversely proportional to their mean reference 
expression, mimicking Poisson variance stabilization; 
$w_g \propto 1/\bar{x}_g$. Weighting scheme assumes Poisson-like mean-variance relationship, which may 
not hold for all feature types (e.g., DNA methylation beta values). \\

\texttt{RLRnnls} &
Ensemble: RLR and NNLS &
Runs both RLR and NNLS independently and selects the estimate with lower 
reconstruction RMSE after column-sum normalization.  This strategy aims to combine the robustness of RLR with the non-negativity guarantees of NNLS, defaulting to NNLS when RLR fails.\\

\bottomrule
\end{tabularx}
\end{table}

All deconvolution options solve a variant of the following problem: given a mixture 
profile $Y_i \in \mathbb{R}^p$ and a reference matrix $X \in \mathbb{R}^{p \times k}$, 
estimate a proportion vector $p_i \in \mathbb{R}^k$ such that:
\[
Y_i \approx X p_i, \quad p_i \geq 0, \quad \sum_{c=1}^{k} p_{ic} = 1.
\]
options differ in how they handle noise, outliers, and reference misspecification (Supplementary Table~\ref{suptab:deconv_methods}).

\subsubsection{Early integration strategies} 
\label{suppmat:integrationEI}

\begin{table}[!ht]
\centering
\small
\caption{Comparison of early integration strategies for multi-omics deconvolution. Each method defines an integration operator $\mathcal{F_\text{early}}$, which is applied prior to deconvolution.
}
\label{supptab:early_integration_summary}

\begin{tabularx}{\linewidth}{p{1.8cm} p{3cm} p{5.2cm} p{5cm}}
\toprule
\textbf{Method} & \textbf{Principle} & \textbf{Key idea} & \textbf{Main limitation} \\
\midrule

\texttt{Concat} &
Raw feature concatenation &
Stacks RNA expression and DNAm level without transformation; preserves original scale and structure &
Sensitive to feature magnitude differences between modalities \\
\midrule
\texttt{ConcatStd} &
Normalized concatenation &
Concatenation followed by sample-wise Gaussianization across features &
Ignores cross-modal correlations and dependancies \\
\midrule
\texttt{MCIA} &
Latent linear embedding  &
Learns a low-dimensional representation via multiblock co-inertia analysis &
Sensitive to non-linear cross-omic relationships \\
\midrule
\texttt{MKL} &
Non-linear kernel embedding &
Constructs a joint non-linear representation via kernel PCA over combined modality-specific kernels &
Sensitive to kernel choice and computationally expensive on large datasets \\
\midrule
\texttt{OT}  &
Optimal transport-based representation &
Aligns RNA and DNAm distributions through optimal transport using uniPort~\cite{cao_unified_2022}, and learns a coupled latent space &
High computational cost and dependence on distance metric quality \\

\bottomrule
\end{tabularx}
\end{table}

\paragraph{Raw feature concatenation (\texttt{Concat}).}
In the case of raw concatenation, the integration operator $\mathcal{F}_{\text{concat}}$ 
is defined as a simple feature-level stacking of both omics without any transformation:
\[
\mathcal{F}_{\text{concat}}:\quad
\begin{cases}
\tilde{Y} =
\begin{bmatrix}
Y^{\text{RNA}} \\
Y^{\text{DNAm}}
\end{bmatrix}, \\[10pt]
\tilde{X} =
\begin{bmatrix}
X^{\text{RNA}} \\
X^{\text{DNAm}}
\end{bmatrix},
\end{cases}
\quad \tilde{Y}, \tilde{X} \in \mathbb{R}^{(p_{\text{RNA}} + p_{\text{DNAm}}) \times n},
\]
where $n = n_{\text{mix}}$ or $n = k$ depending on whether mixture or reference 
samples are considered, and $p_{\text{RNA}}, p_{\text{DNAm}}$ denote the number 
of features in each modality.
No normalization, scaling, or latent transformation is applied beyond the 
concatenation itself.
The resulting representations $\tilde{Y}$ and $\tilde{X}$ are directly used as 
mixture and reference inputs for downstream deconvolution.

\paragraph{Normalized concatenation (\texttt{ConcatStd}).}
In this case, the integration operator $\mathcal{F}_{\text{scale}}$ performs 
feature-level concatenation of RNA-seq and DNAm data, followed by sample-wise 
normalization via a Gaussian CDF mapping.
The concatenation step stacks both modalities along the feature dimension:
\[
\tilde{Y} =
\begin{bmatrix}
Y^{\text{RNA}} \\
Y^{\text{DNAm}}
\end{bmatrix}, \quad
\tilde{X} =
\begin{bmatrix}
X^{\text{RNA}} \\
X^{\text{DNAm}}
\end{bmatrix},
\quad \tilde{Y}, \tilde{X} \in \mathbb{R}^{(p_{\text{RNA}} + p_{\text{DNAm}}) \times n},
\]
where $n = n_{\text{mix}}$ or $n = k$ depending on whether mixture or reference 
samples are considered, and $p_{\text{RNA}}, p_{\text{DNAm}}$ denote the number 
of features in each modality.
Each sample $i$ is then independently normalized across its features and mapped 
to a Gaussian CDF scale: for each feature $j$ of sample $i$,
\[
\mathcal{F}_{\text{scale}}(\tilde{y}_{ij}) =
\Phi\!\left(
\frac{\tilde{y}_{ij} - \mu_i}{\sigma_i}
\right),
\]
where $\mu_i$ and $\sigma_i$ denote the empirical mean and standard deviation of 
sample $i$ computed across all features, and $\Phi(\cdot)$ is the standard normal 
cumulative distribution function. The same sample-wise transformation is applied to $\tilde{X}$,  using the mean and standard deviation of each sample computed 
across its own features.
The resulting representations $\tilde{Y}$ and $\tilde{X}$ are directly used as 
mixture and reference inputs for downstream deconvolution.

\paragraph{Latent linear embedding (\texttt{MCIA}).}
In this case, the integration operator $\mathcal{F}_{\text{mCIA}}$ constructs a joint 
latent representation of RNA-seq and DNAm data using Multiple Co-Inertia Analysis 
(MCIA), implemented via the \textbf{omicade4} 
package~\cite{meng_multivariate_2014}. This method performs a feature-to-latent space 
transformation by seeking a consensus synthetic variable that maximizes the 
co-inertia between modality-specific projections.
First, mixture and reference samples are concatenated within each modality:
\[
W^{\text{RNA}} =
\left[
Y^{\text{RNA}}, X^{\text{RNA}}
\right], \quad
W^{\text{DNAm}} =
\left[
Y^{\text{DNAm}}, X^{\text{DNAm}}
\right] \in \mathbb{R}^{p \times (n_{\text{mix}} + k)},
\]
where columns correspond to samples, $p$ denotes the number of features, and 
$k$ denotes the number of reference cell types, with one reference sample per 
cell type (i.e.\ $n_{\text{ref}} = k$).
MCIA finds a $d$-dimensional synthetic variable $Z_s \in \mathbb{R}^{d \times 
(n_{\text{mix}}+k)}$ and modality-specific projection matrices 
$A^{\text{RNA}}, A^{\text{DNAm}} \in \mathbb{R}^{p \times d}$ by solving:
\[
\max_{A^{\text{RNA}},\, A^{\text{DNAm}},\, Z_s}
\sum_{m \in \{\text{RNA},\text{DNAm}\}}
\text{cov}^2\!\left({A^{(m)}}^\top W^{(m)},\, Z_s\right),
\]
where the maximization is performed iteratively over $d$ orthogonal components. In practice, $d = 10$ components are computed and retained, 
a value fixed empirically and not tuned per dataset.
The latent coordinates are shifted to ensure non-negativity ($Z \geq 0$ coordinate-wise): 
\[
Z = Z_s - \min_{i,j}(Z_s)_{ij}, \quad Z \in \mathbb{R}^{d \times (n_{\text{mix}} + k)},
\]
where $\mathcal{F}_{\text{mCIA}}$ returns the synthetic variable $Z_s$.
The latent representation is then partitioned into mixture and reference components:
\[
Z =
\left[
Z_{\text{mix}}, Z_{\text{ref}}
\right],
\quad
Z_{\text{mix}} \in \mathbb{R}^{d \times n_{\text{mix}}}, \;
Z_{\text{ref}} \in \mathbb{R}^{d \times k}.
\]
Deconvolution is performed sample-wise in the aligned latent space: for each 
mixture sample $i \in \{1, \ldots, n_{\text{mix}}\}$,
\[
z_{\text{mix},i} \approx Z_{\text{ref}} \, p_i,
\quad
p_i \in \mathbb{R}^{k}, \quad p_i \geq 0, \quad \sum_{c=1}^{k} p_{ic} = 1,
\]
where $z_{\text{mix},i}$ is the $i$-th column of $Z_{\text{mix}}$ and $p_i$ is 
the vector of cell-type proportions for sample $i$.

\paragraph{Non-linear kernel embedding (\texttt{MKL}).}
In this case, the integration operator $\mathcal{F}_{\text{kernel}}$ constructs a 
non-linear joint representation of RNA-seq and DNAm data using kernel-based feature 
mapping, implemented via the \textbf{mixKernel} package~\cite{mariette_unsupervised_2018}.
First, mixture and reference samples are concatenated within each modality:
\[
W^{\text{RNA}} =
\left[
Y^{\text{RNA}}, X^{\text{RNA}}
\right], \quad
W^{\text{DNAm}} =
\left[
Y^{\text{DNAm}}, X^{\text{DNAm}}
\right] \in \mathbb{R}^{p \times (n_{\text{mix}} + k)},
\]
where columns correspond to samples, $p$ denotes the number of features, and 
$k$ denotes the number of reference cell types, with one reference sample per 
cell type (i.e.\ $n_{\text{ref}} = k$).
An abundance kernel is then computed independently for each modality:
\[
K^{\text{RNA}}_{ij} = \kappa\!\left(w_i^{\text{RNA}}, w_j^{\text{RNA}}\right),
\quad
K^{\text{DNAm}}_{ij} = \kappa\!\left(w_i^{\text{DNAm}}, w_j^{\text{DNAm}}\right),
\quad
K^{\text{RNA}}, K^{\text{DNAm}} \in \mathbb{R}^{(n_{\text{mix}}+k) \times (n_{\text{mix}}+k)},
\]
where $w_i^{(\cdot)}$ denotes the $i$-th column of the corresponding matrix and 
$\kappa(\cdot, \cdot)$ denotes the abundance kernel function.
The modality-specific kernels are then combined into a single multi-omic kernel 
via STATIS-UMKL optimal weighting~\cite{mariette_unsupervised_2018}:
\[
K = \mathcal{C}\left(K^{\text{RNA}}, K^{\text{DNAm}}\right),
\quad K \in \mathbb{R}^{(n_{\text{mix}}+k) \times (n_{\text{mix}}+k)},
\]
where $\mathcal{C}(\cdot)$ denotes the STATIS-UMKL kernel aggregation operator.
A low-dimensional latent representation is then obtained via kernel PCA:
\[
Z = \mathcal{F}_{\text{KPCA}}(K) - \min_{i,j} \left[\mathcal{F}_{\text{KPCA}}(K)\right]_{ij},
\quad
Z \in \mathbb{R}^{d \times (n_{\text{mix}} + k)},
\]
where $d=10$ is the number of retained components and the shift ensures $Z \geq 0$ 
coordinate-wise.
The latent representation is finally partitioned into mixture and reference components:
\[
Z =
\left[
Z_{\text{mix}}, Z_{\text{ref}}
\right],
\quad
Z_{\text{mix}} \in \mathbb{R}^{d \times n_{\text{mix}}}, \;
Z_{\text{ref}} \in \mathbb{R}^{d \times k}.
\]
Deconvolution is performed sample-wise in the aligned latent space: for each 
mixture sample $i \in \{1, \ldots, n_{\text{mix}}\}$,
\[
z_{\text{mix},i} \approx Z_{\text{ref}} \, p_i,
\quad
p_i \in \mathbb{R}^{k}, \quad p_i \geq 0, \quad \sum_{c=1}^{k} p_{ic} = 1,
\]
where $z_{\text{mix},i}$ is the $i$-th column of $Z_{\text{mix}}$ and $p_i$ is
the vector of cell-type proportions for sample $i$.

\paragraph{Optimal transport-based representation (\texttt{OT}).}

In this case, the integration operator $\mathcal{F}_{\text{OT}}$ aligns RNA-seq and DNAm
profiles using uniPort~\cite{cao_unified_2022}, a variational autoencoder (VAE)
that enforces cross-modal alignment via entropic optimal transport in the latent space.
We consider concatenated RNA and DNAm datasets:
\[
W^{\text{RNA}} =
\left[ Y^{\text{RNA}}, X^{\text{RNA}} \right],
\quad
W^{\text{DNAm}} =
\left[ Y^{\text{DNAm}}, X^{\text{DNAm}} \right] \in \mathbb{R}^{p \times (n_{\text{mix}} + k)},
\]
where columns correspond to samples, $p$ denotes the number of features, and 
$k$ denotes the number of reference cell types, with one reference sample per 
cell type (i.e.\ $n_{\text{ref}} = k$).

Each column $w_i^{\text{RNA}}$ of $W^{\text{RNA}}$ (resp.\ $w_j^{\text{DNAm}}$
of $W^{\text{DNAm}}$) is encoded by the VAE encoder $q_\phi$ into a latent vector:
\[
h_i^{\text{RNA}} = q_\phi(w_i^{\text{RNA}}),
\quad
h_j^{\text{DNAm}} = q_\phi(w_j^{\text{DNAm}}).
\]
Let $\mu_{\text{RNA}}$ and $\mu_{\text{DNAm}}$ denote the empirical distributions
induced by $\{h_i^{\text{RNA}}\}$ and $\{h_j^{\text{DNAm}}\}$.
The entropic optimal transport plan is computed in this latent space:
\[
\pi^\star =
\arg\min_{\pi \in \Pi(\mu_{\text{RNA}}, \mu_{\text{DNAm}})}
\sum_{i,j}
\pi_{ij} \, d\!\left(h_i^{\text{RNA}}, h_j^{\text{DNAm}}\right)
+ \varepsilon \sum_{i,j} \pi_{ij} \log \pi_{ij}.
\]
The regularization parameter $\varepsilon$ is set to the 
default value of uniPort~\cite{cao_unified_2022}, and the model 
is trained for 500 iterations with a batch size adapted to the 
number of available samples ($\text{batch\_size} = \min(50, 
n_{\text{mix}} + n_{\text{ref}})$).
The transport plan $\pi^\star$ is then used to realign the encoder outputs across
modalities, yielding a joint latent representation:
\[
Z = \mathcal{F}_{\text{OT}}\left(W^{\text{RNA}}, W^{\text{DNAm}}\right),
\quad Z \in \mathbb{R}^{d \times (n_{\text{mix}} + k)},
\]
where $Z$ aggregates the OT-aligned encodings of all samples from both modalities.

The latent representation is then partitioned into mixture and reference components:
\[
Z = \left[ Z_{\text{mix}}, Z_{\text{ref}} \right],
\quad
Z_{\text{mix}} \in \mathbb{R}^{d \times n_{\text{mix}}}, \;
Z_{\text{ref}} \in \mathbb{R}^{d \times k},
\]
where $k$ denotes the number of reference cell types.
Deconvolution is performed sample-wise in the aligned latent space: for each 
mixture sample $i \in \{1, \ldots, n_{\text{mix}}\}$,
\[
z_{\text{mix},i} \approx Z_{\text{ref}} \, p_i,
\quad
p_i \in \mathbb{R}^{k}, \quad p_i \geq 0, \quad \sum_{c=1}^{k} p_{ic} = 1,
\]
where $z_{\text{mix},i}$ is the $i$-th column of $Z_{\text{mix}}$ and $p_i$ is 
the vector of cell-type proportions for sample $i$.

\subsubsection{Late integration strategies} 
\label{suppmat:integrationLI}

\begin{table}[!ht]
\centering
\small
\caption{Comparison of late integration strategies for multi-omics deconvolution. Each method combines modality-specific deconvolution outputs via an aggregation operator $\mathcal{F_\text{late}}$.
}
\label{supptab:late_integration_summary}

\begin{tabularx}{\linewidth}{p{1.8cm} p{3cm} p{5.2cm} p{5cm}}
\toprule
\textbf{Method} & \textbf{Principle} & \textbf{Key idea} & \textbf{Main limitation} \\
\midrule

%Only RNA / Only DNAm &
%Single-modality inference &
%Performs deconvolution using a single omic independently &
%Ignores complementary information across modalities \\

\texttt{Mean} &
Uniform averaging &
Averages RNA- and DNAm-based proportion estimates with equal weights &
Assumes equal reliability of both modalities \\
\midrule
\texttt{wMean} &
Error-weighted aggregation &
Weights modality-specific estimates using normalized reconstruction errors &
Depends on accurate RMSE estimation; sensitive to correlation in reference profiles \\
\midrule
\texttt{PriorMean} &
Rule-based selective averaging &
Selects or combines modalities depending on biological context (e.g., cancer vs normal) &
Requires hand-crafted decision rules; requires prior biological knowledge \\

\bottomrule
\end{tabularx}
\end{table}

\paragraph{Uniform averaging (\texttt{Mean}).}
Cell-type proportions are first estimated independently from RNA-seq and DNAm data, yielding $\hat{p}_i^{\text{RNA}}$ and $\hat{p}_i^{\text{DNAm}}$ for sample $i$.

The final estimate is obtained by averaging the two modality-specific predictions:
\[
\hat{p}_i = \frac{1}{2} \left( \hat{p}_i^{\text{RNA}} + \hat{p}_i^{\text{DNAm}} \right).
\]

This strategy assumes equal contribution and reliability of both modalities.

\paragraph{Error-weighted aggregation (\texttt{wMean}).}

For each modality $m \in \{\text{RNA}, \text{DNAm}\}$, the observed bulk signal 
and its reconstruction are first normalized by their respective column sums:
\[
\tilde{Y}^{(m)} = \frac{Y^{(m)}}{\mathbf{1}^\top Y^{(m)}}, 
\quad
\hat{\tilde{Y}}^{(m)} = \frac{X^{(m)} \hat{P}^{(m)}}{\mathbf{1}^\top X^{(m)} \hat{P}^{(m)}},
\]
where $\hat{P}^{(m)} = [\hat{p}_1^{(m)}, \ldots, \hat{p}_{n_{\text{mix}}}^{(m)}]$ 
denotes the matrix of estimated proportions for all samples.
A global reconstruction error is then computed over all samples and features, on the full (unprocessed) feature space:
\[
\text{RMSE}_m = \sqrt{ \frac{1}{G_m \cdot n_{\text{mix}}} 
\sum_{i=1}^{n_{\text{mix}}} \left\| \tilde{Y}_i^{(m)} - 
\hat{\tilde{Y}}_i^{(m)} \right\|_2^2 },
\]
where $G_m$ denotes the number of features in modality $m$ and 
$\tilde{Y}_i^{(m)}$ is the $i$-th column of $\tilde{Y}^{(m)}$.
Normalized weights are then computed inversely proportional to each modality's 
reconstruction error:
\[
w_{\text{RNA}} = \frac{\text{RMSE}^{\text{DNAm}}}{\text{RMSE}^{\text{RNA}} + 
\text{RMSE}^{\text{DNAm}}}, \quad
w_{\text{DNAm}} = \frac{\text{RMSE}^{\text{RNA}}}{\text{RMSE}^{\text{RNA}} + 
\text{RMSE}^{\text{DNAm}}}.
\]
The final estimate is obtained as a weighted combination of the two modalities:
\[
\hat{p}_i = w_{\text{RNA}} \hat{p}_i^{\text{RNA}} + w_{\text{DNAm}} 
\hat{p}_i^{\text{DNAm}}.
\]
This strategy assigns higher weight to the modality with lower reconstruction 
error, reflecting its greater reliability.

Note that, in our workflow implementation of \texttt{wMean} and \texttt{RLRnnls}, the sample-wise normalisation of the profiles before computing the reconstruction error is only approximate (the sample sums are recycled over the rows instead of being applied to each column).

\paragraph{Rule-based selective averaging (\texttt{PriorMean}).}

A baseline estimate is obtained by averaging both modalities:
\[
\tilde{p}_i = \frac{1}{2} \left( \hat{p}_i^{\text{RNA}} + \hat{p}_i^{\text{DNAm}} \right).
\]
To account for modality-specific reliability, a subset of cell types 
$\mathcal{C}_{\text{tumour}} = \{\text{basal}, \text{classic}\}$ 
is estimated using DNAm data only. The final estimate is thus defined as:
\[
\hat{p}_{ic} =
\begin{cases}
\hat{p}_{ic}^{\text{DNAm}} & \text{if } c \in \mathcal{C}_{\text{tumour}}, \\
\tilde{p}_{ic} & \text{otherwise}.
\end{cases}
\]
Finally, proportions are renormalized to satisfy the sum-to-one constraint:
\[
\hat{p}_i = \frac{\hat{p}_i}{\sum_{c=1}^k \hat{p}_{ic}}.
\]

\subsubsection{Uni-omic strategies}

\paragraph{\texttt{onlyDNAm}.}

The final estimate is obtained by selecting the DNAm-based proportions only:
\[
\hat{p}_i = \hat{p}_i^{\text{DNAm}}.
\]

\paragraph{\texttt{onlyRNA}.}

The final estimate is obtained by selecting the RNA-based proportions only:
\[
\hat{p}_i = \hat{p}_i^{\text{RNA}}.
\]

\subsection{Detailed visualisation of performance metrics}

\begin{figure}[p]%
\centering
\includegraphics[width=\linewidth]{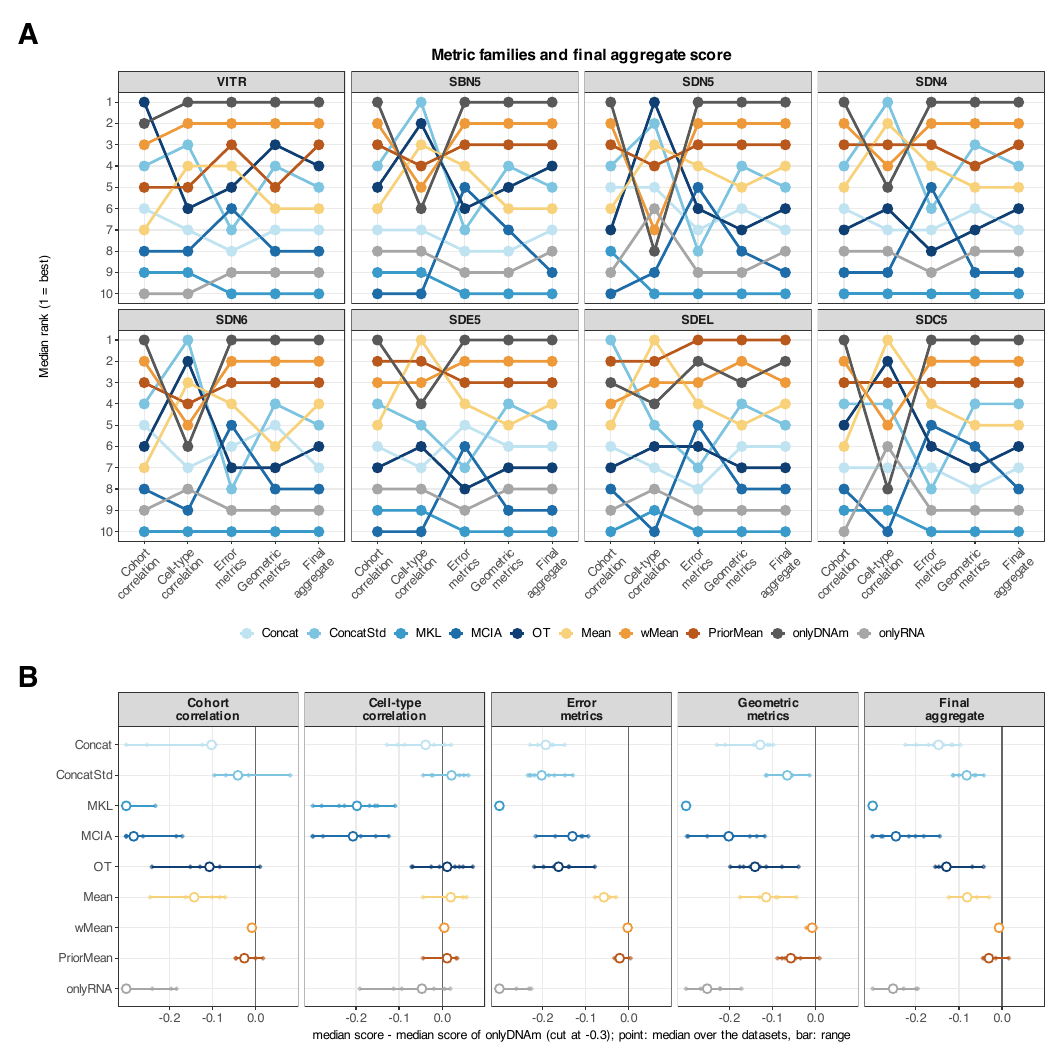}
\caption{\textbf{Strategy rankings within each metric family and for the aggregate score.} In each fully-annotated dataset (VIVO excluded), strategies are ranked separately within each of the four metric families, which contribute equally to the aggregate score: cohort correlation (intra-sample), cell-type correlation (inter-sample), error metrics and geometric metrics, each computed as the geometric mean of its normalized metrics. Rankings under the aggregate score of the twelve metrics are shown for comparison.
\textbf{A.} Rank of each strategy (1 = best), based on its median score over all its pipelines, in each metric family and for the aggregate score; one panel per dataset.
\textbf{B.} Magnitude of the differences underlying these ranks: for each metric family (panels) and strategy (rows), median score of the strategy minus median score of \texttt{onlyDNAm} in the same dataset (vertical line at 0: \texttt{onlyDNAm}). Small points: individual datasets; bar: range across datasets; open circle: median across datasets. Differences below $-0.3$ are drawn at $-0.3$.
For cell-type correlation, the best strategies lie within a few hundredths of \texttt{onlyDNAm}, so that small differences are enough to reorder them in \textbf{A}; in the other metric families, differences are larger and \texttt{onlyDNAm} ranks first in most datasets.}
\label{suppfig:rank4}
\end{figure}

To assess whether onlyDNAm's overall advantage holds uniformly across metric families, we decomposed the final aggregate score into its four constituent families and examined how each method's rank evolves across these families and the final aggregate, for the 8 fully-annotated datasets (Supplementary Figure~\ref{suppfig:rank4}).

\begin{figure}[p]%
\centering
\includegraphics[width=\linewidth]{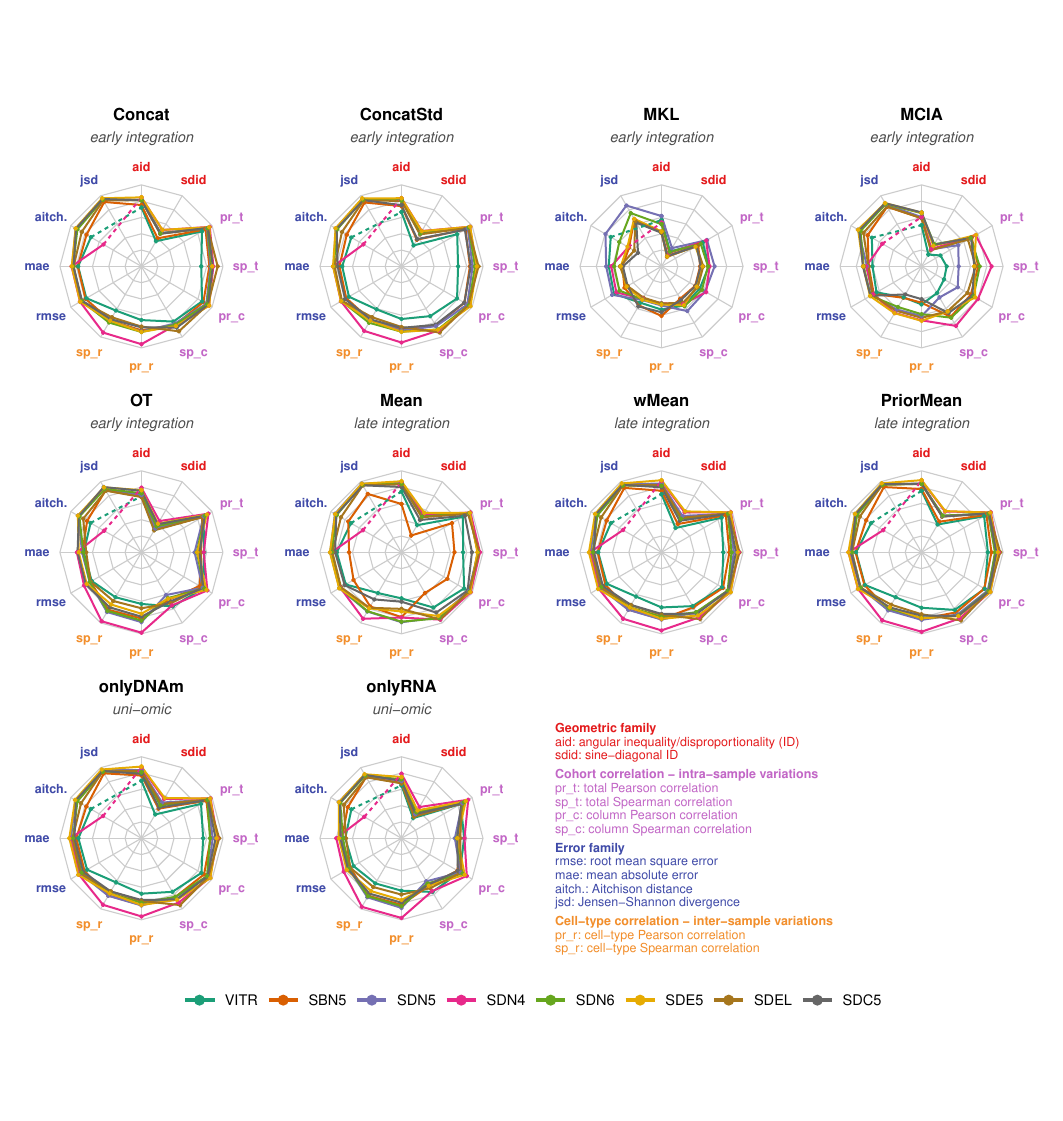}
\caption{\textbf{Per-metric scores of the best pipeline of each strategy.} Radar plots showing per-metric scores of the best-performing pipeline for each integration strategy, on each dataset (except VIVO, which follows a specific scoring procedure).}
\label{suppfig:radar}
\end{figure}

To visualize how performance varies jointly across all twelve individual metrics rather than only through the aggregate score or its four families, we plot the per-metric scores of the best-performing pipeline for each integration strategy on each dataset (Supplementary Figure~\ref{suppfig:radar}).

\begin{figure}[p]%
\centering
\includegraphics[width=\linewidth]{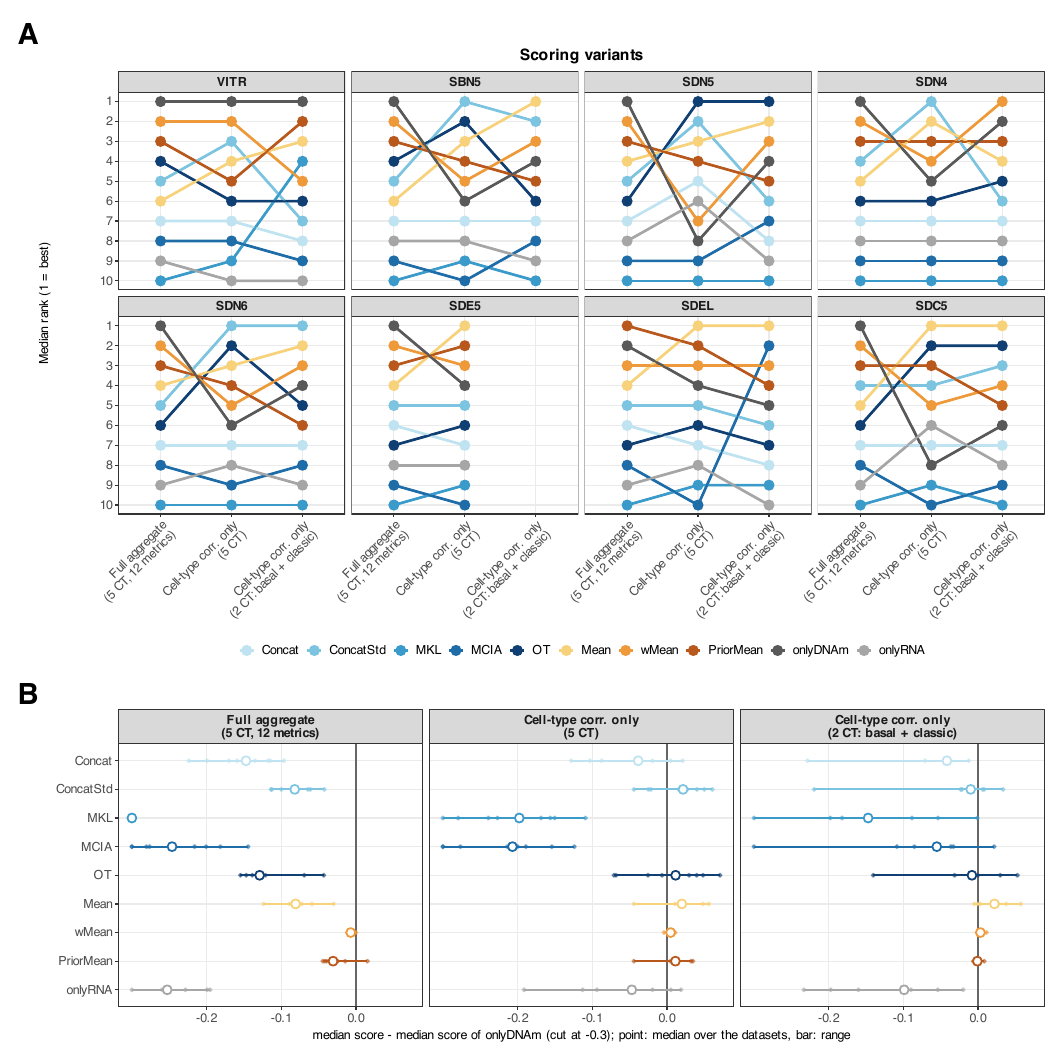}
\caption{\textbf{Strategy rankings under alternative scoring schemes.} Strategies are ranked in each fully-annotated dataset (VIVO excluded) under three scores: the aggregate score of the twelve metrics, cell-type correlation on all five cell types, and cell-type correlation restricted to the two tumour subtypes (basal-like and classical). Comparing the first two isolates the effect of the metric family; comparing the last two isolates the effect of the number of cell types.
\textbf{A.} Rank of each strategy (1 = best), based on its median score over all its pipelines, under each score; one panel per dataset.
\textbf{B.} Magnitude of the differences underlying these ranks: for each score (panels) and strategy (rows), median score of the strategy minus median score of \texttt{onlyDNAm} in the same dataset (vertical line at 0: \texttt{onlyDNAm}). Small points: individual datasets; bar: range across datasets; open circle: median across datasets. Differences below $-0.3$ are drawn at $-0.3$.}
\label{suppfig:rankcor}
\end{figure}

To investigate further the ranking of VIVO, we compared method rankings under the full 12-metric aggregate, the cell-type correlation component alone (still computed on all 5 cell types), and the same cell-type correlation formula restricted to the 2 cell types validated in VIVO (Supplementary Figure~\ref{suppfig:rankcor}). Restricting cell-type correlation from 5 to 2 cell types visibly reshuffles method rankings on several datasets, in a manner distinct from the shift already observed when moving from the full aggregate to cell-type correlation alone. This suggests that VIVO's atypical ranking is not attributable solely to its use of a correlation-only scoring formula, but is further shaped by the specific restriction to the two clinically validated cell types.

\newpage

\subsection{Statistical testing of the median effect}
\label{suppmat:median}

%\subsubsection*{ANOVA modelisation}
%\label{suppmat:anova}

%To quantify the contribution of each module to overall performance, we fitted the 
%following linear model:
%\begin{equation}
 %   s_{ijkl} = \mu + \alpha_i^{\text{prep}} + \beta_j^{\text{fs}} + 
  %  \gamma_k^{\text{deconv}} + \delta_l^{\text{int}} + \varepsilon_{ijkl},
   % \label{eq:anova}
%\end{equation}
%where $s_{ijkl}$ denotes the aggregate score of the combination defined by 
%pre-processing method $i$, feature selection method $j$, deconvolution method $k$, 
%and integration method $l$, and $\varepsilon_{ijkl}$ is a residual term.

%This linear decomposition of aggregate scores by module (Supplementary Methods~\ref{suppmat:anova}, Figure~\ref{fig:interection}) reveals that the dominant factor differs between regimes: the deconvolution algorithm drives late integration performance, while the integration method dominates early integration, largely because latent embedding approaches (\texttt{MCIA}, \texttt{MKL}) perform poorly and inconsistently.

\subsubsection{Bootstrap confidence interval on the median aggregate score}
\label{suppmat:bootstrap}

For the median barplot, we computed, for each (dataset, method) pair, the median aggregate score across all evaluated pipeline configurations within that group. To characterize the stability of this median, we bootstrapped it by resampling, with replacement, the same number of pipeline configurations within each group (2,000 resamples) and recomputing the median each time, error bars show the 2.5th-97.5th percentiles of the resulting bootstrap distribution. Because evaluated pipelines share modules by construction (e.g., many pipelines differ only in a single module) and are therefore not independent draws from an underlying population, this interval should be interpreted as a measure of how stable the median is across the specific space of pipelines we evaluated, rather than a sampling-uncertainty confidence interval over the finite set of biological samples in each dataset.

\subsubsection{Paired Wilcoxon signed-rank test for median aggregate score}
\label{suppmat:wilcoxon}

To test whether the median advantage of \texttt{onlyDNAm} is statistically robust, we held fixed the one pipeline dimension shared by every method: the DNAm-side pre-processing × feature-selection configuration (3×5=15 configurations), and summarized each method's performance for each configuration as the median aggregate score, marginalized over that method's RNA-side choices and its deconvolution (joint, or RNA and DNAm for late integration). This marginalization was necessary because early integration applies a single joint deconvolution step to concatenated features, whereas late integration applies independent per-modality deconvolution steps; these are not the same variable and cannot be forced into a common matching key. For each of the 15 matched configurations, we computed the paired difference in score (integration - DNAm-only) and tested it with a paired Wilcoxon signed-rank test, applying Bonferroni correction within each dataset across its 8 comparisons. We report the effect size as the Hodges-Lehmann estimator (the median of all pairwise averages of the 15 paired differences) together with its associated 95\% confidence interval.

\subsection{Effect of the module options}
\label{suppmat:module}

Let $\mathcal{D}$ be the set of the eight datasets scored with the full set of metrics (VIVO excluded, as its score is computed with a different formula), $d \in \mathcal{D}$, $|\mathcal{D}| = 8$. For a paradigm (early or late integration), $\mathcal{M}$ denotes its integration strategys (early: \texttt{Concat}, \texttt{ConcatStd}, \texttt{MKL}, \texttt{MCIA}, \texttt{OT}; late: \texttt{Mean}, \texttt{wMean}, \texttt{PriorMean}). A pipeline $p$ is defined by its integration strategy $m(p) \in \mathcal{M}$ and by the option $o_j(p)$ it uses for each module $j$ (RNA and DNAm pre-processing, RNA and DNAm feature selection, and deconvolution, split into an RNA and a DNAm module for late integration). We write $s_{p,d}$ for the aggregate score of pipeline $p$ in dataset $d$, and $\bar{s}_p = |\mathcal{D}|^{-1} \sum_{d} s_{p,d}$ for its mean score over the datasets, which is used for display. For a set of pipelines $\mathcal{P}$, $\mathcal{P}_m = \{p \in \mathcal{P} : m(p) = m\}$, and $\mathcal{P}_{m,j=a}$ the pipelines of method $m$ using option $a$ for module $j$.

All tests used the dataset as statistical unit. For each effect, a per-dataset statistic $x_d$ was computed, and the eight values $\{x_d\}_{d \in \mathcal{D}}$ were tested with a two-sided exact Wilcoxon signed-rank test of $H_0$: the distribution of $x_d$ is symmetric around $0$. P-values were adjusted with the Benjamini--Hochberg (BH) procedure within each family of tests defined below; an adjusted $P < 0.05$ was considered significant. With $|\mathcal{D}| = 8$, the smallest attainable P-value is $2/2^8 \approx 0.0078$, so that a significant effect has the same sign in nearly all datasets: the tests assess the reproducibility of an effect across datasets, its magnitude being given by the colour scales.

%Analyses used the aggregate score of every evaluated pipeline in each of the eight datasets scored with the full set of metrics. VIVO was excluded because its score is computed with a different formula (correlation of two cell types only). Early integration pooled the five early methods (\texttt{Concat}, \texttt{ConcatStd}, \texttt{MKL}, \texttt{MCIA}, \texttt{OT}); late integration pooled \texttt{Mean}, \texttt{wMean} and \texttt{PriorMean}. In the violins and heatmaps, each pipeline is represented by its mean score over the eight datasets. All statistical tests used the dataset as the statistical unit ($n = 8$): within each dataset, pipeline-level differences were summarised by their median, and the eight per-dataset values were tested with a two-sided exact Wilcoxon signed-rank test. P-values were adjusted with the Benjamini--Hochberg (BH) procedure, and adjusted $P < 0.05$ was considered significant. With $n = 8$, the smallest attainable two-sided P-value is $2/2^8 = 0.0078$, so a significant result requires a difference of the same sign in nearly all datasets. The tests therefore assess the reproducibility of a difference across datasets, not its magnitude, which is given by the colour scales and medians.

\subsubsection{Percentage of explained variance} For each paradigm, pipeline scores were modelled with the linear model $\mathrm{score} \sim \mathrm{dataset} + (\mathrm{method} + \mathrm{PP_{RNA}} + \mathrm{PP_{DNAm}} + \mathrm{FS_{RNA}} + \mathrm{FS_{DNAm}} + \mathrm{Deconv})^2$, including all main effects and two-way interactions (the deconvolution being split into an RNA and a DNAm factor for late integration). 

For each paradigm, the scores $s_{p,d}$ of all pipelines and datasets were modelled with the linear model
\[
s_{p,d} = \mu + \delta_d + \sum_{f \in \mathcal{F}} \beta_{f}\big(f(p)\big) + \sum_{f < f' \in \mathcal{F}} \gamma_{f f'}\big(f(p), f'(p)\big) + \varepsilon_{p,d},
\]
where $\delta_d$ is a dataset effect, $\mathcal{F}$ the set of factors (the integration strategy and the modules), $\beta_f$ the main effect of factor $f$ and $\gamma_{ff'}$ the two-way interaction between factors $f$ and $f'$. 

The share of variance of each term was computed as its sum of squares divided by the total sum of squares ($\eta^2$, sequential analysis of variance; the nearly complete factorial design makes the shares independent of the order of the terms). For each term $t$, the share of variance is
\[
\eta^2_t = \frac{\mathrm{SS}_t}{\mathrm{SS}_{\mathrm{total}}},
\]
where $\mathrm{SS}_t$ is the sequential (type I) sum of squares of the term and $\mathrm{SS}_{\mathrm{total}} = \sum_{p,d} (s_{p,d} - \bar{s})^2$. Because the pipelines form a nearly complete factorial design, the factors are close to orthogonal and $\eta^2_t$ does not depend on the order of the terms. Ninety-five percent intervals were obtained from $B = 100$ bootstrap resamplings of the datasets: $\mathcal{D}$ was resampled with replacement (a dataset drawn several times being treated as distinct), the model refitted, and the 2.5th and 97.5th percentiles of $\eta^2_t$ reported. No P-values were computed for this model, given the large number of non-independent scores.

\subsubsection{Options within each integration strategy } For each integration strategy, module and option, the colour shows the median score of the pipelines using this option minus the median score of all pipelines of the method. 

For method $m$, module $j$ and option $a$, the colour shows
\[
\Delta_{m,j,a} = \operatorname*{median}_{p \in \mathcal{P}_{m,j=a}} \bar{s}_p \;-\; \operatorname*{median}_{p \in \mathcal{P}_{m}} \bar{s}_p ,
\]
the deviation of the pipelines using option $a$ from all pipelines of the method (positive: improvement). The test used the per-dataset deviations
\[
x_d = \operatorname*{median}_{p \in \mathcal{P}_{m,j=a}} s_{p,d} \;-\; \operatorname*{median}_{p \in \mathcal{P}_{m}} s_{p,d},
\]
with BH correction over all the cells $(m, j, a)$ of the panel.

\subsubsection{RNA $\times$ DNAm interactions within each method.} 

\paragraph{Statistical approach} For method $m$ and a pair of modules $(j_R, j_M)$ acting on the RNA and the DNAm respectively (pre-processing, feature selection and, for late integration, deconvolution), let
\[
\mu^{d}_{ik} = \operatorname*{mean}_{\substack{p \in \mathcal{P}_m \\ o_{j_R}(p) = i,\; o_{j_M}(p) = k}} s_{p,d},
\qquad
\mu^{d}_{i\cdot}, \quad \mu^{d}_{\cdot k}, \quad \mu^{d}_{\cdot\cdot}
\]
be the mean score in dataset $d$ of the pipelines of method $m$ that use the RNA option $i$ and the DNAm option $k$, the RNA option $i$ (any DNAm option), the DNAm option $k$ (any RNA option), and any option, respectively. The interaction contrast of the combination $(i,k)$ in dataset $d$ is
\[
I^{d}_{ik} = \mu^{d}_{ik} - \mu^{d}_{i\cdot} - \mu^{d}_{\cdot k} + \mu^{d}_{\cdot\cdot}.
\]
$I^{d}_{ik} = 0$ when the effect of the combination equals the sum of the effects of its two options (additivity); $I^{d}_{ik} > 0$ (respectively $< 0$) indicates that the combination performs better (respectively worse) than expected from the two options separately. Means are used rather than medians because $I^{d}_{ik}$ is an exact decomposition of a two-way interaction only for means: in a complete design, the contrasts sum to zero over each row and each column, $\sum_i I^{d}_{ik} = \sum_k I^{d}_{ik} = 0$, which does not hold for medians when scores are multimodal, as for early integration. The colour shows $\tilde{I}_{ik} = \operatorname{median}_{d} I^{d}_{ik}$, and the test used $x_d = I^{d}_{ik}$, with BH correction over the cells $(i,k)$ of each heatmap.

\begin{figure}[!ht]%
\centering
\includegraphics[width=\linewidth]{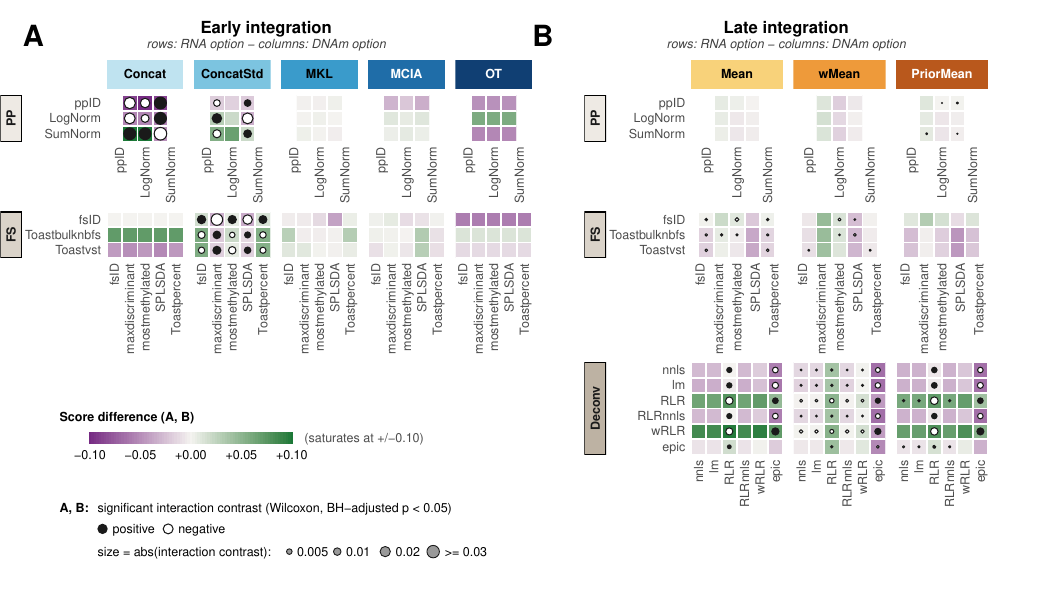}
\caption{\textbf{Combinations of RNA and DNAm options within each integration strategy.}
Each heatmap crosses the options of one module for the RNA (rows) with the options of the same module for the DNAm (columns), within one integration strategy: pre-processing (PP) and feature selection (FS) for the early integration strategys (A), plus the deconvolution of each side (Deconv) for the late integration strategys (B). Pipeline scores are the aggregate score averaged over the the eight fully-annotated datasets (whose aggregate score uses a different formula); the statistical tests use the dataset as the unit (8 datasets).
Colour: median score of the pipelines of the method that use the combination, minus the median score of all the pipelines of the method (green: better than the method median; purple: worse; colours saturate at ±0.10).
Circles: significant interaction between the RNA option and the DNAm option, i.e. a combination that departs from the sum of the effects of its two options. In each dataset, the interaction contrast is mean(combination) - mean(RNA option) - mean(DNAm option) + mean(method); the 8 per-dataset contrasts are tested with a two-sided Wilcoxon signed-rank test, Benjamini-Hochberg adjusted over the cells of the heatmap (adjusted p < 0.05). Filled circle: positive interaction (the combination does better than expected from its two options); open circle: negative interaction. Circle area: absolute value of the interaction contrast (median over the datasets); the largest circle corresponds to contrasts of 0.03 or more. No circle: no significant interaction.}
\label{suppfig:interaction}
\end{figure}

\paragraph{Interpretation} In early integration, RNA $\times$ DNAm interactions are confined to the two concatenation methods (Supplementary Figure~\ref{suppfig:interaction}A) and follow a single mechanism: because both modalities are stacked in one matrix and fitted by a single regression, the modality whose features vary most dominates the fit, and pre-processing and feature selection determine which one it is. Without transformation (\texttt{Concat}), RNA counts exceed DNAm beta values by two to three orders of magnitude and drive the fit. Only \texttt{SumNorm} RNA, which shrinks RNA values below those of DNAm, leaves the DNAm in control; these combinations are the only ones that perform well, and normalising the DNAm as well cancels the gain (interaction $-0.08$). With \texttt{ConcatStd}, the per-sample Gaussian transformation bounds all values in $[0,1]$, so that neither modality is erased, but the same logic holds with smaller effects: DNAm-dominated combinations perform best (+0.05 to +0.07). The weight of each modality then also depends on the number of features kept on each side. Because most CpGs vary little between cell types, keeping all genes lets the RNA dominate even against all CpGs and lowers performance, whereas a 1,000-gene selection leaves the DNAm in control when many CpGs are kept (\texttt{fsID}, \texttt{Toastpercent}). \texttt{MKL}, \texttt{MCIA} and \texttt{OT}, which normalise each modality separately before combining them, show no significant interaction.

In late integration, RNA and DNAm choices are almost additive (Supplementary Figure~\ref{suppfig:interaction}B). The RNA deconvolution sets the level of performance for \texttt{Mean} and \texttt{PriorMean}, whereas for \texttt{wMean}, which down-weights a poor RNA side, the DNAm deconvolution matters most. Pre-processing and feature selection interactions are negligible (all contrasts below 0.002), and deconvolution interactions remain small (compared with the main effect of the RNA deconvolution (up to 0.11); they do not change the optimal choice of robust regression for both modalities.

\subsection{Top-performing pipelines}

\subsubsection{Best pipelines of each strategy}

\begin{figure}[!ht]
  \begin{center}
  \includegraphics[width=0.9\textwidth]{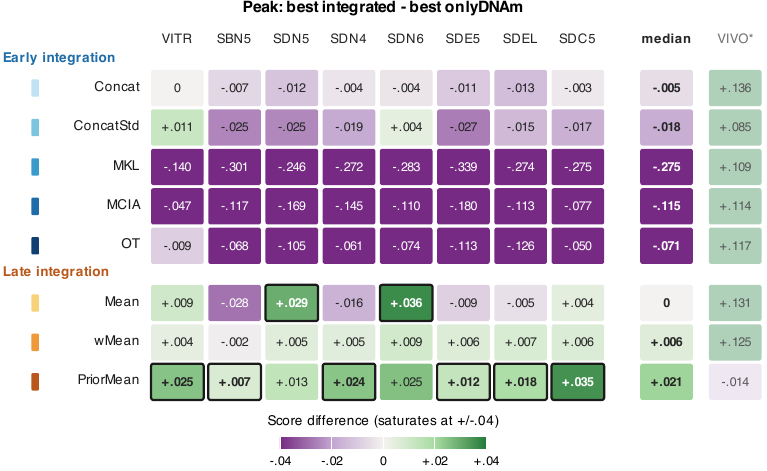}
  \end{center}
  \caption{\textbf{ Peak score of each integration strategy compared with the best onlyDNAm pipeline.} 
  For each integration strategy (rows; early integration above, late integration below) and each dataset (columns), score of the best pipeline of the method in the dataset minus the score of the best onlyDNAm pipeline of the same dataset (green: the integration strategy does better; purple: worse; colours saturate at ±0.04; the values are written in the cells). Outlined, in bold: best integration strategy of each dataset. Median: median of the differences over the 8 benchmark datasets. VIVO* is shown apart and faded: its aggregate score uses a different formula (cell-type correlations only), and it is not included in the median.
The maxima are taken over grids of very different sizes (90 pipelines for onlyDNAm, 810 per early integration strategy and 4,860 per late integration strategy in each dataset), which favours the integration strategys: a difference close to 0 means that the much larger search space does not yield a better peak score.}
  \label{suppfig:peakscore}
\end{figure}

\begin{figure}[p]
  \begin{center}

  \includegraphics[width=0.9\textwidth]{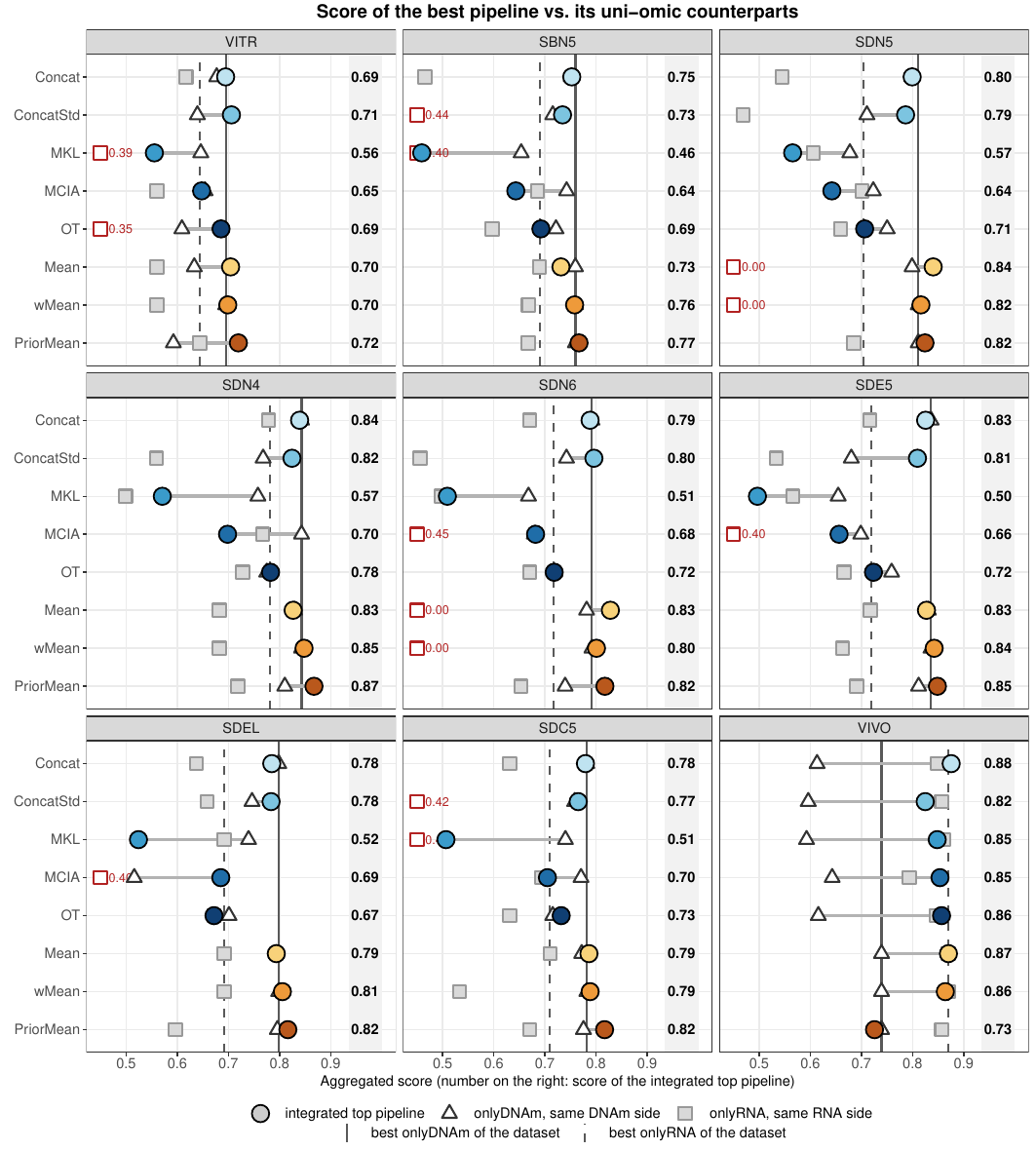}
  \end{center}
  \caption{\textbf{Best pipeline of each integration strategy compared with its uni-omic counterparts.}
One panel per dataset; one row per integration strategy (early integration: Concat to OT; late integration: Mean to PriorMean).
\textbf{Circle:} best pipeline of the method in the dataset; its score is also written in the grey band on the right.
\textbf{Triangle:} onlyDNAm pipeline with the same DNAm side as this pipeline (same pre-processing, feature selection and deconvolution of the DNAm).
\textbf{Square:} onlyRNA pipeline with the same RNA side.
\textbf{Grey segment:} joins the best pipeline to its matched onlyDNAm pipeline; it is the score gained (or lost) by adding the RNA to this DNAm side.
\textbf{Red open square at the left edge:} the matched onlyRNA pipeline scores below the axis (0.45); its score is written next to it. A score of 0.00 means that one normalised metric of this onlyRNA pipeline is at its worst value, which sets the geometric mean to 0. 
\textbf{Solid vertical line:} best onlyDNAm pipeline of the dataset.
\textbf{Dashed vertical line:} best onlyRNA pipeline of the dataset.
A circle to the right of its triangle means that integration improved on the same DNAm side. A circle to the right of the solid line means that the best pipeline of the method beats every onlyDNAm pipeline of the dataset.
VIVO uses a different aggregate score (cell-type correlations of basal-like and classical only), so its panel is not comparable with the others.
}
  \label{suppfig:topuniomic}
\end{figure}

The following figures focus on the best pipeline of each method in each dataset, that is, on what each strategy can reach when its modules are tuned on the dataset itself. Supplementary Figure~\ref{suppfig:peakscore} compares this peak score with the best onlyDNAm pipeline of the same dataset. \texttt{PriorMean} is the only method that reaches or exceeds it in all eight benchmark datasets (median gain +0.021), and \texttt{Mean} does so in SDN5 and SDN6. Supplementary Figure~\ref{suppfig:topuniomic} breaks this comparison down for each best pipeline. It places the pipeline next to the onlyDNAm and onlyRNA pipelines built with the same DNAm and RNA sides, which separates the gain brought by integration from the quality of the uni-omic modules it starts from. Supplementary Figure~\ref{suppfig:bestpip} shows the modules of these best pipelines. Their composition changes from one dataset to the next for most methods, which is why we also evaluated generalist pipelines, applied unchanged to every dataset.

\subsubsection{Stability of top-performing pipelines}
\label{sec:stability_top_pipelines}

\begin{figure}[!ht]
  \begin{center}

  \includegraphics[width=0.9\textwidth]{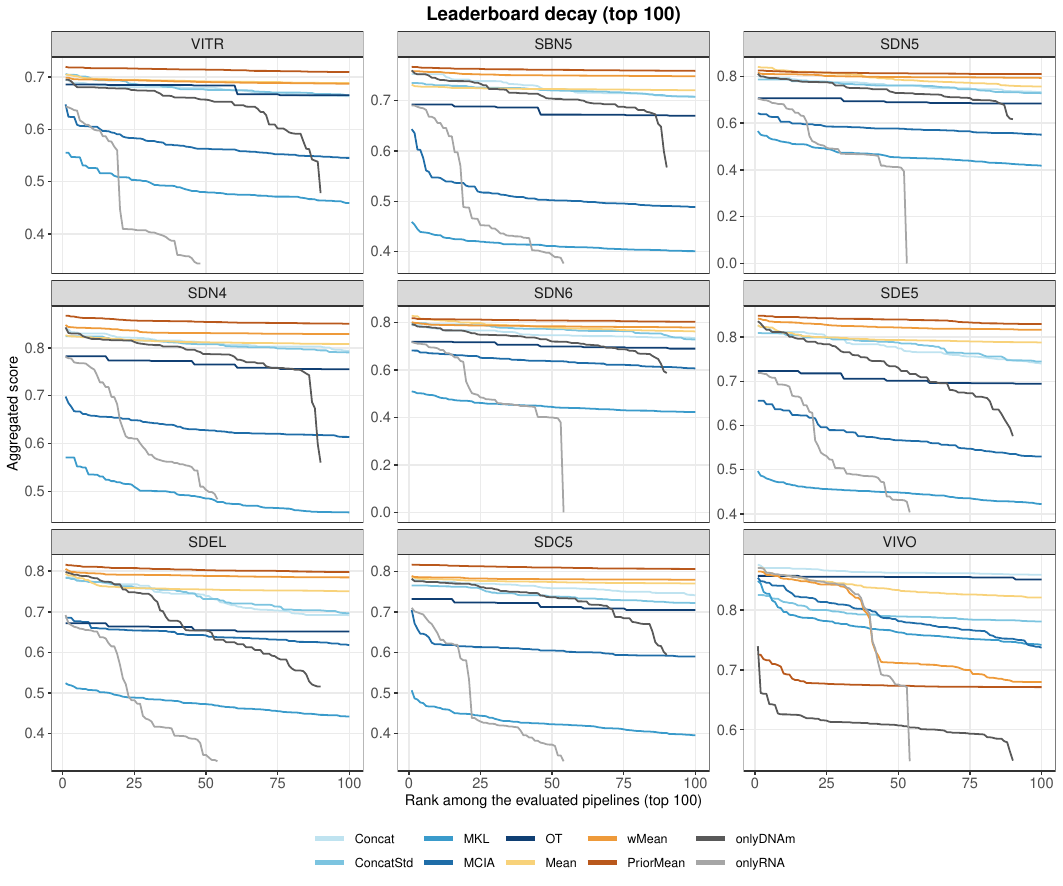}
  \end{center}
  \caption{\textbf{Top 100 Leaderboard-decay curves.} For each dataset and
    integration strategy (including the onlyDNAm and onlyRNA uni-omic
    baselines), evaluated pipelines are ranked by aggregate score
    (rank~1 = best) and plotted against their score, restricted to the
    top 100 ranks. A flat curve indicates many pipelines achieve
    near-equivalent performance to the top-ranked one (robust
    plateau); a steep initial drop indicates the top-ranked pipeline
    is an outlier relative to the rest of the search space.}
  \label{suppfig:leaderboard_decaytop100}
\end{figure}

The single best-performing pipeline highlighted per method and dataset is one pipeline selected from a very large search space; with such a large number of comparisons, some configuration is nearly guaranteed to score well by search-space noise alone, independent of any systematic property of that pipeline. To
assess whether each highlighted ``winner'' reflects a genuinely robust
configuration rather than a lucky draw, we ranked all evaluated
pipelines within each (dataset, method) group by aggregate score and
examined how performance decays beyond rank~1 (Supplementary Figure~\ref{suppfig:leaderboard_decaytop100}). 

\begin{figure}[!ht]
  \begin{center}

  \includegraphics[width=0.9\textwidth]{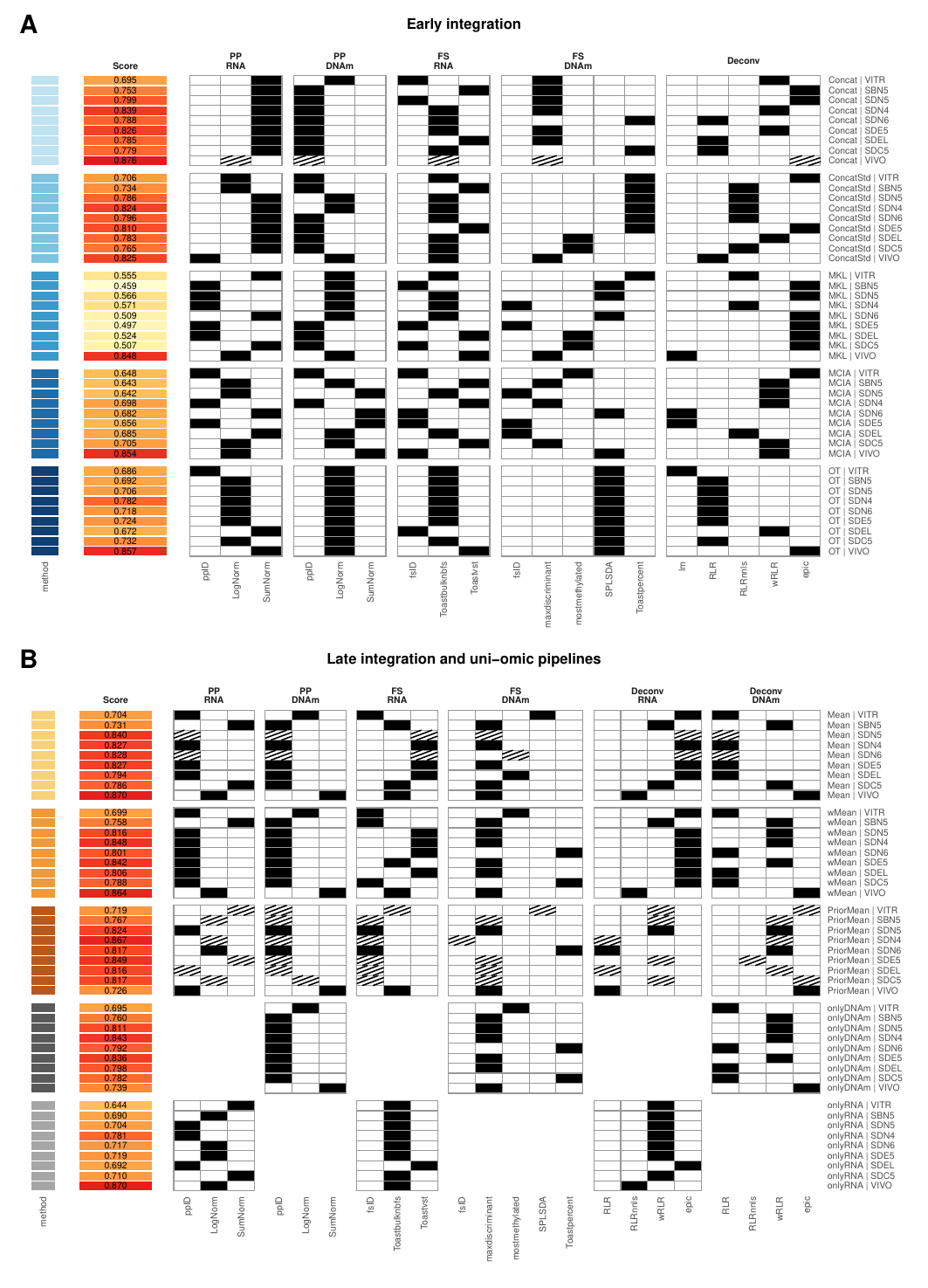}
  \end{center}
  \end{figure}

  \begin{figure}
  \caption{\textbf{Best pipeline of each method, re-optimised in each dataset.}
\textbf{(A)} Early integration strategys. \textbf{(B)} Late integration strategys and uni-omic pipelines (onlyDNAm, onlyRNA).
Each row is the best pipeline of one method in one dataset (label on the right: method \textbar{} dataset), so that the modules can change from one dataset to the next. Unlike the generalist pipelines of Figure~6, these pipelines are re-optimised in each dataset.
\textbf{Left strip:} colour of the method.
\textbf{Score:} aggregate score of the pipeline; the colour scale is the same in A and B and spans the range of the scores shown.
\textbf{Module columns:} pre-processing (PP), feature selection (FS) and deconvolution (Deconv) of each omic. In A, the deconvolution is a single joint deconvolution of the integrated data; in B, each omic has its own deconvolution. A black cell is the option chosen; only the options chosen at least once are shown. A module that does not apply (the RNA side of onlyDNAm, the DNAm side of onlyRNA) is left blank.
\textbf{Hatched cells:} the best pipeline of the dataset over all the methods (early, late and uni-omic).
VIVO uses a different aggregate score (cell-type correlations of basal-like and classical only), so its scores are not comparable with those of the other datasets.
}
  \label{suppfig:bestpip}
\end{figure}

\subsection{Generalist pipelines}
\label{suppmat:fixed_recipes}

The best pipeline of a strategy in a given dataset (Supplementary Figure~\ref{suppfig:bestpip}) is selected on that dataset and is therefore optimistic: its modules change from one dataset to the next, and part of its advantage reflects the selection itself. To compare strategies without this bias, we selected for each strategy a single \emph{generalist pipeline}, i.e.\ one combination of modules (pre-processing, feature selection, deconvolution and integration) applied unchanged to every dataset, and evaluated it both on the benchmark datasets and on data not used for its selection.

Let $\mathcal{D}$ be the eight fully-annotated datasets (VIVO excluded). For each dataset $d$, let $u_d = \max\left(s^{\texttt{onlyDNAm}}_{d}, s^{\texttt{onlyRNA}}_{d}\right)$ be the score of the best uni-omic pipeline of $d$, re-optimised in $d$. The relative gain of a generalist pipeline $r$ in dataset $d$ is $g_{r,d} = (s_{r,d} - u_d)/u_d$, and the generalist pipeline of a strategy is the one that maximises the mean relative gain
\[
\bar{g}_r = \frac{1}{|\mathcal{D}|} \sum_{d \in \mathcal{D}} g_{r,d}
\]
among the pipelines of this strategy evaluated in all eight datasets. Relative gains, rather than raw scores, give each dataset the same weight whatever its difficulty. The same criterion was applied to the five strategies compared in Figure~\ref{fig:validation}: \texttt{onlyDNAm}, \texttt{onlyRNA}, and the best integration strategy of each family, \texttt{ConcatStd} (early), \texttt{Mean} and \texttt{PriorMean} (late).

\section{Computational costs}
\label{suppmat:compucost}

All computations were performed on the GRICAD Kraken high-performance computing cluster (https://gricad.univ-grenoble-alpes.fr), which is supported by Grenoble research communities.
The computational pipeline is structured as a collection of discrete tasks that vary considerably in their resource requirements. Tasks are heterogeneous by design: at one extreme, trivial operations such as returning an identity matrix require negligible computation; at the other, memory-intensive dataframe operations can demand significant CPU time and RAM. This heterogeneity must be kept in mind when interpreting raw task counts.
The minimal pipeline (the smallest complete set of tasks required to reproduce the core results of this paper) consists of 1,073,340 tasks. This figure is a strict lower bound, assuming no failures, reruns, or exploratory work.
In practice, the total number of tasks executed over the course of this project was an order of magnitude larger, reaching at least 11,881,512 tasks. This inflation reflects failed jobs requiring resubmission, parameter sweeps, prototyping runs, and results that were ultimately not included in the paper.
To estimate CPU consumption, we rely on the empirical throughput of the main production runs, in which approximately 6,000,000 tasks were processed over three weeks (504 hours of wall time) across 3 nodes of 192 cores each (a total of 290,304 core-hours). This yields an average of approximately 2.9 CPU-minutes per task (~0.048 core-hours), taken across the full heterogeneous task mix.
Applying this rate, the minimal pipeline requires an estimated ~51,000 core-hours, while the full scope of computation (including failures, prototyping, and non-reported results) accounts for upwards of 575,000 core-hours.

%\newpage

\end{document}